\documentclass[aps,tightenlines,12pt,preprint,showkeys,nofootinbib]{revtex4} 
\newcommand{\ARNPS}[1]{}
\newcommand{\ARXIV}[1]{#1}

\ARXIV{\setcounter{tocdepth}{2}}

\newcommand{\figwidth}{0.65\textwidth} 

\ARNPS{\bibliographystyle{arnuke_revised}}
\ARNPS{\usepackage{amstext}} 
\ARXIV{\bibliographystyle{utphys}} 
\ARXIV{\usepackage{hyperref}}      

\usepackage{graphicx} 

\ARXIV{\setcounter{tocdepth}{2}}

\ARNPS{\newcommand{\alt}{\mathrel{\hbox{\rlap{\hbox{\lower4pt\hbox{$\sim$}}}\hbox{$<$}}}}}
\ARNPS{\newcommand{\agt}{\mathrel{\hbox{\rlap{\hbox{\lower4pt\hbox{$\sim$}}}\hbox{$>$}}}}}

\newcommand{\ttwo}[2]{\begin{minipage}{.9in}\noindent #1 \\
    \vspace*{-.05in} #2\end{minipage}} 

\newcommand{\mweak}{m_{\text{weak}}}
\newcommand{\mgut}{m_{\text{GUT}}}
\newcommand{\mplanck}{M_{\text{Pl}}}

\newcommand{\ifb}{\text{fb}^{-1}}

\newcommand{\mev}{\text{MeV}}
\newcommand{\gev}{\text{GeV}}
\newcommand{\tev}{\text{TeV}}

\newcommand{\eg}{{\em e.g.}}
\newcommand{\ie}{{\em i.e.}}

\newcommand{\Eqref}[1]{Equation~(\ref{#1})}
\newcommand{\eqref}[1]{Eq.~(\ref{#1})}
\newcommand{\eqsref}[2]{Eqs.~(\ref{#1}) and (\ref{#2})}

\newcommand{\secref}[1]{Sec.~\ref{sec:#1}}
\newcommand{\secsref}[2]{Secs.~\ref{sec:#1} and \ref{sec:#2}}
\newcommand{\figref}[1]{Fig.~\ref{fig:#1}}

\newcommand{\tableref}[1]{Table~\ref{table:#1}}

\newcommand{\Bino}{\tilde{B}}
\newcommand{\Wino}{\tilde{W}}
\newcommand{\Higgsino}{\tilde{H}}

\newcommand{\slepton}{\tilde{l}}

\newcommand{\squark}{\tilde{q}}
\newcommand{\gluino}{\tilde{g}}

\newcommand{\caln}{{\cal{N}}}
\newcommand{\calnmax}{{\cal{N}_{\text{max}}}}
\newcommand{\met}{\mbox{\ensuremath{\slash\kern-.7emE_{T}}}}
\newcommand{\rpv}{R_p \!\!\!\!\!\! \slash}
\newcommand{\LambdaQCD}{\Lambda_{\text{QCD}}}

\begin{document}



\ARNPS{\jname{Ann.~Rev.~Nucl.~Part.~Sci.}
\jyear{2013}
\jvol{63}
\ARinfo{AR Info}}

\ARXIV{\preprint{UCI-TR-2013-01}}

\title{\ARXIV{\vspace*{.8in}}Naturalness and the Status of Supersymmetry}

\markboth{Feng}{Naturalness and the Status of Supersymmetry}

\ARXIV{\author{\vspace*{.3in} Jonathan L.~Feng \vspace*{.1in}} 
\affiliation{Department of Physics and Astronomy\\
University of California, Irvine, CA 92697, USA \vspace*{.5in}}}

\ARNPS{\author{Jonathan L.~Feng
\affiliation{Department of Physics and Astronomy, University of 
California, Irvine, CA 92697, USA}} }


\begin{abstract}
\ARXIV{\vspace*{.3in}} For decades, the unnaturalness of the weak
scale has been the dominant problem motivating new particle physics,
and weak-scale supersymmetry has been the dominant proposed solution.
This paradigm is now being challenged by a wealth of experimental
data.  In this review, we begin by recalling the theoretical
motivations for weak-scale supersymmetry, including the gauge
hierarchy problem, grand unification, and WIMP dark matter, and their
implications for superpartner masses. These are set against the
leading constraints on supersymmetry from collider searches, the Higgs
boson mass, and low-energy constraints on flavor and CP violation.  We
then critically examine attempts to quantify naturalness in
supersymmetry, stressing the many subjective choices that impact the
results both quantitatively and qualitatively.  Finally, we survey
various proposals for natural supersymmetric models, including
effective supersymmetry, focus point supersymmetry, compressed
supersymmetry, and $R$-parity-violating supersymmetry, and summarize
their key features, current status, and implications for future
experiments.
\ARXIV{\vspace*{.6in}}
\end{abstract}

\maketitle

\ARXIV{\newpage}

\ARXIV{\tableofcontents}

\section{INTRODUCTION}

Good physical theories are expected to provide {\em natural}
explanations of experimental data and observations.  Although
physicists disagree about the definition of ``natural,'' the idea that
the criterion of naturalness exists and is a useful pointer to deeper
levels of understanding has a long and storied history.  In 1693, for
example, when asked by clergyman Robert Bentley to explain how the law
of universal gravitation was consistent with a static universe, Isaac
Newton wrote~\cite{Newton:1756aa}:
\begin{quotation}
That there should be a central particle, so accurately placed in the
middle, as to be always equally attracted on all sides, and thereby
continue without motion, seems to me a supposition fully as hard as to
make the sharpest needle stand upright on its point upon a
looking-glass.
\end{quotation}
Newton went on to conclude that this unnatural state of affairs could
be taken as evidence for an infinite universe with initial conditions
set by a divine power.  Three hundred years later, the assumption of a
static universe appears quaint, but we are no closer to a natural
explanation of our accelerating universe than Newton was to his static
one.  More generally, the image of a needle balanced upright on a
mirror remains the classic illustration of a possible, but unnatural,
scenario that cries out for a more satisfactory explanation, and the
notion of naturalness continues to play an important role in many
areas of physics.

In particle physics today, the role of naturalness is nowhere more
central than in the statement of the gauge hierarchy problem, the
question of why the weak scale $\mweak \sim 0.1 - 1~\tev$ is so far
below the (reduced) Planck scale $\mplanck = \sqrt{ \hbar c / (8 \pi
  G_N) } \simeq 2.4 \times 10^{18}~\gev$.  For many years, this has
been the dominant problem motivating proposals for new particles and
interactions.  Chief among these is supersymmetry, which solves the
gauge hierarchy problem if there are supersymmetric partners of the
known particles with masses not far above the weak scale.  This has
motivated searches for superpartners at colliders, in low-energy
experiments, and through astrophysical observations.  So far, however,
no compelling evidence for weak-scale supersymmetry has been found,
and recent null results from searches at the Large Hadron Collider
(LHC) have disappointed those who find supersymmetry too beautiful to
be wrong and led its critics to declare supersymmetry dead.

In this article, we review the status of weak-scale supersymmetry at
this brief moment in time when a Higgs-like particle has been
discovered at the 8 TeV LHC, and the LHC has entered a two-year
shutdown period before turning on again at its full center-of-mass
energy.  The field of supersymmetry phenomenology is vast, and we will
necessarily review only a small subset of its many interesting
aspects.  As we will see, however, weak-scale supersymmetry is neither
ravishingly beautiful (and hasn't been for decades), not is it
excluded by any means; the truth lies somewhere in between.  The goal
of this review is to understand the extent to which naturalness and
experimental data are currently in tension and explore models that
resolve this tension and their implications for future searches.

We begin in \secref{theoretical} with a brief discussion of some of
the longstanding theoretical motivations for weak-scale supersymmetry
and their implications for superpartner masses.  We then discuss some
of the leading experimental constraints on weak-scale supersymmetry in
\secref{experimental}.  In \secref{quantifying}, we critically review
attempts to quantify naturalness.  Naturalness is a highly contentious
subject with many different approaches leading to disparate
conclusions.  We will highlight some of the assumptions that are often
implicit in discussions of naturalness and discuss the various
subjective choices that impact the conclusions, both qualitatively and
quantitatively.

With all of these considerations in hand, we then turn in
\secref{models} to an overview of model frameworks that have been
proposed to reconcile naturalness with current experimental
constraints, summarizing their key features, current status, and
implications for future searches.  As a rough guide to the discussion,
these models and the problems they attempt to address are shown in
\tableref{summary}.  We conclude in \secref{conclusions}.

\begin{table}[tbp]
\begin{minipage}{\columnwidth}
\begin{tabular}{lcccccc} \hline\hline
\rule[1mm]{0mm}{5mm}
& \ttwo{Effective}{SUSY} & \ttwo{Focus Point}{SUSY}
& \ttwo{Compressed}{SUSY} &  \ttwo{$R_p$-Violating}{SUSY}
\vspace*{.1in} \\ \hline
\hspace*{.02in} Naturalness \rule[3mm]{0mm}{5mm}
& $\surd$ & $\surd$ & $\surd$ & $\surd$ \vspace*{.2in} \\
\ttwo{Grand}{Unification}
& $\surd$ & $\surd$ & $\surd$ & $\surd$ \vspace*{.12in} \\
\ttwo{WIMP}{Dark Matter}
& $\surd$ & $\surd$ & $\surd$ &         \vspace*{.12in} \\
\ttwo{LHC Null}{Results}
& $\surd$ & $\surd$ & $\surd$ & $\surd$ \vspace*{.12in} \\
\hspace*{.02in} Higgs Mass
&         & $\surd$ &         &         \vspace*{.2in} \\
\ttwo{Flavor/CP}{Constraints}
& $\surd$ & $\surd$ &         &         \vspace*{.12in} \\
\hline \hline
\end{tabular}
\end{minipage}
\caption{Some of the supersymmetric models discussed in this review,
  the virtues they are intended to preserve, and the constraints they
  are designed to satisfy, with varying degrees of success.  For the
  rationale behind the check marks, see Secs.~\ref{sec:theoretical},
  \ref{sec:experimental}, and \ref{sec:models} for discussions of the
  virtues, constraints, and models, respectively.}
\vspace{0.2cm}
\label{table:summary}
\end{table}

\section{THEORETICAL MOTIVATIONS}
\label{sec:theoretical}

To review the status of supersymmetry, we should begin by recalling
the problems it was meant to address.
Supersymmetry~\cite{Golfand:1971iw,Volkov:1973ix,Wess:1974tw} has
beautiful mathematical features that are independent of the scale of
supersymmetry breaking.  In addition, however, there are three
phenomenological considerations that have traditionally been taken as
motivations for weak-scale supersymmetry: the gauge hierarchy problem,
grand unification, and WIMP dark matter.  In this section, we review
these and their implications for superpartner masses.

\subsection{The Gauge Hierarchy Problem}
\label{sec:gaugehierarchyproblem}

\subsubsection{The Basic Idea}

The gauge hierarchy problem of the standard
model~\cite{Weinberg:1975gm,Susskind:1978ms,'tHooft:1980xb}
and its possible resolution through weak-scale
supersymmetry~\cite{Davier:1979hr,Veltman:1980mj,Witten:1981nf,Kaul:1981wp}
are well-known.  (For reviews and discussion, see, \eg,
Refs.~\cite{Martin:1997ns,Drees:2004jm,Baer:2006rs,Dine:2007zp,Vissani:1997ys}.)
The standard model includes a fundamental, weakly-coupled, spin-0
particle, the Higgs boson.  Its bare mass receives large quantum
corrections.  For example, given a Dirac fermion $f$ that receives its
mass from the Higgs boson, the Higgs mass is
\begin{equation}
m_h^2 \approx m_{h\, 0}^2 - \frac{\lambda_f^2}{8 \pi^2} N_c^f  
\int^\Lambda \frac{d^4 p}{p^2} 
\approx m_{h\, 0}^2 + \frac{\lambda_f^2}{8 \pi^2} 
N_c^f \Lambda^2 \ ,
\label{smhiggs}
\end{equation}
where $m_h \approx 125~\gev$ is the physical Higgs boson
mass~\cite{:2012gk,:2012gu}, $m_{h\, 0}$ is the bare Higgs mass, and
the remaining term is $m_{h\, \text{1-loop}}^2$, the 1-loop
correction.  The parameters $\lambda_f$ and $N_c^f$ are the Yukawa
coupling and number of colors of fermion $f$, $\Lambda$ is the largest
energy scale for which the standard model is valid, and subleading
terms have been neglected. For large $\Lambda$, the bare mass and the
1-loop correction must cancel to a large degree to yield the physical
Higgs mass.  Attempts to define naturalness quantitatively will be
discussed in detail in \secref{quantifying}, but at this stage, a
simple measure of naturalness may be taken to be
\begin{equation}
\caln^0 \equiv \frac{m_{h\, \text{1-loop}}^2}{m_h^2} \ .
\end{equation}
For $\Lambda \sim \mplanck$ and the top quark with $\lambda_t \simeq
1$, \eqref{smhiggs} implies $\caln^0 \sim 10^{30}$, \ie, a fine-tuning
of 1 part in $10^{30}$.

Supersymmetry moderates this fine-tuning.  If supersymmetry is exact,
the Higgs mass receives no perturbative corrections.  With
supersymmetry breaking, the Higgs mass becomes
\begin{equation}
m_h^2 \approx m_{h\, 0}^2 + \frac{\lambda_f^2}{8 \pi^2} N_c^f  
\left(m_{\tilde{f}}^2 - m_f^2 \right) 
\ln \left( \Lambda^2 / m_{\tilde{f}}^2 \right) \ ,
\label{mhsusy}
\end{equation}
where $\tilde{f}$ is the superpartner of fermion $f$.  The quadratic
dependence on $\Lambda$ is reduced to a logarithmic one, and even for
$\Lambda \sim \mplanck$, the large logarithm is canceled by the loop
suppression factor $1/(8 \pi^2)$, and the Higgs mass is natural,
provided $m_{\tilde{f}}$ is not too far above $m_h$.  Requiring a
maximal fine-tuning $\caln^0_{\text{max}}$, the upper bound on
sfermion masses is
\begin{equation}
m_{\tilde{f}} \alt 800~\gev \,
\frac{1.0}{\lambda_f}
\Biggl[ \frac{3}{N_c^f} \Biggr]^{\frac{1}{2}}
\Biggl[ \frac{70}{\ln ( \Lambda^2 / m_{\tilde{f}}^2 )} 
\Biggr]^{\frac{1}{2}}
\Biggl[ \frac{\caln^0_{\text{max}}}{100} \Biggr]^{\frac{1}{2}} ,
\label{roughbound}
\end{equation}
where $\lambda_f$ and $N_c^f$ have been normalized to their top quark
values, the logarithm has been normalized to its value for $\Lambda
\sim \mplanck$, and $\caln^0_{\text{max}}$ has been normalized to 100,
or 1\% fine-tuning.

\subsubsection{First Implications}
\label{sec:first}

Even given this quick and simple analysis, \eqref{roughbound} already
has interesting implications:
\begin{itemize}
\setlength{\itemsep}{1pt}\setlength{\parskip}{0pt}\setlength{\parsep}{0pt}
\item Naturalness constraints vary greatly for different
  superpartners.  As noted as early as 1985~\cite{Drees:1985jx}, the
  1-loop contributions of first and second generation particles to the
  Higgs mass are suppressed by small Yukawa couplings.  For the first
  generation sfermions, naturalness requires only that they be below
  $10^4~\tev$!  In fact, this upper bound is strengthened to $\sim
  4~\tev - 10~\tev$ by considerations of $D$-term and 2-loop effects,
  as discussed in \secref{natbounds}.  Nevertheless, it remains true
  that {\em without additional theoretical assumptions, there is no
    naturalness reason to expect first and second generation squarks
    and sleptons to be within reach of the LHC}.

\item Naturalness bounds on superpartner masses are only challenged by
  LHC constraints for large $\Lambda$, such as $\Lambda \sim \mplanck$
  or $\Lambda \sim \mgut \simeq 2 \times 10^{16}~\gev$, the grand
  unified theory (GUT) scale.  For low $\Lambda$, the loop suppression
  factor is not compensated by a large logarithm, and naturalness
  constraints are weakened by as much as an order of magnitude.  For
  example, even for top squarks, for low $\Lambda$ such that $\ln (
  \Lambda^2 / m_{\tilde{f}}^2 ) \sim 1$, the naturalness bound for
  $\caln^0_{\text{max}} = 1$ becomes $m_{\tilde{t}} \alt 700~\gev$,
  beyond current LHC bounds, and for $\caln^0_{\text{max}} =100$, the
  bound is $m_{\tilde{t}} \alt 7~\tev$, far above even the reach of
  the 14 TeV LHC.  This is as expected for a 1-loop effect.  The
  heuristic expectation that ${\cal O}(1)$ fine-tuning requires
  $m_{\tilde{t}} \sim m_h$ assumes implicitly that the 1-loop
  suppression factor is compensated by a large logarithm.

Of course, supersymmetry makes it possible to contemplate a
perturbative theory all the way up to $\mgut$ or $\mplanck$, and grand
unification, radiative electroweak symmetry breaking, and other key
virtues of supersymmetry make this highly motivated.  There are
therefore strong reasons to consider $\Lambda \sim \mgut, \mplanck$.
This observation, however, suggests that if the enhancement from large
logarithms may somehow be removed, supersymmetric theories with
multi-TeV superpartners may nevertheless be considered natural; this
is the strategy of models that will be discussed in
\secref{focuspoint}.

\item Last, all naturalness bounds depend on what level of fine-tuning
  is deemed acceptable, with mass bounds scaling as
  $\sqrt{\caln^0_{\text{max}}}$.  This is an irreducible subjectivity
  that must be acknowledged in all discussions of naturalness.

There is a sociological observation perhaps worth making here,
however.  In the past, in the absence of data, it was customary for
some theorists to ask ``What regions of parameter space are most
natural?''~and demand fine-tuning of, say, less than 10\%
($\caln^0_{\text{max}} = 10$).  This requirement led to the promotion
of models with very light superpartners, and heightened expectations
that supersymmetry would be discovered as soon as the LHC began
operation.

In retrospect, however, this history has over-emphasized light
supersymmetry models and has little bearing on the question of whether
weak-scale supersymmetry is still tenable or not.  As with all
questions of this sort, physicists vote with their feet.  Rather than
asking ``What regions of parameter space are most natural?'', a more
telling question is, ``If superpartners were discovered, what level of
fine-tuning would be sufficient to convince you that the gauge
hierarchy problem is solved by supersymmetry and you should move on to
researching other problems?''  An informal survey of responses to this
question suggests that values of $\caln^0_{\text{max}} = 100, 1000$,
or even higher would be acceptable. {}From this perspective, the
normalization of $\caln^0$ in \eqref{roughbound} is reasonable, and
current bounds from the LHC do not yet preclude a supersymmetric
solution to the hierarchy problem, especially given the many caveats
associated with attempts to quantify naturalness, which will be
discussed in \secref{quantifying}.

\end{itemize}

\subsection{Grand Unification}
\label{sec:gauge}

The fact that the standard model particle content fits neatly into
multiplets of larger gauge groups, such as SU(5), SO(10), or E$_6$, is
striking evidence for
GUTs~\cite{Pati:1973uk,Georgi:1974sy,Fritzsch:1974nn,Gursey:1975ki}.
In the standard model, the strong, weak, and electromagnetic gauge
couplings do not unify at any scale.  However, in the minimal
supersymmetric standard model (MSSM), the supersymmetric extension of
the standard model with minimal field content, the gauge coupling
renormalization group equations (RGEs) are modified above the
superpartner mass scale.  With this modification, if the superpartners
are roughly at the weak scale, the gauge couplings meet at $\mgut
\simeq 2 \times 10^{16}~\gev$, further motivating both grand
unification and weak-scale
supersymmetry~\cite{Dimopoulos:1981zb,Dimopoulos:1981yj,%
  Sakai:1981gr,Ibanez:1981yh,Einhorn:1981sx}.

Gauge coupling unification is sensitive to the superpartner mass
scale, since this governs when the RGEs switch from non-supersymmetric
to supersymmetric.  However, the sensitivity to the superpartner mass
scale is only logarithmic.  Furthermore, full SU(5) multiplets of
superpartners, such as complete generations of squarks and sleptons,
may be heavy without ruining gauge coupling unification.  Note,
however, that the MSSM particles do not completely fill SU(5)
multiplets; in particular, the Higgs bosons must be supplemented with
Higgs triplets. One might therefore hope that the masses of SU(2)
doublet Higgsinos might be stringently constrained by gauge coupling
unification, but even this is not the case.  A full justification
requires a complete discussion of GUTs and proton
decay~\cite{Mohapatra:1999vv,Nath:2006ut}, but roughly speaking, heavy
sfermions suppress the leading contributions to proton decay, and
there is sufficient freedom in threshold corrections from the
GUT-scale spectrum to allow unification even for relatively heavy
Higgsinos; see, \eg, Refs.~\cite{Feng:2000bp,Arvanitaki:2012ps}.

In summary, grand unification is a significant motivation for
supersymmetry, but gauge coupling unification is a blunt tool when it
comes to constraining the superpartner mass scale. Note, however, that
the relations imposed by grand unification may have a strong impact on
naturalness bounds, either weakening or strengthening them
significantly; see \secref{natbounds}.

\subsection{Dark Matter}
\label{sec:WIMP}

Supersymmetry provides an excellent WIMP dark matter candidate when
the neutralino is the lightest supersymmetric particle
(LSP)~\cite{Goldberg:1983nd,Ellis:1983ew}.  Neutralinos naturally
freeze out
with approximately the correct thermal relic density.  This density is
inversely proportional to the thermally-averaged annihilation cross
section, which, on dimensional grounds, is inversely proportional to
the superpartner mass scale squared:
\begin{equation}
\Omega_{\chi} \propto \frac{1}{\langle \sigma v \rangle} \propto
\tilde{m}^2 \ .
\end{equation}
The requirement $\Omega_{\chi} \le 0.23$ therefore places an upper
bound on the superpartner mass scale $\tilde{m}$.

Unfortunately, when the constants of proportionality are included, the
upper bounds for some types of neutralinos are far above current LHC
sensitivities.  For example, for mixed
Bino-Higgsino~\cite{Mizuta:1992qp,Cirelli:2007xd,Feng:2000gh} and pure
Wino-like~\cite{Hisano:2006nn} neutralino dark matter, the upper
bounds are
\begin{eqnarray}
 m_{\tilde{B}-\tilde{H}} &<& 1.0~\tev \nonumber \\
 m_{\tilde{W}} &<& 2.7-3.0~\tev \ .
\label{LSPlimits}
\end{eqnarray}
Such neutralinos may be produced in the cascade decays of squarks and
gluinos, but this is model-dependent.  The model-independent search
strategy is to consider Drell-Yan production of neutralino pairs with
a radiated jet or photon, which contributes to mono-jet and
mono-photon searches~\cite{Birkedal:2004xn,Feng:2005gj,%
  Beltran:2010ww,Goodman:2010yf,Bai:2010hh}.
The limits in \eqref{LSPlimits} are far above current LHC
sensitivities~\cite{Chatrchyan:2012me,ATLAS:2012ky}.  The
spin-independent and spin-dependent scattering cross sections of such
neutralinos are also consistent with current bounds from direct search
experiments~\cite{Hisano:2012wm}.

Of course, dark matter may be composed of other particles, such as
axions, sterile neutrinos, hidden sector dark matter, or
gravitinos~\cite{Feng:2010gw}.  There is no requirement that
superpartners be light in these dark matter scenarios. In fact, some
scenarios in which gravitinos are the dark matter provide motivation
for extremely heavy superpartners, which freeze out with $\Omega \gg
0.23$, but then decay to gravitinos with $\Omega_{\tilde{G}} \simeq
0.23$~\cite{Feng:2012rn}.

In summary, the requirement of WIMP dark matter provides upper bounds
on superpartner masses, but these upper bounds are high and far beyond
the reach of current colliders.  In addition, the dark matter doesn't
have to be made of supersymmetric WIMPs.  As with the case of grand
unification, the possibility of WIMP dark matter is a significant
virtue of weak-scale supersymmetry, but it does not provide stringent
upper bounds on superpartner masses.

\section{EXPERIMENTAL CONSTRAINTS}
\label{sec:experimental}

\subsection{Superpartner Searches at Colliders}

The search for weak-scale supersymmetry has been ongoing for decades
at many colliders.  Before the 2013-14 shutdown, however, the LHC
experiments ATLAS and CMS each collected luminosities of more than
$5~\ifb$ at $\sqrt{s} = 7~\tev$ and $20~\ifb$ at $\sqrt{s} = 8~\tev$,
and the resulting LHC limits supersede previous collider constraints
in almost all scenarios.  We will therefore confine the discussion to
LHC results and focus on a small subset that is particularly relevant
for the following discussion.  For a summary of pre-LHC constraints,
see Ref.~\cite{Feng:2009te}, and for the full list of LHC analyses,
see Refs.~\cite{ATLASSUSY,CMSSUSY}.

\subsubsection{Gluinos and Squarks}

The greatest mass reach at the LHC is for strongly-interacting
particles, such as gluinos and squarks, which are produced through $pp
\to \gluino \gluino, \gluino \squark, \squark \squark$.  The limits
depend, of course, on the decays.  Limits in the $(m_{\tilde{g}},
m_{\tilde{q}})$ plane, assuming the decays $\tilde{g} \to q \bar{q}
\chi$ and $\tilde{q} \to q \chi$, leading to $\text{jets} + \met$, are
shown in \figref{LHCoverview}. The results imply $m_{\squark} \agt
1.3~\tev$ for decoupled gluinos, $m_{\gluino} \agt 1.2~\tev$ for
decoupled squarks, and $m_{\gluino} = m_{\squark} \agt 1.5~\tev$ in
the degenerate case. Note, however, that the squarks appearing in this
analysis are squarks of the first two generations.  The lightest
neutralino is assumed massless, but top and bottom squarks, as well as
all other superpartners, are assumed heavy and decoupled.

\begin{figure}
\includegraphics[width=0.48\columnwidth]{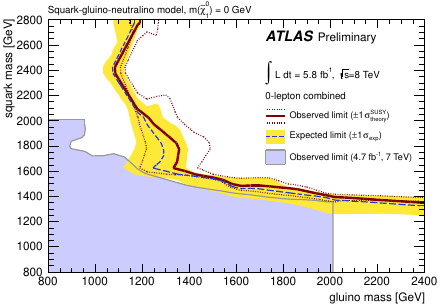} 
\hfil
\includegraphics[width=0.48\columnwidth]{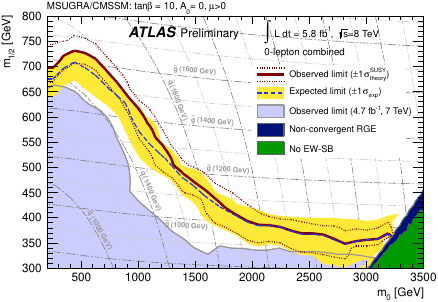}
\caption{Constraints on gluinos and first and second generation
  squarks from ATLAS at the LHC with $L = 5.8~\ifb$ and $\sqrt{s} =
  8~\tev$~\cite{ATLAS:2012ona}. Left: Limits in the $(m_{\gluino},
  m_{\squark})$ plane from $pp \to \gluino \gluino, \gluino \squark,
  \squark \squark$ followed by $\gluino \to q \bar{q} \chi$ and
  $\squark \to q \chi$, leading to $\text{jets} + \met$.  The analysis
  assumes $m_{\squark} \equiv m_{\tilde{u}_{L,R}} =
  m_{\tilde{d}_{L,R}} = m_{\tilde{s}_{L,R}} = m_{\tilde{c}_{L,R}}$,
  $m_{\chi} = 0$, and that all other superpartners, including the top
  and bottom squarks, are very heavy. The shaded region boundaries at
  $m_{\gluino}, m_{\squark} = 2~\tev$ are artifacts of the previous 7
  TeV analysis.  Right: Limits from the $\text{jets} + \met$ search in
  the $(m_0, M_{1/2})$ plane of mSUGRA, with $\tan\beta = 10$,
  $A_0 = 0$, and $\mu > 0$. }
\label{fig:LHCoverview}
\end{figure}

The resulting bounds may also be applied to the framework of minimal
supergravity (mSUGRA), also know as the constrained MSSM.  This
framework has 4 continuous parameters defined at the GUT scale (a
unified scalar mass $m_0$, a unified gaugino mass $M_{1/2}$, a unified
tri-linear scalar coupling $A_0$, the ratio of Higgs boson vacuum
expectation values $\tan \beta \equiv \langle H^u_0 \rangle / \langle
H^d_0 \rangle$), and one discrete choice, the sign of the Higgsino
mass parameter $\mu$.  Constraints in this model parameter space are
shown in \figref{LHCoverview}.  In the limit of heavy sfermions (large
$m_0$), the $\text{jets} + \met$ search implies $m_{\gluino} \agt
1.0~\tev$.

\subsubsection{Top and Bottom Squarks}

The constraints of \figref{LHCoverview} might appear to require all
superpartners to be above the TeV scale.  As noted in
\secref{gaugehierarchyproblem}, however, naturalness most stringently
constrains the top and bottom squarks, but allows effectively
decoupled first and second generation squarks, exactly the opposite of
the assumptions made in deriving these bounds.  It is therefore
important to consider other analyses, including searches for light top
and bottom squarks.  Results from such searches are shown in
\figref{LHCstop}.  Limits from direct stop pair production followed by
$\tilde{t} \to t \chi$ are shown, as are limits from gluino pair
production followed by $\gluino \to \tilde{t}^* \bar{t} \to t \bar{t}
\chi$, which is the dominant decay mode if stops are significantly
lighter than all other squarks.  In the case of stop pair production,
we see that stops as light as 500 GeV are allowed for massless
neutralinos, and much lighter stops are allowed if one approaches the
kinematic boundary $m_{\tilde{t}} - m_{\chi} = m_t$.  In the case of
gluino pair production, the bound is $m_{\gluino} \agt 1.1~\tev$ for
$m_{\chi} = 0$, but again, there are allowed regions with much lighter
gluinos near the kinematic boundary $m_{\gluino} - m_{\chi} = 2 m_t$.
Searches for light stops in other channels, as well as searches for
bottom squarks, yield roughly similar
constraints~\cite{ATLASSUSY,CMSSUSY}.

\begin{figure}
\includegraphics[width=0.48\columnwidth,height=0.40\columnwidth]{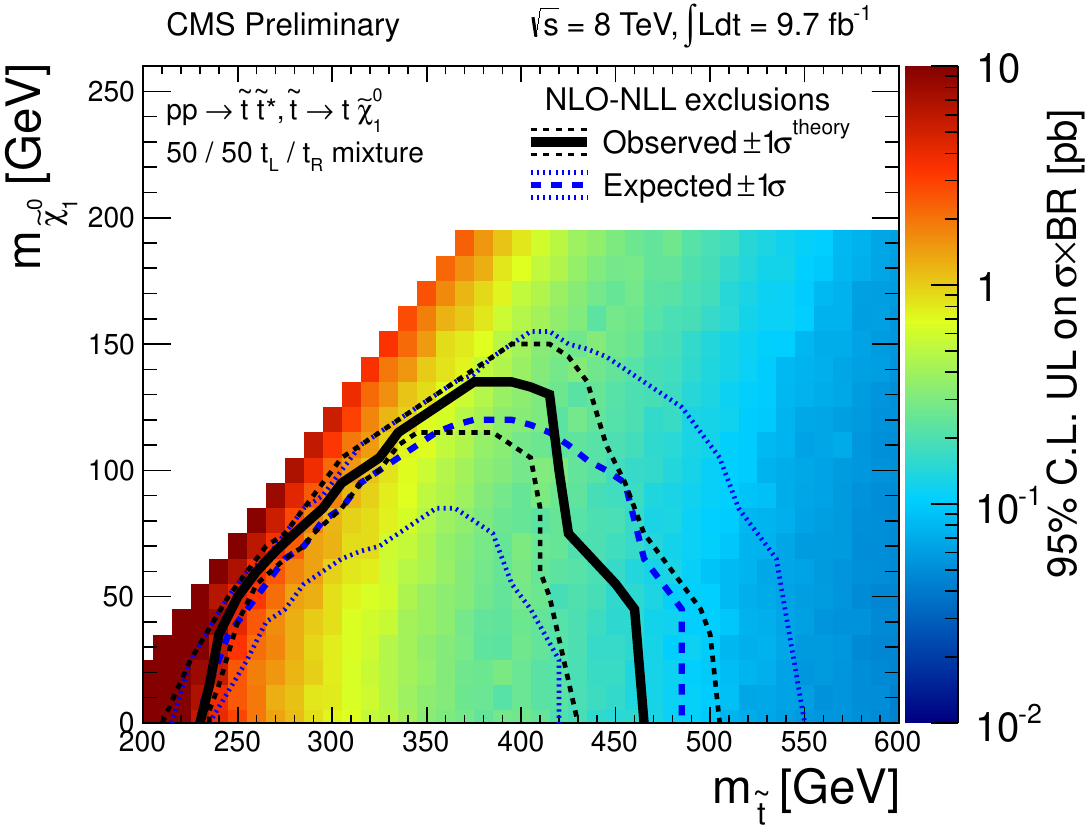} 
\hfil
\includegraphics[width=0.48\columnwidth]{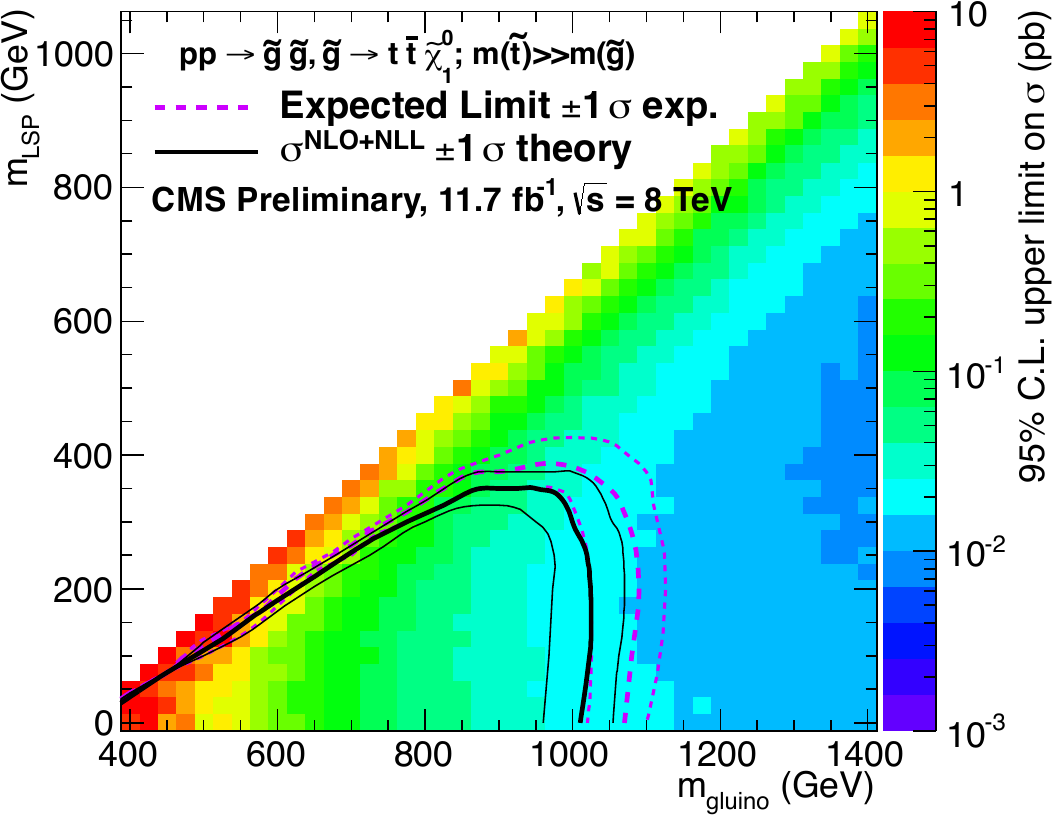}
\caption{Constraints on gluinos and top squarks from CMS at the LHC.
  Left: Limits from $L = 9.7~\ifb$ at $\sqrt{s} = 8~\tev$ in the
  $(m_{\tilde{t}}, m_{\chi})$ plane from $pp \to \tilde{t}
  \tilde{t}^*$, followed by $\tilde{t} \to t \chi$, leading to the
  signature of $l + \text{$b$-jet}+ \met$~\cite{CMS-PAS-SUS-12-023}.
  Right: Limits from $L = 11.7~\ifb$ at $\sqrt{s} = 8~\tev$ in the
  $(m_{\gluino}, m_{\chi})$ plane from $pp \to \gluino \gluino$,
  followed by $\gluino \to t \bar{t} \chi$, leading to signatures of
  $\text{$N_j$ jets} + \text{$N_b$ $b$-jets} + \met$, where $2 \le N_j
  \le 3$ or $N_j \ge 4$, and $N_b = 0, 1, 2, 3$, or $\ge
  4$~\cite{CMSstop}. }
\label{fig:LHCstop}
\end{figure}

\subsubsection{$R$-Parity Violation}
\label{sec:rpLHC}

The search results presented so far require missing transverse energy.
Although WIMP dark matter motivates this possibility, large $\met$ is
far from a requirement of supersymmetry, and $\met$ signals may be
degraded in a number of ways, for example, by compressed superpartner
spectra, a possibility discussed in \secref{compressed}.

Perhaps the most dramatic way is with $R$-parity ($R_p$)
violation. When the standard model is extended to include
supersymmetry, there are many new gauge-invariant, renormalizable
interactions.  If any one of these is present, all superpartners
decay, effectively eliminating the $\met$ signature.  These
$R_p$-violating (RPV) terms arise from superpotentials of the form
\begin{equation}
W_{\rpv} = \lambda_{ijk} L_i L_j E_k + \lambda'_{ijk} L_i Q_j D_k +
\lambda''_{ijk} U_i D_j D_k + \mu_i L_i H_u \ ,
\label{rpterms}
\end{equation}
where the first three types of terms are categorized as leptonic,
semi-leptonic, and hadronic, and $i, j, k = 1,2,3$ are generation
indices.  If any of these couplings is non-zero, all gauginos may
decay to three standard model fermions.

The most difficult case for the LHC is hadronic RPV.  In
\figref{LHCnomet}, we show results from $pp \to \gluino \gluino$
followed by the RPV decay $\gluino \to qqq$ through a $\lambda''$
operator~\cite{ATLAS:2012dp}.  The resulting bound on the gluino mass
is 670 GeV, far weaker than in cases where the gluino cascade decay
includes significant $\met$.

\begin{figure}
\includegraphics[width=0.48\columnwidth]{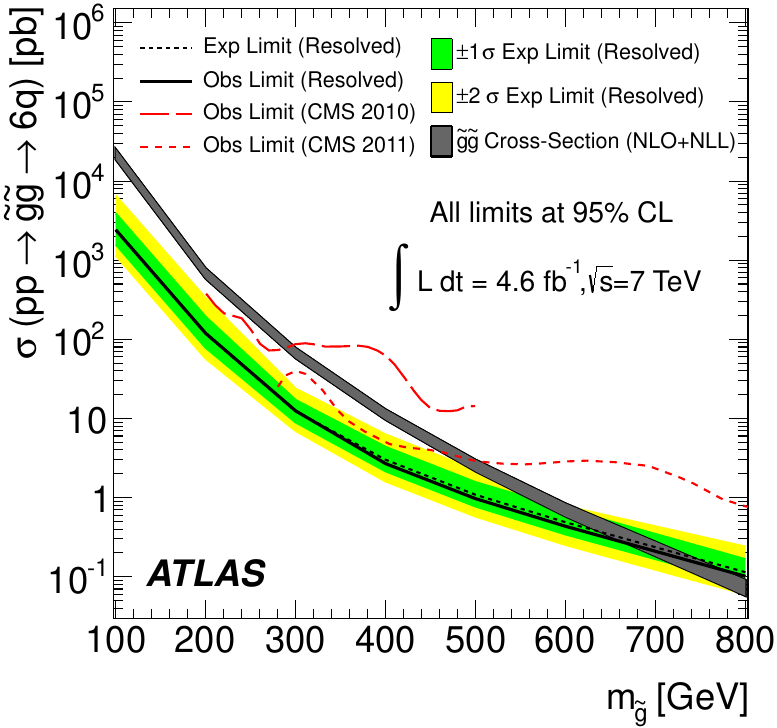}
\caption{Constraint on gluinos in supersymmetry with $R_p$ violation
  from ATLAS at the LHC with $L = 4.6~\ifb$ and $\sqrt{s} =
  7~\tev$~\cite{ATLAS:2012dp}.  The constraint is on gluino pair
  production $pp \to \gluino \gluino$ followed by the hadronic RPV
  decay $\gluino \to qqq$, leading to 6 jets with no $\met$, and
  implies $m_{\gluino} > 670~\gev$. }
\label{fig:LHCnomet}
\end{figure}

\subsubsection{Sleptons, Charginos, and Neutralinos}

Finally, the mass reach for superpartner searches is, of course,
greatly reduced for uncolored superpartners.  In \figref{LHCweak}, we
show constraints from CMS on Drell-Yan slepton pair production and
associated chargino-neutralino pair production~\cite{CMSweak}.  The
limits are impressive, as they extend LEP bounds of $\tilde{m} \agt
100~\gev$ to much higher masses, requiring $m_{\tilde{e}},
m_{\tilde{\mu}} \agt 275~\gev$ and $m_{\chi^{\pm}_1} = m_{\chi^0_2}
\agt 330~\gev$ for $m_{\chi} = 0$.  Note, however, that these limits
do not apply to staus, they degrade significantly for larger
$m_{\chi}$ and more degenerate spectra, and they bound superpartner
masses that are not highly constrained by naturalness in the absence
of additional theoretical assumptions.

\begin{figure}
\includegraphics[width=0.48\columnwidth]{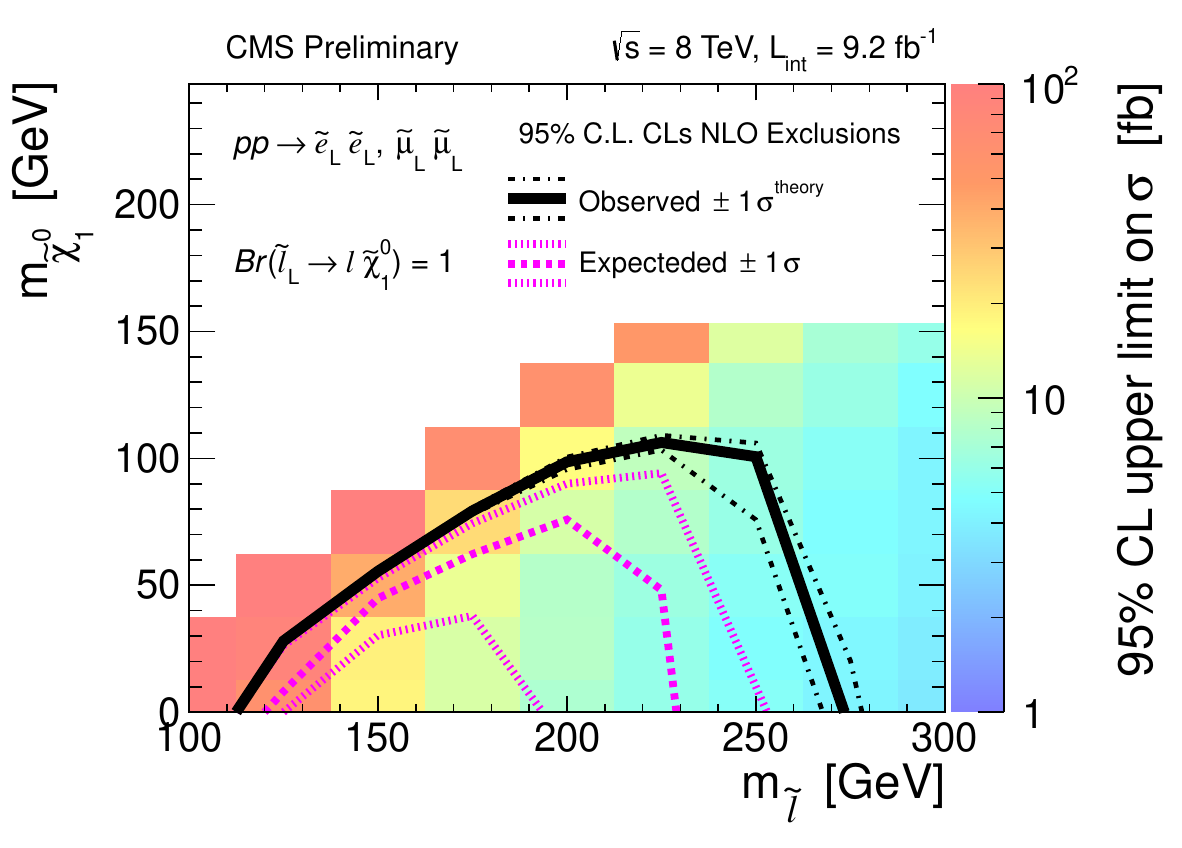} 
\hfil
\includegraphics[width=0.48\columnwidth]{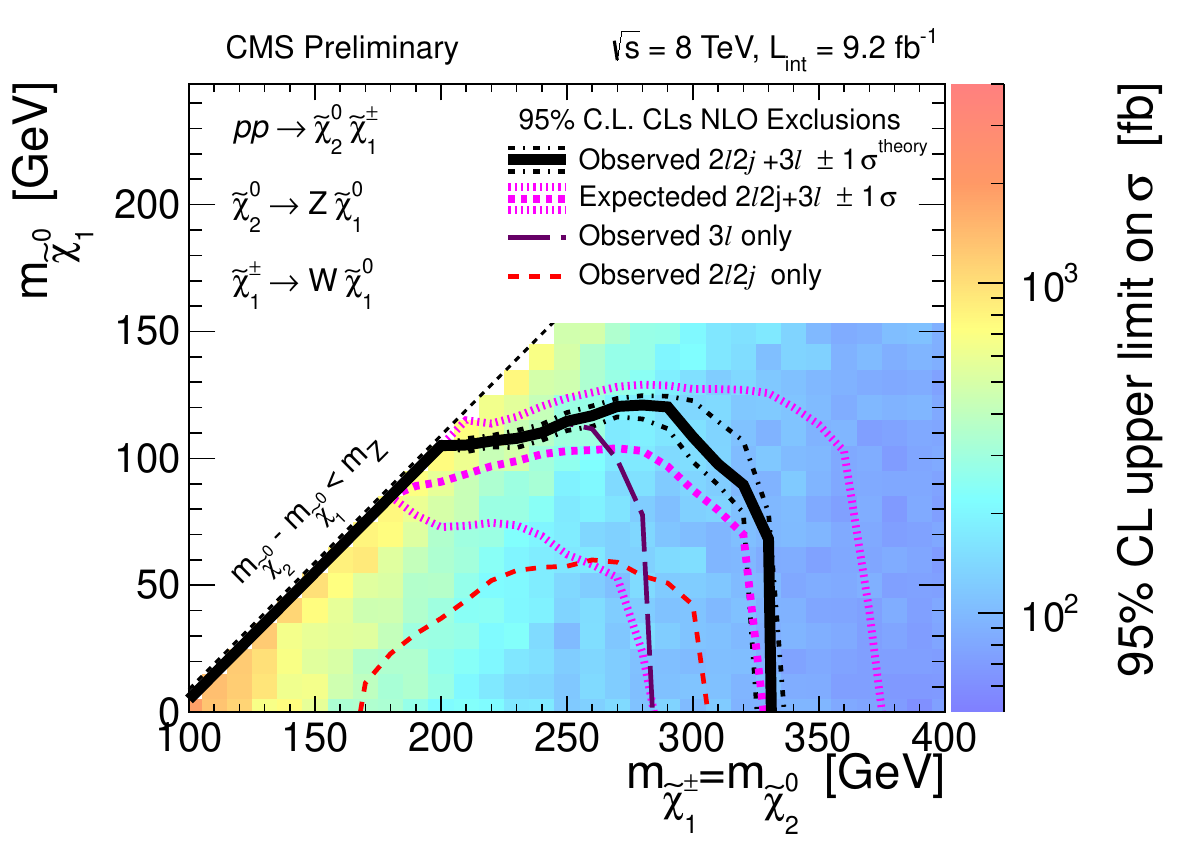}
\caption{Constraints on sleptons and electroweak gauginos from
  Drell-Yan production at CMS at the LHC with $L = 9.2~\ifb$ and
  $\sqrt{s} = 8~\tev$~\cite{CMSweak}.  Left: Limits in the
  $(m_{\tilde{l}}, m_{\chi})$ plane from $pp \to \tilde{l}_L
  \tilde{l}_L^*$, where $l = e, \mu$, followed by $\tilde{l}_L \to l
  \chi$, leading to $2l + \met$ events.  Right: Limits in the
  $(m_{\chi^{\pm}_1} = m_{\chi^0_2}, m_{\chi})$ plane from $pp \to
  \chi^{\pm}_1 \chi^0_2$, followed by $\chi^{\pm}_1 \to W \chi$ and
  $\chi^0_2 \to Z \chi$, leading to $2j\, 2l + \met$ and $3l + \met$
  events. }
\label{fig:LHCweak}
\end{figure}

\subsection{The Higgs Boson}
\label{sec:higgs}

The Higgs boson, or at least an eerily similar particle, has been
discovered at the LHC~\cite{:2012gk,:2012gu}.  Constraints on the
Higgs boson mass and the signal strength in the $h \to \gamma \gamma$
and $h \to ZZ^* \to 4l$ channels are shown in \figref{higgsmass}.
Early hints of slight inconsistencies between the mass measurements
and signal strengths in various channels have now largely disappeared.
At ATLAS, the $\gamma \gamma$ mass is slightly larger than the $ZZ^*$
mass, and both signal strengths are slightly above SM
expectations. None of these discrepancies is significant, however, and
the results of the two experiments are also quite consistent, as
evident in \figref{higgsmass}.

\begin{figure}
\includegraphics[width=0.48\columnwidth,height=0.4\columnwidth]{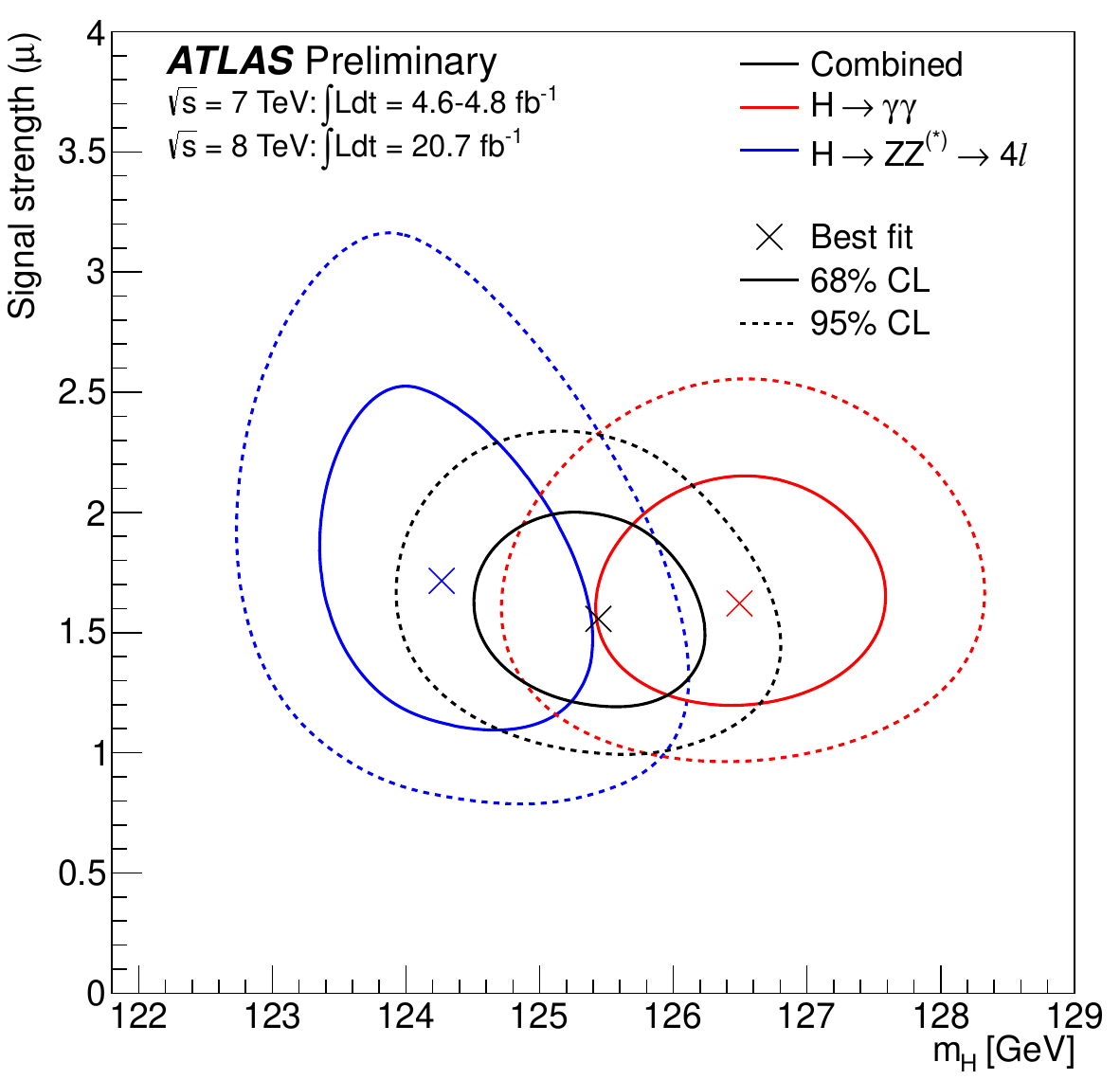} 
\hfil
\includegraphics[width=0.48\columnwidth,height=0.4\columnwidth]{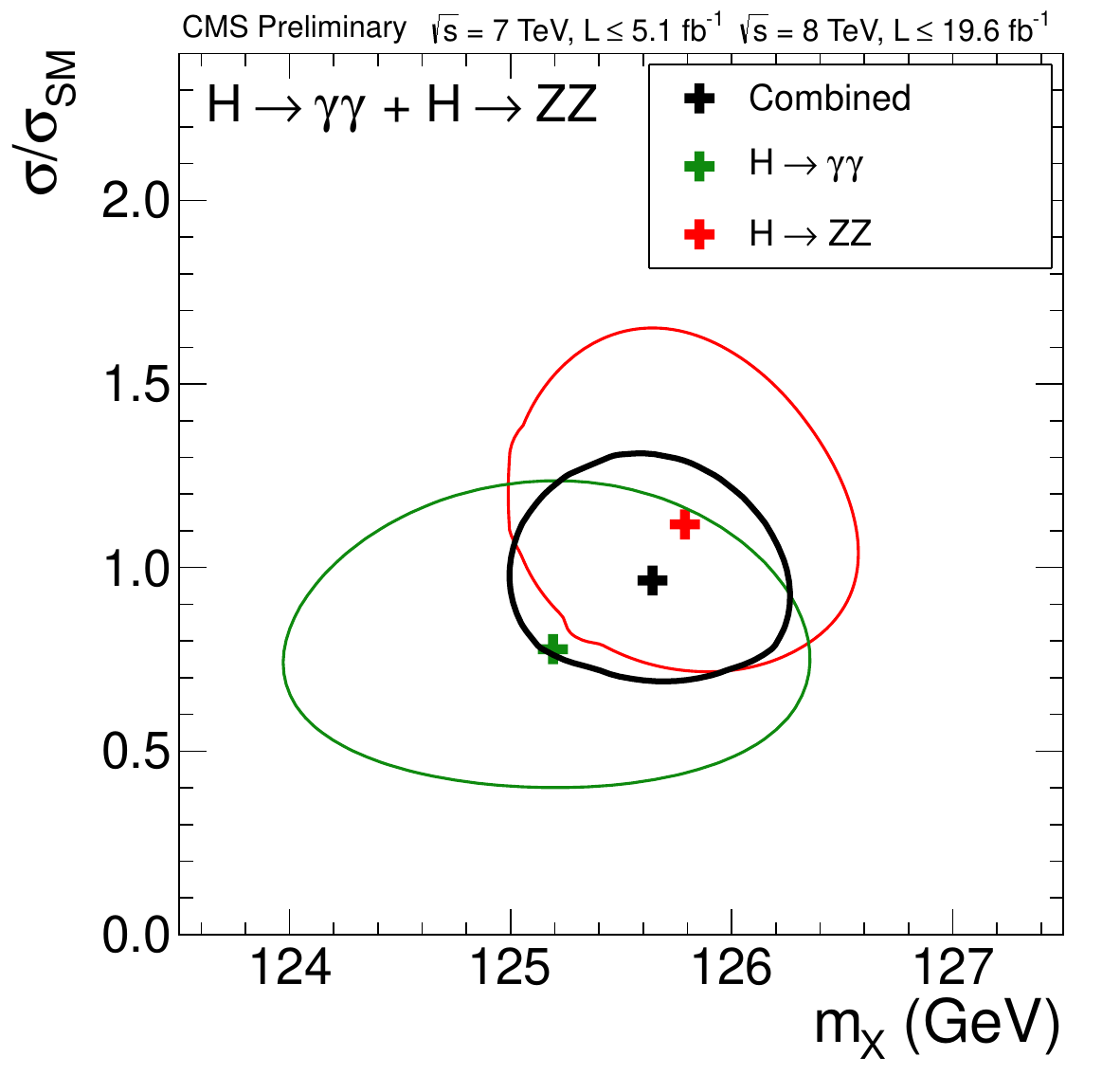}
\caption{Constraints in the $(m_h, \sigma/\sigma_{\text{SM}})$ plane
  for the $h \to \gamma \gamma$ and $h \to ZZ^* \to 4l$ channels from
  ATLAS~\cite{ATLAS-CONF-2013-014} (left) and
  CMS~\cite{CMS-PAS-HIG-13-005} (68\% CL) (right).  }
\label{fig:higgsmass}
\end{figure}

At present, the most pressing concern for supersymmetry is
simply the Higgs boson mass.  In the MSSM the Higgs boson is
generically light, since the quartic coupling in the scalar potential
is determined by the electroweak gauge couplings.  Indeed, the tree
level value $m_h({\text{tree}})= m_Z | \cos 2 \beta |$ cannot exceed
the $Z$ boson mass.  However, 
the Higgs mass may be raised significantly by
radiative corrections~\cite{Okada:1990vk,Haber:1990aw,Ellis:1990nz}.
For moderate to large $\tan\beta$, a 2-loop expression for the Higgs
mass is~\cite{Carena:1995wu,Carena:2000dp}
\begin{eqnarray}
m_h^2 &\approx& m_Z^2 \cos^2 2 \beta 
+ \frac{3 m_t^4}{2 \pi^2 v^2} \Biggl\{ \ln \frac{M_S^2}{m_t^2}
+ \frac{X_t^2}{M_S^2}
\Biggl( 1 - \frac{X_t^2}{12 M_S^2} \Biggr) \Biggr. 
\nonumber
\\
&& + \frac{1}{16 \pi^2} \Biggl( \frac{3 m_t^2}{v^2} - 32 \pi \alpha_s
\Biggr)
\Biggl[ \frac{ 2 X_t^2}{M_S^2}
\Biggl( 1 - \frac{X_t^2}{12 M_S^2} \Biggr) 
\ln \frac{M_S^2}{m_t^2} 
+ \Biggl( \ln \frac{M_S}{m_t^2} \Biggr)^2 \Biggr] 
\Biggr\} \ ,
\label{higgs}
\end{eqnarray}
where $v \simeq 246~\gev$, $M_S \equiv \sqrt{m_{\tilde{t}_1}
  m_{\tilde{t}_2}}$, $X_t \equiv A_t - \mu \cot\beta$ parameterizes
the stop left-right mixing, and $\alpha_s \approx 0.12$.  Several
codes incorporate
2-loop~\cite{Heinemeyer:1998yj,Allanach:2001kg,Djouadi:2002ze}, or
even 3-loop~\cite{Harlander:2008ju,Kant:2010tf}, corrections.

\begin{figure}
\includegraphics[width=0.48\columnwidth,height=0.36\columnwidth]{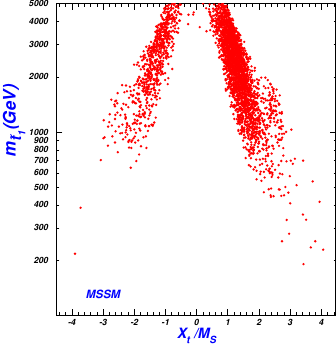}
\caption{Values of top squark parameters that give $123~\gev < m_h <
  127~\gev$ in viable MSSM models~\cite{Cao:2012fz}.  The parameters
  are $m_{\tilde{t}_1}$, the mass of the lighter stop, and $X_t/M_S$,
  where $X_t \equiv A_t - \mu \cot \beta$ parameterizes left-right stop
  mixing and $M_S \equiv \sqrt{m_{\tilde{t}_1} m_{\tilde{t}_2}}$ is the
    geometric mean of the physical stop masses.}
\label{fig:higgsmasstheory}
\end{figure}

\Eqref{higgs} has several interesting features.  First, increasing
$\tan\beta$ increases the tree-level Higgs mass; this effect saturates
for $\tan\beta \agt 13$, where the tree-level mass is within 1 GeV of
its maximum.  Second, the Higgs boson mass may be greatly increased
either by large stop mixing ($X_t \approx \pm \sqrt{6} M_S$) or by
heavy stops ($M_S \gg m_t$).  Numerical results are shown in
\figref{higgsmasstheory}.  For negligible stop mixing, stop masses
$M_S \agt 4~\tev$ are required to give a consistent Higgs boson
mass. For near-maximal mixing, sub-TeV values of $M_S$ are possible,
but such large mixing is highly fine-tuned with respect to the $A_t$
parameter~\cite{Hall:2011aa}.  The generic lesson to draw is that the
measured Higgs mass favors stop masses well above a TeV.

At present, the Higgs mass measurement is at least as significant a
challenge to naturalness as the absence of superpartners at the LHC.
First, the Higgs mass can only be raised to $\sim 125~\gev$ by raising
the masses of superpartners that couple strongly to the Higgs.  But it
is exactly these particles that, at least at first sight, must be
light to preserve naturalness. Second, because the Higgs mass is only
logarithmically sensitive to the top squark mass, it has tremendous
reach, favoring, in the no-mixing case, stop masses that are far above
current LHC bounds and even challenging for all proposed future
colliders.

\subsection{Flavor and CP Violation}

Bounds on low-energy flavor and CP violation stringently constrain all
proposals for new physics at the weak scale.  For supersymmetry, these
longstanding constraints are extremely stringent and are {\em a
  priori} a strong challenge to naturalness.  The constraints on
supersymmetry may be divided into two qualitatively different classes.

\subsubsection{Flavor-Violating Constraints}

The first are those that require flavor violation.  Supersymmetry
breaking generates sfermion masses that generically violate both
flavor and CP.  For example, for the left-handed down-type squarks,
the mass matrix $m^2_{ij}$, where $i, j = \tilde{d}_L, \tilde{s}_L,
\tilde{b}_L$, generically has off-diagonal entries that mediate flavor
violation and complex entries that violate CP. Flavor and CP violation
may also arise from all of the other mass matrices, as well as from
the supersymmetry-breaking $A$-terms.

The constraints from low-energy flavor violation have been analyzed in
many works.  In Ref.~\cite{Gabbiani:1996hi}, for example, constraints
are derived by requiring that the supersymmetric box diagram
contributions to meson mass splittings not exceed their observed
values, and the supersymmetric penguin diagram contributions to
radiative decays $l_i \to l_j \gamma$ not exceed current bounds.  A
small sample of these results include
\begin{eqnarray}
\left[ \frac{12~\tev}{m_{\squark}} \right] ^2
\Biggl| \text{Re} \left( 
\frac{m_{\tilde{d}\tilde{s}}^2}{m_{\squark}^2}
\right) ^2 \Biggr|
& \alt & \frac{\Delta m_K}{3.49 \times 10^{-12}~\mev} \\
\left[ \frac{16~\tev}{m_{\squark}} \right] ^2
\Biggl| \text{Re} \left( 
\frac{m_{\tilde{u}\tilde{c}}^2}{m_{\squark}^2}
\right) ^2 \Biggr|
& \alt & \frac{\Delta m_D}{1.26 \times 10^{-11}~\mev} \\
\left[ \frac{5.4~\tev}{m_{\squark}} \right] ^2
\Biggl| \text{Re} \left( 
\frac{m_{\tilde{d}\tilde{b}}^2}{m_{\squark}^2}
\right) ^2 \Biggr|
& \alt & \frac{\Delta m_B}{3.38 \times 10^{-10}~\mev} \\
\left[ \frac{160~\tev}{m_{\squark}} \right] ^2
\Biggl| \text{Im} \left( 
\frac{m_{\tilde{d}\tilde{s}}^2}{m_{\squark}^2}
\right) ^2 \Biggr|
& \alt & \frac{\epsilon_K}{2.24 \times 10^{-3}} \\
\left[ \frac{2.4~\tev}{m_{\slepton}} \right] ^4
\left| \frac{m_{\tilde{e}\tilde{\mu}}^2}{m_{\slepton}^2} \right|^2
& \alt & \frac{B(\mu \to e \gamma)}{2.4 \times 10^{-12}} \\
\left[ \frac{150~\gev}{m_{\slepton}} \right] ^4
\left| \frac{m_{\tilde{e}\tilde{\tau}}^2}{m_{\slepton}^2} \right|^2
& \alt & \frac{B(\tau \to e \gamma)}{3.3 \times 10^{-8}} \\
\left[ \frac{140~\gev}{m_{\slepton}} \right] ^4
\left| \frac{m_{\tilde{\mu}\tilde{\tau}}^2}{m_{\slepton}^2} \right|^2
& \alt & \frac{B(\tau \to \mu \gamma)}{4.4 \times 10^{-8}} \ ,
\end{eqnarray}
where the constraints apply to both left- and right-handed fermions
and arise from the indicated observables, which have been normalized
to current values~\cite{Beringer:1900zz}.  Here $m_{\squark}$ and
$m_{\slepton}$ are average masses of the relevant squark and slepton
generations, and we have set $m_{\gluino} = m_{\squark}$ and
$m_{\tilde{\gamma}} = m_{\slepton}$.  For ${\cal O}(1)$ flavor
violation, low-energy constraints require that the first and second
generation sfermions have masses $\agt 10~\tev$, and if there are
additionally ${\cal O}(1)$ phases, the down-type squarks must have
masses $\agt 100~\tev$. In contrast, constraints from processes
involving third generation squarks and sleptons are generally much
less severe, and are typically satisfied for sub-TeV masses.

\subsubsection{Electric Dipole Moments}
\label{sec:EDMs}

The second class of constraints arises from flavor-conserving, but
CP-violating, observables, namely the electric dipole moments (EDMs)
of the electron and neutron.  There are well-known frameworks, \eg,
gauge-mediated supersymmetry breaking~\cite{Dine:1981za,%
  Dimopoulos:1981au,Nappi:1982hm,AlvarezGaume:1981wy,Dine:1994vc,%
  Dine:1995ag} and anomaly-mediated supersymmetry
breaking~\cite{Randall:1998uk,Giudice:1998xp}, in which the sfermion
mass matrices are essentially diagonal, and all of the
flavor-violating observables discussed above may be suppressed. Even
in these frameworks, however, the gaugino masses $M_i$, $A$-terms, and
the $\mu$ and $B$ parameters may have CP-violating phases, and these
generate potentially dangerous contributions to the EDMs.

The EDMs of the electron and neutron are generated by penguin diagrams
with gauginos, Higgsinos and sfermions in the loop.  The dominant
diagram involves Wino-Higgsino mixing.  The electron EDM is $d_e$ and
the neutron EDM is, assuming the naive quark model, $d_n = (4 d_d -
d_u)/3$.  The electron and down quark EDMs are particularly dangerous
in supersymmetry, as they are enhanced for large $\tan\beta$, and have
the form~\cite{Moroi:1995yh}
\begin{equation}
d_f \sim e \frac{g_2^2}{64 \pi^2} m_f \frac{|\mu M_2|}{m_{\tilde{f}}^4}
\tan\beta \sin \theta_{\text{CP}} \ ,
\end{equation}
where $f = e, d$, $m_{\tilde{f}}$ is the mass scale of the heaviest
superpartners in the loop, which we take to be $\tilde{f}$, and
$\theta_{\text{CP}} \equiv \text{Arg} (\mu M_a B^* )$ is the
CP-violating phase.  Given the $\tan\beta$-enhanced EDMs, and setting
$m_d = 5~\mev$, the EDM constraints are
\begin{eqnarray}
\left( \frac{2.5~\tev}{m_{\tilde{l}}} \right)^2 
\frac{|\mu M_2|}{m_{\tilde{l}}^2} \frac{\tan\beta}{10}
\frac{\sin \theta_{\text{CP}}}{0.1} 
& \alt & \frac{d_e}{1.05 \times 10^{-27} \ e \, \text{cm}} \\
\left( \frac{1.7~\tev}{m_{\tilde{q}}} \right)^2 
\frac{|\mu M_2|}{m_{\tilde{q}}^2} \frac{\tan\beta}{10}
\frac{\sin \theta_{\text{CP}}}{0.1} 
& \alt & \frac{d_n}{2.9 \times 10^{-26} \ e \, \text{cm}} \ ,
\end{eqnarray}
where the electron and neutron EDMs are normalized to their current
bounds~\cite{Beringer:1900zz}.

The EDM constraints are extremely robust.  The CP-violating phase can
be suppressed only by a mechanism that correlates the phases of the
supersymmetry-breaking gaugino masses, $B$, and the
supersymmetry-preserving $\mu$ parameter. In many frameworks, such as
gauge-mediated supersymmetry breaking, it is already challenging to
generate $\mu$ and $B$ parameters of the correct magnitude, much less
to correlate their phases with the gaugino masses, and, of course,
some CP-violation is desirable to generate the matter--anti-matter
asymmetry of the universe.  Although CP-conserving mechanisms have
been proposed~\cite{Dimopoulos:1995kn,Dine:1996xk,Moroi:1998km}, they
are typically far from the simple and elegant ideas proposed to
eliminate flavor violation.  In the absence of such mechanisms, the
EDM constraints require multi-TeV first generation superpartners to be
consistent with ${\cal O}(0.1)$ phases.

\subsection{Hints of New Physics}

In addition to constraints excluding large effects from new physics
beyond the standard model, there are also experimental signals that
may be taken as indications for new physics.  Chief among these is the
anomalous magnetic moment of the muon, $a_{\mu} \equiv (g_{\mu}-2)/2$,
where the final measurement from the Brookhaven E821
experiment~\cite{Bennett:2006fi} disagrees with standard model
predictions by 2.6$\sigma$ to
$3.6\sigma$~\cite{Davier:2010nc,Jegerlehner:2011ti}.  If supersymmetry
is to resolve this discrepancy, the mass of the lightest observable
superpartner, either a chargino or a smuon, must
satisfy~\cite{Feng:2001tr}
\begin{equation}
m_{\text{LOSP}} < 480~\gev 
\left[ \frac{ \tan\beta}{50} \right]^{\frac{1}{2}}
\left[ \frac{ 287 \times 10^{-11}}{\Delta a_{\mu}}
  \right]^{\frac{1}{2}} \ ,
\end{equation}
where $\Delta a_{\mu}$ has been normalized to the current discrepancy.

The anomalous magnetic moment of the muon is not the only potential
signal for new physics, however.  For example, $A_{\text{FB}}^b$, the
forward-backward asymmetry in $Z \to b \bar{b}$ deviates from the
standard model prediction by 2.8$\sigma$~\cite{ALEPH:2005ab}, the
Higgs signal strength in $\gamma \gamma$ is 1$\sigma$ to 2$\sigma$ too
large, and the various Higgs mass measurements discussed above
disagree with each other at the 1$\sigma$ to 3$\sigma$ level.

In this review, as tempting as it is be optimistic, we do not consider
these results to be compelling evidence for new physics.  Of course,
if well-motivated supersymmetric models elegantly explain a
tantalizing anomaly, that should be noted, but here we will take a
more conservative view and will not require supersymmetry to resolve
these tentative disagreements between experiment and the standard
model.

\section{QUANTIFYING NATURALNESS}
\label{sec:quantifying}

We now return to naturalness and discuss attempts to quantify it in
more detail.  All such attempts are subject to quantitative
ambiguities. However, this fact should not obscure the many
qualitative differences that exist in naturalness prescriptions
proposed in the literature. In this section, we begin by describing a
standard prescription for quantifying naturalness.  We then critically
review some of the many alternative prescriptions that have been
proposed, stressing the qualitative differences and their
implications.  After this lengthy discussion has highlighted the many
caveats in any attempt to quantify naturalness, we present some
naturalness bounds on superpartner masses that may serve as a rough
guide as we turn to models in \secref{models}.

\subsection{A Naturalness Prescription}
\label{sec:prescription}

We begin by describing a general five-step prescription for assigning
a numerical measure of naturalness to a given supersymmetric model.
So that clarity is not lost in abstraction, we also show how it is
typically applied to mSUGRA, as implemented in software programs, such
as SoftSUSY~\cite{Allanach:2001kg}.

\begin{itemize}
\setlength{\itemsep}{1pt}\setlength{\parskip}{0pt}\setlength{\parsep}{0pt}

\item {\em Step 1: Choose a framework with input parameters $P_i$.} In
  mSUGRA, the input parameters are $\{ P_i \} = \{ m_0, M_{1/2}, A_0,
  \tan\beta, \text{sign}(\mu) \}$.

\item {\em Step 2: Specify a model.} A model is specified by choosing
  values for the input parameters and using experimental data and RGEs
  to determine all the remaining parameters.  One key constraint on
  the weak-scale parameters is the relation
\begin{equation}
m_Z^2 = 2 \frac{ m_{H_d}^2 - m_{H_u}^2 \tan^2 \beta}{\tan^2 \beta - 1} 
- 2 \mu^2 \ ,
\label{mZ}
\end{equation}
suitably improved to include subleading corrections.

\item {\em Step 3: Choose a set of fundamental parameters $a_i$.}
  These parameters are independent and continuously variable; they are
  not necessarily the input parameters. In mSUGRA, a common choice is
  the GUT-scale parameters $\{ a_i \} = \{ m_0, M_{1/2}, A_0, B_0,
  \mu_0 \}$.

\item {\em Step 4: Calculate the sensitivity parameters $\caln_i$.}
  These parameters are~\cite{Ellis:1986yg,Barbieri:1987fn}
\begin{equation}
\caln_i
\equiv \left| \frac{\partial \ln m_Z^2} {\partial \ln a_i^2} \right| 
= \left| \frac{a_i^2}{m_Z^2} 
\frac{\partial m_Z^2} {\partial a_i^2} \right| \ . 
\label{calni}
\end{equation}
They measure the sensitivity of the weak scale, represented by the $Z$
mass, to variations in the fundamental parameters.

\item {\em Step 5: Determine the overall measure of naturalness $\caln
  \equiv \text{max} \{ \caln_i \}$.}  In mSUGRA, the overall measure
  of naturalness is, then, $\caln \equiv \text{max} \{\caln_{m_0},
  \caln_{M_{1/2}}, \caln_{A_0}, \caln_{B_0}, \caln_{\mu_0} \}$.

\end{itemize}

\subsection{Subjective Choices}
\label{sec:subjective}

There are many subjective choices and caveats associated with each of
the steps outlined in \secref{prescription}.  Here we highlight some
of these for each step in turn.

\subsubsection{Choosing a Framework}

This initial step is absolutely crucial, as all naturalness studies
are inescapably model-dependent. In any supersymmetry study, some
fundamental framework must be adopted. In studies of other topics,
however, there exists, at least in principle, the possibility of a
model-independent study, where no correlations among parameters are
assumed. This model-independent study is the most general possible, in
that all possible results from any other (model-dependent) study are a
subset of the model-independent study's results. In studies of
naturalness, however, the correlations determine the results, and
there is no possibility, even in principle, of a model-independent
study in the sense described above.

Given this caveat, there are two general approaches, each with their
advantages and disadvantages.  The first is a bottom-up approach, in
which one relaxes as many theoretical assumptions as is reasonable in
the hope that one might derive some generic insights.  The drawback to
this approach is that, since generic weak-scale supersymmetry is
excluded by experimental constraints, we expect there to be structure
in the supersymmetry-breaking parameters, which implies correlations,
which impact naturalness.  Ignoring these correlations is analogous to
ignoring constraints from, say, the CPT theorem, allowing the electric
charges of the electron and positron to be independent parameters, and
concluding that the neutrality of positronium is incredibly
fine-tuned.  Of course, for supersymmetry, we do not know what the
underlying correlations are, but we know there are some, and the only
assumption that is guaranteed to be wrong is that the
supersymmetry-breaking parameters are completely uncorrelated.

The second approach is a top-down approach, in which one takes various
theoretical frameworks seriously and analyzes their naturalness
properties, incorporating all the assumed correlations of the
framework.  The hope is that by examining various frameworks in
sufficient detail and sampling enough of them, one can derive new
insights to resolve known problems.  The disadvantage here, of course,
is that it is unlikely that any of the known frameworks correctly
captures all the correlations realized in nature.

\subsubsection{Specifying a Model}  

As noted above, it is important to include subleading corrections to
the tree-level expression for $m_Z^2$.  For example, it is important
to use 2-loop RGEs and 1-loop threshold corrections, decouple
superpartners at their masses, and minimize the electroweak potential
at an appropriate scale (typically the geometric mean of the stop
masses).  The tree-level expression of \eqref{mZ} is very useful to
obtain an intuitive understanding of many naturalness results, but it
does not capture many dependencies, especially in the case of heavy
superpartners.

\subsubsection{Choosing a Set of Fundamental Parameters}  
\label{sec:choosing}

Many naturalness studies differ at this step.  As an example, let's
consider mSUGRA.  The choice given above follows from the view that
GUT-scale parameters are more fundamental than weak-scale parameters
and that sensitivity of the weak scale to variations in any of the
parameters $m_0$, $M_{1/2}$, $A_0$, $B_0$, and $\mu_0$ is an
indication of fine-tuning.

Another choice is simply $\{ a_i \}$ = $\{ \mu_0 \}$.  The advantage
of this choice is that it is extremely simple to implement.  The $\mu$
parameter is (barely) multiplicatively renormalized, and so $c_{\mu_0}
\equiv \partial \ln m_Z^2 / \partial \ln \mu_0^2 = \partial \ln m_Z^2
/ \partial \ln \mu^2 \simeq 2 \mu^2 / m_Z^2$.  With this choice,
naturalness is, therefore, deemed equivalent to low $\mu$.  Some
string-inspired models in which all squarks masses are $\sim 10~\tev$
are claimed to be natural based on this
prescription~\cite{Feldman:2011ud}.

Such claims are subject to caveats, however.  Given our current
understanding, the $\mu$ parameter is typically assumed to have an
origin separate from the supersymmetry-breaking parameters.  It is
therefore reasonable to assume that it is not correlated with other
parameters, and so low $\mu$ is a necessary condition for naturalness.
(Note, however, that the discussion of EDMs in \secref{EDMs} provides
a counterargument.) Much more problematic, however, is that low $\mu$
is certainly not a sufficient condition for naturalness.  In the
moderate to large $\tan\beta$ limit, \eqref{mZ} becomes $m_Z^2 \approx
- 2 m_{H_u}^2 - 2 \mu^2$.  It is certainly possible for $m_{H_u}^2$ to
be small as the result of large cancellations.  In this case, $\mu$
will be small.  But this does not imply there is no fine-tuning: the
relation $a - b - c = 1$ with $a = 1,000,000$, $b = 999,998$, and $c =
1$ is fine-tuned, despite the fact that $c$ is small.  Claims that
such theories are natural are implicitly assuming that some
unspecified correlation explains the large cancellation that yields
low $m_{H_u}^2$.

A third possible choice for the fundamental parameters is to include
not only the dimensionful supersymmetry-breaking parameters, but all
of the parameters of the standard model.  Some naturalness studies
include these~\cite{Barbieri:1987fn,Ross:1992tz,deCarlos:1993yy,%
  Anderson:1994tr,Romanino:1999ut}, while others do
not~\cite{Ellis:1986yg,Ciafaloni:1996zh,Bhattacharyya:1996dw,%
  Chan:1997bi,Barbieri:1998uv,Chankowski:1998xv}.  {}From a low-energy
point of view, one should include all the parameters of the
Lagrangian.  However, by assuming some underlying high energy
framework and defining our parameters at $\mgut$, we have already
abandoned a purely low-energy perspective.  Once we consider the
high-energy perspective, the case is not so clear. For example, the
top Yukawa coupling $y_t$ may be fixed to a specific value in a sector
of the theory unrelated to supersymmetry breaking.  An example of this
is weakly-coupled string theory, where $y_t$ may be determined by the
correlator of three string vertex operators and would therefore be
fixed to some discrete value determined by the compactification
geometry.  The fact that all of the Yukawa couplings are roughly 1 or
0 helps fuel such speculation.\footnote{Of course, one might argue
  that some more fundamental theory will fix all parameters, including
  those that break supersymmetry.  There are no known examples,
  however.}  In such a scenario, it is clearly inappropriate to vary
$y_t$ continuously to determine the sensitivity of the weak scale to
variations in $y_t$.  Dimensionless couplings may also be effectively
fixed if they run to fixed points.  Other such scenarios are discussed
in Ref.~\cite{Feng:2000bp}. In the end, it is probably reasonable to
consider the fundamental parameters both with and without the
dimensionless parameters and see if any interesting models emerge.
Note that the question of which parameters to include in the $\{ a_i
\}$ is independent of which parameters have been measured; see
\secref{calculating}.

A final alternative approach is to choose the fundamental parameters
to be weak-scale parameters.  This is perhaps the ultimate bottom-up
approach, and it has the advantage of being operationally simpler than
having to extrapolate to the GUT or Planck scales.  However, as noted
above, many of the motivating virtues of supersymmetry are tied to
high scales, and some structure must exist if weak-scale supersymmetry
is to pass experimental constraints.  Working at the weak-scale
ignores such structure. It is possible, however, to view sensitivity
to variations in electroweak parameters as a lower bound on
sensitivities to variations in high-scale parameters, as they neglect
large logarithm-enhanced terms; see, \eg, Ref.~\cite{Baer:2012cf}.

\subsubsection{Calculating the Sensitivity Parameters} 
\label{sec:calculating}

Alternative choices, sometimes found in the literature, are $\caln_i
\equiv \left| \partial \ln m_Z^2 / \partial \ln a_i \right|$ or
$\caln_i \equiv \left| \partial \ln m_Z / \partial \ln a_i^2 \right|$.
There is little reason to choose one over the others, except in the
case of scalar masses, where $m_0^2$ is the fundamental parameter, not
$m_0$ ($m_0^2$ may be negative, for example).  In any case, these
definitions differ by factors of only 2 or 4, which should be ignored.
This is easier said than done: for example, one definition may yield
$\caln_i = 20$, or ${\cal O}(10)\%$ fine-tuning, while the other
definition yields $\caln_i = 80$ or ${\cal O}(1)\%$ fine-tuning,
leading to a rather different impression. Such examples serve as
useful reminders to avoid grand conclusions based on hard cutoffs in
naturalness measures.

There are other caveats in defining the sensitivity parameters.  The
role of the sensitivity coefficients is to capture the possibility of
large, canceling contributions to $m_Z^2$.  In principle, it is
possible to have a contribution to $m_Z^2$ that is small, but rapidly
varying, or large, but slowly varying.  It is also possible that
$m_Z^2$ is insensitive to variations of any single parameter, but
highly sensitive to variations in a linear combination of parameters.
In all of these cases, the sensitivity coefficients are highly
misleading, and these possibilities again serve as reminders of how
crude naturalness analyses typically are.

Finally, some studies have advocated alternative definitions of
sensitivity parameters that incorporate experimental
uncertainties. For example, some authors have proposed that the
definition of \eqref{calni}, be replaced
by~\cite{Ciafaloni:1996zh,Barbieri:1998uv}
\begin{equation}
\caln_i^{\text{exp}} \equiv 
\left| \frac{\Delta a_i^2}{m_Z^2} 
\frac{\partial m_Z^2} {\partial a_i^2}
\right| \ ,
\label{calniexp}
\end{equation}
where $\Delta a_i^2$ is the experimentally allowed range of
$a_i^2$. The intent of this definition is to encode the idea that
naturalness is our attempt to determine which values of parameters are
most likely to be realized in nature.

To contrast this definition with the conventional definition,
consider, for example, the hypothetical scenario in which our
theoretical understanding of supersymmetry has not improved, but the
$\mu$ parameter is measured to be $10^9~\gev$ with very high
accuracy. With the standard definition of \eqref{calni}, this model is
fine-tuned, but with \eqref{calniexp}, it is not.  In our view,
\eqref{calniexp} encodes an unconventional view of naturalness.
Naturalness is not a measure of our experimental knowledge of the
parameters of nature. Rather it is a measure of how well a given
theoretical framework explains the parameters realized in nature. It
is perfectly possible for values of parameters realized in nature to
be unnatural --- this is what the gauge hierarchy problem is!  --- and
once a parameter's value is reasonably well-known, naturalness cannot
be increased (or decreased) by more precise measurements.

\subsubsection{Determining the Overall Measure of Naturalness}
\label{sec:overall}

There are many possible ways to combine the $\caln_i$ to form a single
measure of naturalness.  A simple variation, advocated by some
authors, is to add the $\caln_i$ in quadrature.

There are also reasons to consider normalizing the $\caln_i$ either
before or after combining them.  The rationale for this is that in
certain cases, all possible choices of a fundamental parameter may
yield large sensitivities.  A well-known example of this is the
hierarchy between $\mplanck$ and $\LambdaQCD \sim \mplanck
e^{-c/g^2}$, which is often considered the textbook example of how to
generate a hierarchy naturally. The related sensitivity parameter,
\begin{equation}
c_g \equiv \left| \frac{\partial \ln \LambdaQCD} 
{\partial \ln g} \right| 
= \ln ( \mplanck^2 / \LambdaQCD^2) \sim 90\ ,
\label{cg}
\end{equation}
however, is large.  The authors of
Refs.~\cite{deCarlos:1993yy,Anderson:1994tr,Anderson:1994dz} have
argued that in such cases, only {\em relatively} large sensitivities
should be considered fine-tuned, and conclusions based on sensitivity
parameters consistently overestimate the degree of fine-tuning
required.  These authors propose replacing the sensitivity parameters
$\caln_i$ defined above, with fine-tuning parameters, defined as
$\gamma_i \equiv \caln_i / \bar{\caln}_i$, where $\bar{\caln}_i$ is an
average sensitivity.  These $\gamma_i$ are then combined to form an
overall measure of naturalness.

Unfortunately, the averaging procedure brings additional
complications.  If it is done only over a subspace of parameter space,
it may mask important features~\cite{Feng:2000bp}, and so it should be
carried out over the entire parameter space, which is computationally
intensive.  In addition, it requires defining a measure on the
parameter space and defining its boundaries.  These additional
complications have dissuaded most authors from including an averaging
procedure.  Nevertheless, many would agree that the sensitivity
parameters should, in principle, be normalized in some way, and the
naturalness parameter derived from un-normalized sensitivity
parameters exaggerates the fine-tuning required for a given model.

\subsection{Naturalness Bounds}
\label{sec:natbounds}

We now derive upper bounds on superpartner masses from naturalness
considerations.  Given all the caveats of \secref{subjective}, it
should go without saying that these should be considered at most as
rough guidelines.  The goal here is to give a concrete example of how
naturalness bounds may be derived, compare these with the other
theoretical and experimental constraints discussed in
\secsref{theoretical}{experimental}, and provide a starting point for
the discussion of models in \secref{models}.

We will consider a bottom-up approach, following the general
prescription of \secref{prescription}.  We consider a model defined at
the GUT-scale with input parameters $P_i = M_1, M_2, M_3, m_{H_u},
m_{H_d}, m_{Q_3}, m_{U_3}, m_{D_3}, A_t, \ldots, \text{sign}(\mu)$.
These include the gaugino masses $M_i$, the soft SUSY-breaking scalar
masses, and the $A$-terms, all treated as independent.  The weak-scale
value of $|\mu|$ is determined by $m_Z$.  The fundamental parameters
are taken to be the GUT-scale values of the input parameters, with
$\text{sign}(\mu)$ replaced by the GUT-scale value of $\mu$.
Sensitivity parameters are defined as in \eqref{calni}, and the
overall naturalness parameter is defined as the largest one.

The weak-scale values of supersymmetry-breaking parameters may be
determined analytically or numerically in terms of their GUT-scale
values~\cite{Ibanez:1983di,Ibanez:1984vq,Abe:2007kf,Martin:2007gf,%
  Antusch:2012gv}. Recent analyses for $\tan\beta = 10$ and using 1-
and 2-loop RGEs find~\cite{Abe:2007kf,Martin:2007gf}
\begin{eqnarray}
M_1 (\mweak) &=& 0.41 M_1 \label{M1} \\
M_2 (\mweak) &=& 0.82 M_2 \label{M2} \\
M_3 (\mweak) &=& 2.91 M_3 \label{M3} \\
- 2 \mu^2 (\mweak) &=& - 2.18 \mu^2 \label{mu} \\
- 2 m_{Hu}^2(\mweak) &=& 3.84 M_3^2 + 0.32 M_3 M_2 + 0.047 M_1 M_3 
- 0.42 M_2^2  \nonumber \\
&& + 0.011 M_2 M_1 - 0.012 M_1^2 - 0.65 M_3 A_t - 0.15 M_2 A_t \nonumber \\
&& - 0.025 M_1 A_t + 0.22 A_t^2 + 0.0040 M_3 A_b \nonumber \\
&& - 1.27 m_{H_u}^2 - 0.053 m_{H_d}^2 \nonumber \\
&& + 0.73 m_{Q_3}^2 + 0.57 m_{U_3}^2 
+ 0.049 m_{D_3}^2 - 0.052 m_{L_3}^2 + 0.053 m_{E_3}^2 \nonumber \\
&& + 0.051 m_{Q_2}^2 - 0.110 m_{U_2}^2
+ 0.051 m_{D_2}^2 - 0.052 m_{L_2}^2 + 0.053 m_{E_2}^2 \nonumber \\
&& + 0.051 m_{Q_1}^2 - 0.110 m_{U_1}^2
+ 0.051 m_{D_1}^2 - 0.052 m_{L_1}^2 + 0.053 m_{E_1}^2 \ ,
\label{Hu} 
\end{eqnarray}
where all the parameters on the right-hand sides of these equations
are GUT-scale parameters.  The RGEs mix the parameters.  Although
$H_u$ does not couple to gluinos directly, the gluino mass enters the
squark mass RGEs and the squark masses enter the $H_u$ RGE, and so
$m_{H_u}^2 (\mweak)$ depends on the gluino mass $M_3$.  For the first
and second generation sfermions, their Yukawa couplings are so small
that their main impact on the Higgs potential is through hypercharge
$D$-term contributions or, if GUT or other boundary conditions cause
these terms to vanish, through 2-loop effects in the $H_u$
RGE~\cite{Dimopoulos:1995mi,Pomarol:1995xc}.

The naturalness prescription of \secref{prescription} is applicable to
complete models, but we may derive rough bounds on individual
superpartner masses by neglecting other parameters when deriving the
bound on a given superpartner mass. As an example, keeping only the
$M_3^2$ term in \eqref{Hu}, we find
\begin{equation}
\caln_{M_3} \equiv 
\Biggl| \frac{ \partial \ln m_Z^2}{\partial \ln M_3^2} \Biggr|
\approx \frac{M_3^2}{m_Z^2} 
\Biggl| \frac{ \partial [- 2 m_{H_u}^2 ( \mweak ) 
- 2 \mu^2( \mweak ) ] } {\partial M_3^2} \Biggr|
= 3.84 \frac{M_3^2}{m_Z^2} \ .
\end{equation}
Requiring $\caln_{M_3} < \calnmax$ and using \eqref{M3}, we can derive
a naturalness bound on the physical gluino mass $m_{\gluino} \approx
M_3 (\mweak)$.  Proceeding in a similar way for all the parameters,
and using $m_{Q_3}^2 (\mweak) = 0.885 m_{Q_3}^2 + \ldots$, $m_{U_3}^2
(\mweak) = 0.770 m_{U_3}^2 + \ldots$, and $m_{\tilde{f}}^2 (\mweak)
\approx m_{\tilde{f}}^2 + \ldots$ for all other
sfermions~\cite{Abe:2007kf}, we find
\begin{eqnarray}
m_{\Higgsino} &\alt& 640~\gev \left( \calnmax / 100 \right)^{1/2} \\
m_{\Bino} &\alt& 3.4~\tev \left( \calnmax / 100 \right)^{1/2} \\
m_{\Wino} &\alt& 1.2~\tev \left( \calnmax / 100 \right)^{1/2} \\
m_{\gluino} &\alt& 1.4~\tev \left( \calnmax / 100 \right)^{1/2} \\
m_{\tilde{t}_L,\tilde{b}_L} &\alt& 1.0~\tev \left( \calnmax / 100
\right)^{1/2} \label{tbL} \\
m_{\tilde{t}_R} &\alt& 1.1~\tev \left( \calnmax / 100 \right)^{1/2} 
\label{tR} \\
m_{\tilde{b}_R} &\alt& 4.1~\tev \left( \calnmax / 100 \right)^{1/2}
\label{bR} \\
m_{\tilde{\tau}_L,\tilde{\nu}_{\tau}} 
&\alt& 4.0~\tev \left( \calnmax / 100 \right)^{1/2} \\
m_{\tilde{\tau}_R} &\alt& 4.0~\tev \left( \calnmax / 100 \right)^{1/2} \\
m_{\tilde{c}_L,\tilde{s}_L,\tilde{u}_L,\tilde{d}_L} &\alt& 4.0~\tev \left( \calnmax / 100 \right)^{1/2} \\
m_{\tilde{c}_R, \tilde{u}_R} &\alt& 2.7~\tev \left( \calnmax / 100 \right)^{1/2} \\
m_{\tilde{s}_R, \tilde{d}_R} &\alt& 4.0~\tev \left( \calnmax / 100 \right)^{1/2} \\
m_{\tilde{\mu}_L,\tilde{\nu}_{\mu}, \tilde{e}_L, \tilde{\nu}_{e}} 
&\alt& 4.0~\tev \left( \calnmax / 100 \right)^{1/2} \\
m_{\tilde{\mu}_R,\tilde{e}_R} &\alt& 4.0~\tev \left( \calnmax / 100
\right)^{1/2} \label{lR} \ .
\end{eqnarray}

If the standard model particles are unified into GUT multiplets at the
GUT scale, the correlations lead to significantly different
conclusions.  For example, assuming $M_{1/2} = M_3 = M_2 = M_1$,
$m_{\text{\bf 10}_i} = m_{Q_i} = m_{U_i} = m_{E_i}$ and $m_{\text{\bf
    5}_i} = m_{D_i} = m_{L_i}$, where $i = 1, 2$, we find, within
the accuracy of these numerical results,
\begin{eqnarray}
-2 m_{H_u}^2 \! ( \mweak ) = 3.79 M_{1/2}^2 \!
+ 0.0071 m_{\text{\bf 10}_2}^2 \! \! + 0.0013 m_{\text{\bf 5}_2}^2 \! \!
+ 0.0071 m_{\text{\bf 10}_1}^2 \! \! + 0.0013 m_{\text{\bf 5}_1}^2 \!
\! + \! \ldots ,
\end{eqnarray}
implying
\begin{eqnarray}
m_{\Bino} &\alt& 190~\gev \left( \calnmax / 100 \right)^{1/2} \\
m_{\Wino} &\alt& 380~\gev \left( \calnmax / 100 \right)^{1/2} \\
m_{\gluino} &\alt& 1.4~\tev \left( \calnmax / 100 \right)^{1/2} \\
m_{\tilde{u}_L, \tilde{d}_L, \tilde{c}_L, \tilde{s}_L, 
\tilde{u}_R, \tilde{c}_R, \tilde{e}_R, \tilde{\mu}_{R}}
&\alt& 11~\tev \left( \calnmax / 100 \right)^{1/2} \label{secondGUT}  \\
m_{\tilde{d}_R, \tilde{s}_R, \tilde{\nu}_e, \tilde{e}_L, 
\tilde{\nu}_{\mu}, \tilde{\mu}_L}
&\alt& 25~\tev \left( \calnmax / 100 \right)^{1/2} \ . \label{firstGUT}
\end{eqnarray}

These results may be understood as follows: The $\tilde{t}_L$,
$\tilde{b}_L$, and $\tilde{t}_R$ masses enter the $H_u$ RGE through
top Yukawa couplings, and their bounds in \eqsref{tbL}{tR} are
consistent with those of \eqref{roughbound}.  For the other sfermions,
the naturalness constraints are weaker.  Generically, these masses
enter the $H_u$ RGE dominantly through hypercharge $D$-terms, and so
one expects constraints on them to be weaker by a factor of
$\sqrt{\alpha_{y_t} / \alpha_1} = y_t / g_1 \sim 3$, consistent with
Eqs.~(\ref{bR})--(\ref{lR}). 

In the case of GUTs, for the gauginos, the most stringent bound is
from the gluino, with the Wino and Bino bounds following from the
relations $M_{\Bino} : M_{\Wino} : M_{\gluino} \approx 1 : 2 : 7$.
For the scalars, in GUTs the masses enter only through two-loop terms,
and so the constraints are weaker by a factor of $\sqrt{4\pi /
  \alpha_1} \sim 10$, as seen in
\eqsref{secondGUT}{firstGUT}~\cite{Dimopoulos:1995mi,Pomarol:1995xc}.
Note that the GUT correlations greatly strengthen the naturalness
bounds on Binos and Winos, but greatly weaken the bounds on first and
second generation scalars: the choice of underlying framework can have
enormous qualitative implications for naturalness bounds.

\section{MODEL FRAMEWORKS}
\label{sec:models}

We now discuss a few classes of models that have been proposed to
relieve the tension between the various constraints discussed so far.
To set the stage, we present all of the theoretical and experimental
constraints discussed in this review in \figref{spectrum_constraints}.

\begin{figure}
\includegraphics[width=0.78\columnwidth]{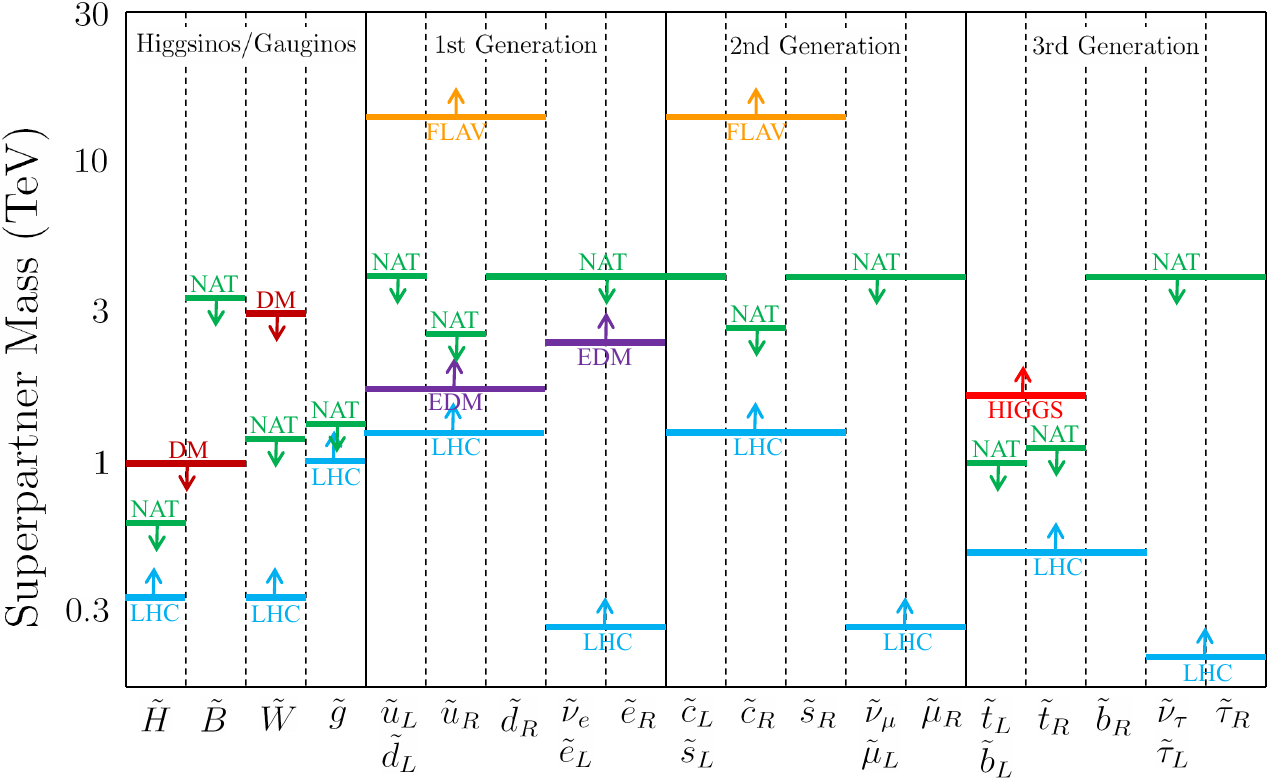}
\caption{A sample of constraints on the superpartner spectrum from
  naturalness (NAT), dark matter (DM), collider searches (LHC), the
  Higgs boson mass (HIGGS), flavor violation (FLAV), and EDM
  constraints (EDM).  The constraints assume a moderate value of $\tan
  \beta = 10$.  The naturalness constraints derive from a bottom-up
  analysis and scale as $( \calnmax / 100 )^{1/2}$, where $\calnmax$
  is the maximally allowed naturalness parameter; see
  \secref{quantifying}.  All of the constraints shown are merely
  indicative and subject to significant loopholes and caveats; see the
  text for details.  }
\label{fig:spectrum_constraints}
\end{figure}

\subsection{Effective Supersymmetry}
\label{sec:effective}

As evident from \figref{spectrum_constraints}, naturalness most
stringently restricts the masses of scalars with large Yukawa
couplings, since these are most strongly coupled to the Higgs sector.
At the same time, experimental constraints are most stringent for
scalars with small Yukawa couplings, since light fermions are most
easily produced and studied. This suggests that light fermions have
heavy superpartners and vice versa, which provides a promising way to
reconcile naturalness with flavor and CP
constraints~\cite{Drees:1985jx,Dine:1990jd,Dimopoulos:1995mi,%
  Pomarol:1995xc}.  A representative spectrum for such models, known
as ``effective supersymmetry''~\cite{Cohen:1996vb} or, alternatively,
``more minimal supersymmetry'' or ``inverted hierarchy models,'' is
shown in \figref{spectrum_effective}.

\begin{figure}
\includegraphics[width=0.78\columnwidth]{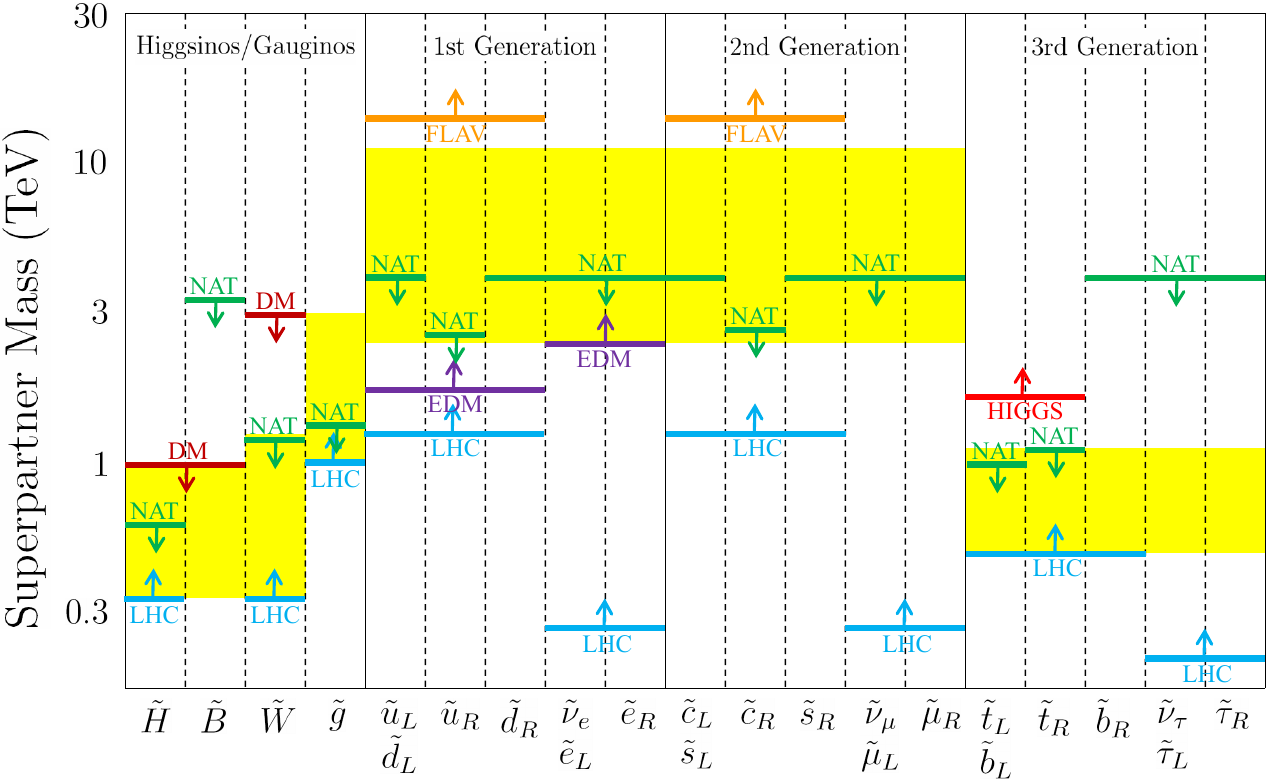}
\caption{Example superpartner mass ranges for effective supersymmetry
  (shaded) with constraints as given in \figref{spectrum_constraints}.
  Heavy and degenerate first and second generation sfermions satisfy
  flavor and EDM constraints, and light third generation sfermions
  satisfy naturalness.  The Higgs mass constraint requires either
  near-maximal stop mixing, or physics beyond the MSSM.  }
\label{fig:spectrum_effective}
\end{figure}

The effective supersymmetry spectrum may be realized in many
ways~\cite{Cohen:1996vb,Dvali:1996rj,Binetruy:1996uv,%
  Mohapatra:1996in,Nelson:1997bt,Agashe:1998zz,Kaplan:1998jk,%
  Hisano:1998tm,Kaplan:1999iq,Everett:2000hb,Feng:1998iq,%
  Bagger:1999ty,Bagger:1999sy,Krippendorf:2012ir,Badziak:2012rf}.  For
example, if there is an extra anomalous or non-anomalous U(1) gauge
group under which the first two generations are charged, but the third
generation is neutral, the first two generation sfermions may receive
additional contributions to their mass.  The U(1) symmetry may also
allow fermion masses for the third generation, but forbid masses for
the first and second, naturally explaining the inverted hierarchy
structure.  Alternatively, the split sparticle spectrum may be
radiatively generated~\cite{Feng:1998iq,Bagger:1999ty,Bagger:1999sy}.
All sparticle masses may begin at $\sim 10~\tev$ at the GUT scale, but
for particular GUT-scale boundary conditions, those with large Yukawa
couplings may be driven to low values at the weak scale.  In these
scenarios, the large Yukawa coupling produces both heavy fermions and
light sfermions, again naturally explaining the inverted hierarchy
structure.

Effective supersymmetry predates not only the Higgs discovery, but
even the most stringent LEP limits on the Higgs mass, and was not
originally intended to explain the large Higgs boson mass.  As
discussed in \secref{higgs}, the Higgs boson mass may be consistent
with sub-TeV stops, but only in the highly fine-tuned case when there
is large left-right mixing; for a recent discussion of this in the
context of effective supersymmetry, see
Ref.~\cite{Badziak:2012rf}. Alternatively, more minimal supersymmetry
may be made less minimal by adding extra fields to raise the Higgs
mass; see, \eg, Refs.~\cite{Barger:2006dh,Ellwanger:2009dp}.
Effective supersymmetry with this extension has attracted renewed
attention, sometimes under the confusingly generic moniker ``natural
supersymmetry,'' as a strategy to reconcile naturalness with LHC
constraints~\cite{Kitano:2006gv}.

In effective supersymmetry, the first and second generation sfermions
are beyond the reach of the LHC, but gluinos, stops, and sbottoms may
be within reach.  The most promising collider signals are therefore
direct stop and sbottom squark production, or gluinos with top- and
bottom-rich cascade decays~\cite{Baer:2010ny}. Effects in low-energy
$B$ physics may also be accessible~\cite{Cohen:1996sq}.  Finally, new
particles added to raise the Higgs mass may have associated signals.

\subsection{Focus Point Supersymmetry}
\label{sec:focuspoint}

In focus point
supersymmetry~\cite{Feng:1999mn,Feng:1999zg,Feng:2000bp}, correlations
between parameters allow sparticle masses to be larger than their
naive naturalness bounds.  A representative spectrum with heavy
scalars is given in \figref{spectrum_fp}.  Heavy first and second
generation scalars suppress flavor and CP violation, as in effective
supersymmetry.  In contrast to effective supersymmetry, however, the
third generation is also heavy, naturally raising the Higgs mass to
within current bounds.  There are many theoretical reasons for
expecting scalar superpartners to be heavier than the gauginos.  For
example, such a hierarchy follows from an approximate U(1)$_R$
symmetry, which suppresses the gaugino masses (and $A$-terms) but not
the scalar masses.  It also results if there are no singlet
supersymmetry-breaking fields~\cite{Randall:1998uk,Giudice:1998xp}.
Note that gaugino masses enter the scalar mass RGEs, but scalar masses
do not enter the gaugino mass RGEs; the hierarchy $m_0 \gg M_{1/2}$ is
therefore stable under RGE evolution, whereas $M_{1/2} \gg m_0$ is
not.

\begin{figure}
\includegraphics[width=0.78\columnwidth]{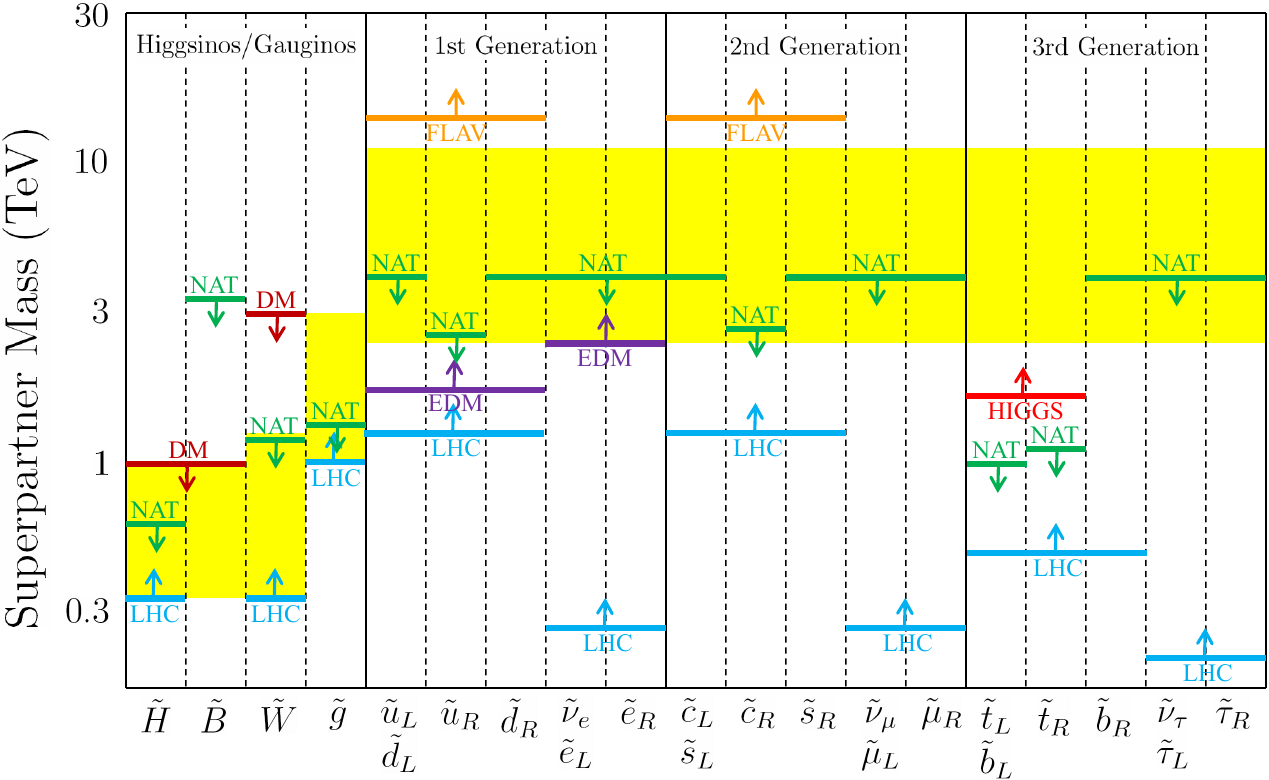}
\caption{Example superpartner mass ranges for focus point
  supersymmetry (shaded) with constraints as given in
  \figref{spectrum_constraints}.  Heavy and degenerate first and
  second generation sfermions satisfy flavor and EDM constraints and
  heavy third generation sfermions raise the Higgs mass, while
  naturalness is preserved despite heavy stops by correlations between
  GUT-scale parameters. }
\label{fig:spectrum_fp}
\end{figure}
 
The obvious difficulty is that heavy top squarks naively contradict
naturalness. In focus point supersymmetry, correlations in GUT-scale
parameters are invoked to alleviate this fine-tuning.  A simple
example is evident from \eqref{Hu}.  The weak-scale value of
$m_{H_u}^2(\mweak)$ is highly sensitive to the GUT-scale values of
$m_{H_u}^2$, $m_{Q_3}^2$, and $m_{U_3}^2$, but if these have a unified
value $m_0^2$ at the GUT scale, $- 2 m_{Hu}^2(\mweak) = - 1.27
m_{H_u}^2 + 0.73 m_{Q_3}^2 + 0.57 m_{U_3}^2 + \ldots = 0.03 m_0^2 +
\ldots$, and the weak scale becomes highly insensitive to variations
in these GUT-scale parameters, even if they are large.  The reasoning
here is similar to that leading to natural $\sim 10~\tev$ first and
second generation sfermions with $\calnmax \sim 100$ in the GUT case
analyzed in \secref{natbounds}. 

This behavior may be understood as a property of the RGEs.  The
$m_{H_u}^2$ RGEs in a focus point model are shown in \figref{fpRGEs}.
The RG trajectories have a focus point at the weak scale, and so the
weak-scale value of $m_{H_u}^2$ is insensitive to variations in the
GUT-scale parameters.  The weak scale still receives quadratic
contributions from heavy stops, but the large logarithm enhancement
from RG evolution in \eqref{roughbound} is absent, reducing the
fine-tuning associated with multi-TeV stops by a factor of $\sim
\ln(\mgut^2/\mweak^2) \sim 60$.  Such focusing occurs if the GUT-scale
parameters satisfy~\cite{Feng:1999mn}
\begin{equation}
( m_{H_u}^2, m_{\tilde{t}_R}^2, m_{\tilde{t}_L}^2 ) \propto
(1, 1+x, 1-x)
\label{moderatetb}
\end{equation}
for moderate values of $\tan\beta$, and
\begin{equation}
( m_{H_u}^2, m_{\tilde{t}_R}^2, m_{\tilde{t}_L}^2,
m_{\tilde{b}_R}^2, m_{H_d}^2 ) \propto
(1, 1+x, 1-x, 1+x-x', 1+x' )
\label{largetb}
\end{equation}
for large values of $\tan\beta$, where $x$ and $x'$ are arbitrary
constants.  Note that the scale at which focusing occurs is sensitive
to dimensionless couplings, particularly the top Yukawa $y_t$.  As
discussed in \secref{choosing}, one may include $y_t$ as a fundamental
parameter or not.  If it is included, $\caln_{y_t}$ is large for large
superpartner masses, but it is large throughout parameter
space~\cite{Feng:1999zg}.  If one adopts the averaging procedure
described in \secref{overall} to identify only relatively large
sensitivities, the effect of including $y_t$ as a fundamental
parameter is greatly moderated.

A universal scalar mass obviously satisfies both
\eqsref{moderatetb}{largetb}, and the large $m_0$ region of mSUGRA has
become the canonical example of focus point
supersymmetry~\cite{Feng:1999mn,Feng:1999zg,Chan:1997bi}.  Focus point
supersymmetry is, however, a far more general phenomenon, as one may
postulate many relations between the GUT-scale parameters to reduce
the fine-tuning in \eqref{Hu}. For example, considering the $M_2^2$,
$M_2 M_3$, and $M_3^2$ terms, one finds focusing for $M_3/M_2 \approx
0.3, -0.4$ at the GUT scale, allowing large, non-universal gaugino
masses to be natural~\cite{Abe:2007kf,Horton:2009ed,%
  Younkin:2012ui,Antusch:2012gv,Yanagida:2013ah}.  Focusing may also
be found in models with right-handed neutrinos~\cite{Asano:2011kj} and
large $A$-terms~\cite{Feng:2012jfa}, and may emerge from the boundary
conditions of mirage
mediation~\cite{Choi:2005hd,Kitano:2005wc,Lebedev:2005ge} or be
enforced by a symmetry~\cite{Yanagida:2013ah}.

\begin{figure}
    \includegraphics[width=\figwidth]{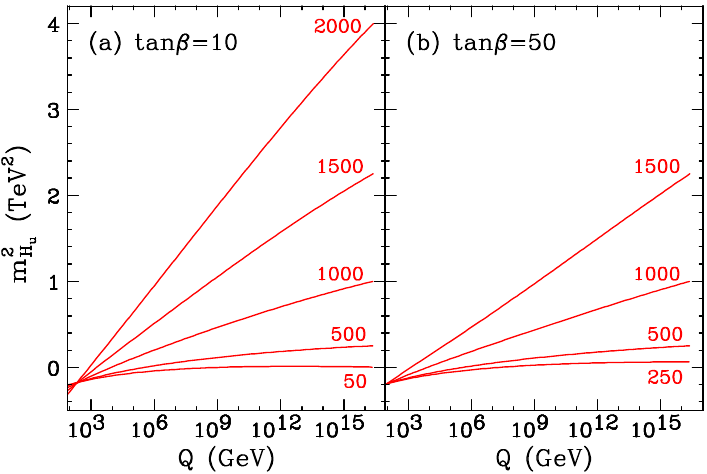}
\caption{The RG evolution of $m_{H_u}^2$ in the focus point region of
  mSUGRA for $\tan\beta = 10$ (left) and 50 (right), several values of
  $m_0$ (as shown in GeV), $M_{1/2} = 300~\gev$, and $A_0 = 0$. For
  both values of $\tan\beta$, $m_{H_u}^2$ exhibits an RG focus point
  near the weak scale, implying that the weak scale is insensitive to
  variations in the GUT-scale supersymmetry-breaking
  parameters~\cite{Feng:1999zg}. }
\label{fig:fpRGEs}
\end{figure}

In the most studied focus point supersymmetry models, all scalars are
heavy, but typically the stops are slightly lighter.  They may be
produced in future LHC runs, or may be beyond reach, but light enough
to enhance the top content of gluino decays.  The most promising LHC
signals are therefore again direct stop production with cascade decays
through charginos and neutralinos, or gluino production, followed by
top- and bottom-rich cascade decays.  As the first generation scalars
are heavy but not extremely heavy, there may also be a signal in EDMs.
Last, the prospects for WIMP dark matter detection are extremely
promising in focus point models~\cite{Feng:2000gh,Feng:2000zu}.  In
particular, focus point supersymmetry predicts a mixed Bino-Higgsino
neutralino with a spin-independent proton cross sections typically
above the zeptobarn level, which should be probed in the coming few
years.

Last, note that the large logarithm enhancement may also be eliminated
by adding additional particles.  This is the approach of an entirely
different class of models, typically called ``supersoft
supersymmetry''~\cite{Fox:2002bu}, where the MSSM is extended to
include a gauge adjoint chiral superfield for each gauge group,
providing an interesting alternative strategy for reconciling
naturalness with experimental
constraints~\cite{Kribs:2007ac,Kribs:2012gx}.

\subsection{Compressed Supersymmetry}
\label{sec:compressed}

In many supersymmetric models, there is a large mass splitting between
the gluino and squarks at the top of the spectrum and the lighter
superparters at the bottom.  This reduces the naturalness of these
models in two ways.  First, the large mass splittings imply that the
gluino and squark cascade decays produce energetic particles and large
$\met$, leading to distinct signals and strong bounds on sparticle
masses. Second, lower bounds on the masses of the lighter sparticles
imply stringent lower bounds on gluino and squark masses, which
decreases naturalness.

The superpartner spectrum may be much more degenerate, however.  This
has been explored in the context of ``compressed
supersymmetry''~\cite{Martin:2007gf}, in which there are small
splittings between colored superpartners and an LSP neutralino.  For
the reasons given above, this leads to weaker bounds on sparticle
masses and provides an interesting approach to developing viable and
natural models~\cite{Kane:1998im,BasteroGil:1999gu}.

There are well-motivated reasons to expect large mass splittings.  RG
evolution drives up colored sparticles masses relative to uncolored
ones.  For example, assuming gaugino mass unification at the GUT
scale, Eqs.~(\ref{M1})--(\ref{M3}) imply $|M_1| : |M_2| : |M_3|
\approx 1 : 2 : 7$ at the weak scale.  This is not a strict prediction
of GUT models, however~\cite{Ellis:1984bm}. For example, if gaugino
masses are generated not by gauge singlet $F$-terms, but by a {\bf 75}
multiplet of SU(5), group theoretic factors imply $|M_1| : |M_2| :
|M_3| \approx 5 : 3 : 1$ at the GUT scale, leading to $|M_1| : |M_2| :
|M_3| \approx 5 : 6 : 6$ and a highly degenerate spectrum at the weak
scale~\cite{Anderson:1996bg}. Note that the $M_3^2$ and $M_2^2$ terms
enter with opposite signs in \eqref{Hu}, and so when $|M_2|$ is a
little larger than $|M_3|$ at the GUT scale, these terms partially
cancel and naturalness is improved by essentially the same mechanism
discussed in \secref{focuspoint} for focus point scenarios with
non-universal gaugino masses.

A representative spectrum is shown in \figref{spectrum_compressed}.
The virtue of compressed supersymmetry is that it decreases the
tension between naturalness and LHC superpartner search bounds.  A
shortcoming of these models is that the light spectrum exacerbates
problems with flavor and CP violation.  In particular, $\sim 100~\gev$
superpartners generically require $\phi_{\text{CP}} \alt 10^{-4} -
10^{-3}$ to satisfy EDM constraints, and so these models require some
additional mechanism to suppress CP violation.  In addition, the
problem of obtaining a 125 GeV Higgs boson mass is present in
compressed supersymmetry if the stops are light.  As in the case of
effective supersymmetry, physics beyond the
MSSM~\cite{Barger:2006dh,Ellwanger:2009dp} is required to raise the
Higgs mass to its measured value, bringing with it additional
complications.

\begin{figure}
\includegraphics[width=0.78\columnwidth]{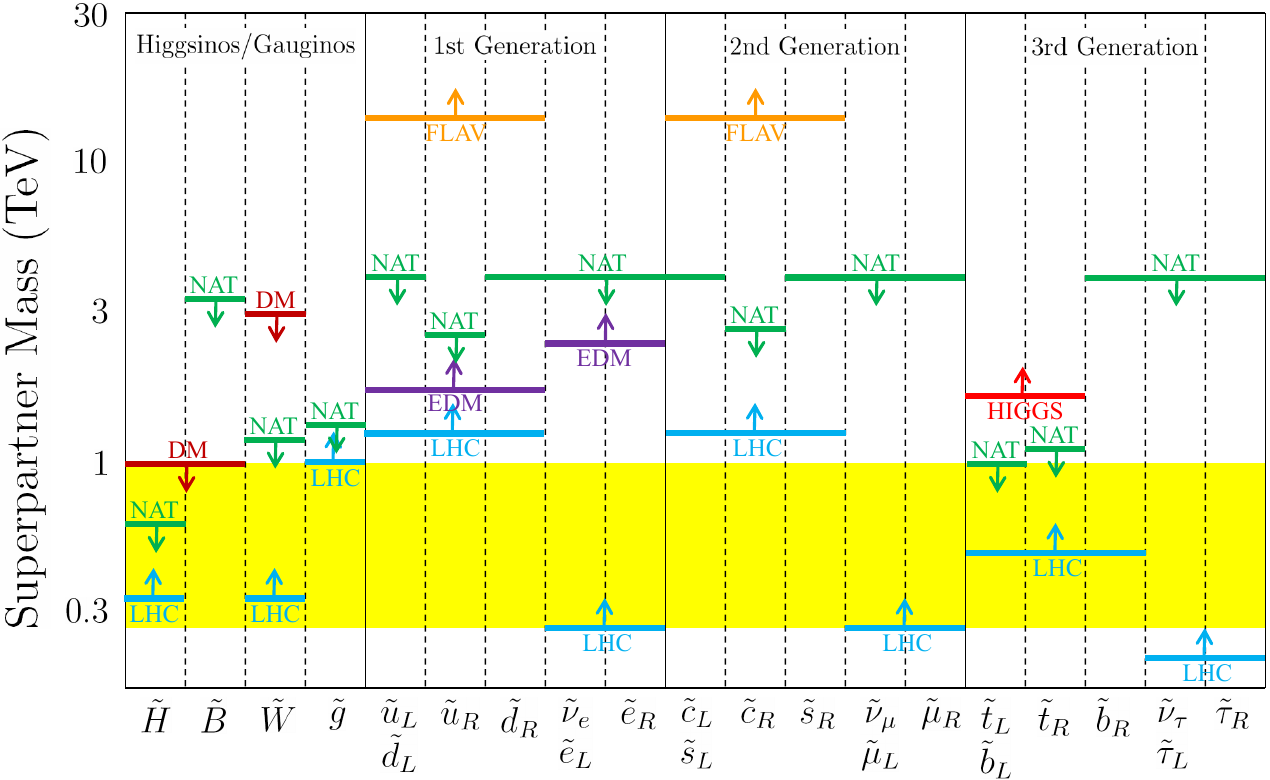}
\caption{Example superpartner mass ranges for compressed supersymmetry
  and RPV supersymmetry (shaded) with constraints as given in
  \figref{spectrum_constraints}.  Light sfermions preserve naturalness
  and evade LHC bounds because $\met$ signals are degraded by
  superpartner degeneracies in compressed supersymmetry or by LSP
  decays in RPV supersymmetry.  These models require mechanisms to
  eliminate flavor violation and reduce CP-violating phases to ${\cal
    O}(10^{-3}) - {\cal O}(10^{-4})$, and also require near-maximal
  stop mixing or physics beyond the MSSM to raise the Higgs mass.  In
  addition, in RPV supersymmetry, there is no WIMP dark matter
  candidate.}
\label{fig:spectrum_compressed}
\end{figure}

The collider signals of compressed supersymmetry have been explored in
a number of studies~\cite{Baer:2007uz,Martin:2008aw,LeCompte:2011cn,%
  LeCompte:2011fh,Rolbiecki:2012gn,Belanger:2012mk,Dreiner:2012gx}.
The relevant signals depend on the degree of compression.  For
$m_{\tilde{t}} - m_{\chi} < m_t$, stops may dominantly decay via
$\tilde{t} \to b W^+ \chi$ or even $\tilde{t} \to c \chi$ or
$\tilde{t} \to b f \bar{f}' \chi$, leading to softer leptons and a
suppressed multi-lepton rate~\cite{Baer:2007uz,Rolbiecki:2012gn}.  For
even greater degeneracies, the only possible signals are
monophotons~\cite{Belanger:2012mk} and monojets~\cite{Dreiner:2012gx}.
At present these searches imply $m_{\gluino} \agt 500~\gev$.
Implications for neutralino dark matter have been explored in
Refs.~\cite{Martin:2007gf,Baer:2007uz,Martin:2007hn}.

Finally, there are many other models in which the $\met$ signal is
reduced.  Interesting possibilities in which cascade decays go through
hidden sectors include hidden valley
models~\cite{Strassler:2006im,Strassler:2006qa} and stealth
supersymmetry~\cite{Fan:2011yu}.

\subsection{$R$-Parity-Violating Supersymmetry}
\label{sec:rparity}

As discussed in \secref{rpLHC}, the characteristic $\met$ collider
signal of supersymmetry may also be degraded in the presence of
$R$-parity violation.  If any of the superpotential terms of
\eqref{rpterms} is non-zero, all superpartners decay, and, provided
the decay length is not too long, supersymmetric particles do not
escape the detector.  The phenomenology of RPV supersymmetry has been
studied for a long time~\cite{Hall:1983id,Ellis:1984gi,Barger:1989rk},
but it has recently attracted renewed attention as a way to make light
superpartners viable, and thereby reduce fine-tuning.

In general, once one allows $R_p$ violation, one opens a Pandora's box
of possibilities.  There are few principled ways to violate $R$-parity
conservation.  The RPV couplings cannot all be sizable.  In fact,
there are stringent bounds on individual RPV couplings, and even more
stringent bounds on products of pairs of
couplings~\cite{Allanach:1999ic,Barbier:2004ez}; in particular, if any
lepton number-violating coupling and any baryon number-violating
coupling are both non-zero, proton decay sets extremely stringent
constraints.

If theory is any guide, one might expect that the RPV couplings follow
the pattern of the $R_p$-conserving couplings, with those involving
the third generation the biggest, the second generation smaller, and
the first smaller still. Realizations of this hypothesis have been
presented in Refs.~\cite{Csaki:2011ge,Krnjaic:2012aj,%
  Bhattacherjee:2013gr,Franceschini:2013ne,Csaki:2013we}, where models
of $R_p$ violation based on the principle of minimal flavor violation
lead to scenarios in which only the hadronic RPV terms
$\lambda''_{ijk} U_i D_j D_k$ are sizable, with $\lambda''_{323}$
typically the largest.  Such models somewhat moderate the naturalness
motivation for $R_p$ violation, as they imply that, say, the LSP
neutralino decays dominantly to top and bottom quarks, leading to
$b$-jets, leptons, and $\met$ from neutrinos, all distinctive
characteristics that one was hoping to avoid.  Nevertheless, such RPV
signals likely do reduce LHC limits somewhat, and, of course, one may
always ignore theoretical bias and consider, say, $\lambda''_{112}$
couplings that would lead to decays to light flavor and pure jet
signals, such as those discussed in \secref{rpLHC} and
\figref{LHCnomet}.

In summary, as with compressed supersymmetry, $R_p$ violation provides
another possibility for reducing the distinctiveness of supersymmetry
signals at colliders and potentially improving the naturalness of
viable models.  The shortcomings, however, are also similar: if all of
the superpartners are light, the Higgs boson is generically too light,
requiring physics beyond the MSSM, and the EDM constraints are
generically not satisfied, requiring yet more structure to remove
troubling CP-violating phases.  In RPV supersymmetry, one also loses
the motivation of WIMP dark matter, although the gravitino or other
candidates may play this role.

\section{CONCLUSIONS}
\label{sec:conclusions}

\noindent {\em PARABLE. Some children notice that a soap bubble's
  width and height are remarkably similar. They get excited when they
  find that this can be explained by surface tension and rotational
  symmetry.  Later, with amazing experiments, they find that the width
  and height are not identical, but differ by 1 part in
  $10^{28}$. They remember, however, that wind can distort the shape
  of the bubble and calculate that, given typical winds, one would
  expect differences of 1 part in $10^{30}$. Some of the children
  become despondent and wonder how such a beautiful solution could be
  so wrong; others consider alternative explanations; others postulate
  that bubbles can be any shape, but only nearly spherical ones are
  compatible with the presence of children; and others study the
  wind.} \\

Supersymmetry has long been the leading candidate for new physics at
the weak scale.  In this review, we have evaluated its current status
in light of many theoretical and experimental considerations.

The leading theoretical motivations for weak-scale supersymmetry are
naturalness, grand unification, and WIMP dark matter.  Each of these
prefers supersymmetry breaking at the weak scale, but each argument is
subject to caveats outlined in \secref{theoretical}.  Of course, taken
as a whole, these continue to strongly motivate supersymmetry.

Current experimental constraints are discussed in
\secref{experimental} and summarized in \figref{spectrum_constraints}.
For some varieties of supersymmetry models, the LHC now requires
superpartner masses well above 1 TeV, but there are also
well-motivated examples in which superpartners may be significantly
lighter without violating known bounds.  The 125 GeV Higgs boson mass
prefers heavy top squarks in the MSSM, and longstanding flavor and CP
constraints strongly suggest multi-TeV first and second generation
sfermions.  We have especially emphasized the robustness of the EDM
constraints, which are present even in flavor-conserving theories.  In
the absence of a compelling mechanism for suppressing CP violation,
the EDM constraints require first generation sfermions to be well
above the TeV scale.  Against the backdrop of these indirect
constraints, LHC bounds on supersymmetry are significant because they
are direct, but they are hardly game-changing. One may like
supersymmetry or not, but to have thought it promising in 2008 and to
think it much less promising now is surely the least defensible
viewpoint.

In \secref{quantifying}, we have critically examined attempts to
quantify naturalness.  There are many studies embodying philosophies
that differ greatly from each other.  We have expressed reservations
about some, but for many, one can only acknowledge the subjective
nature of naturalness and make explicit the underlying assumptions.
Very roughly speaking, however, current bounds are beginning to probe
naturalness parameters of $\caln \sim 100$, corresponding to gluino
masses of 1 TeV.

In \secref{models}, we have described a few of the leading frameworks
that attempt to preserve naturalness in viable models, giving their
key features and implications for experimental searches.  Their
primary motivations are summarized in \tableref{summary}, and their
rough implications for superpartner spectra are given in
Figs.~\ref{fig:spectrum_effective}, \ref{fig:spectrum_fp}, and
\ref{fig:spectrum_compressed}.  Although supersymmetry does not work
``out of the box,'' these models provide longstanding (pre-LHC) and
well-motivated frameworks that remain viable and preserve naturalness
at the 1\% level.

In summary, weak-scale supersymmetry is neither unscathed, nor is it
dead.  The true status is somewhere in between, and requires a nuanced
view that incorporates at least some of the many caveats and
subtleties reviewed here.  Thankfully, the status quo will not last
long, given expected experimental progress on many fronts.  In
particular, after the two-year shutdown from 2013-14, the LHC is
currently expected to begin running again at $\sim 13~\tev$ in 2015,
with initial results available by Summer 2015, and $100~\ifb$ of data
analyzed by 2018.  Such a jump in energy and luminosity will push the
reach in gluino and squark masses from around 1 TeV to around 3-4 TeV,
and probe models that are roughly an order of magnitude less natural.
Given these exciting prospects for drastically improved sensitivity to
supersymmetry or other new physics at the weak scale, patience is a
virtue.  In the grand scheme of things, we will soon know.

\section*{ACKNOWLEDGMENTS}

I am grateful to colleagues and collaborators for sharing their
insights through the years, and especially Markus Luty, Steve Martin,
Ann Nelson, Michael Ratz, and Xerxes Tata for reading a draft of this
manuscript and suggesting important improvements.  I also thank the
organizers of seminars and colloquia at the University of Chicago,
Argonne, Fermilab, UC Berkeley, and SLAC for invitations to give talks
that formed the basis for some of this review.  This work was
supported in part by NSF Grant No.~PHY--0970173 and a Simons
Foundation Fellowship.

\providecommand{\href}[2]{#2}\begingroup\raggedright\endgroup

\ARNPS{

LIST OF IMPORTANT ACRONYMS USED IN THE TEXT

\begin{itemize}
\item EDM: electric dipole moment
\item GUT: grand unified theory
\item LHC: Large Hadron Collider
\item LSP: lightest supersymmetric particle
\item MSSM: minimal supersymmetric standard model
\item $\met$: missing transverse energy
\item mSUGRA: minimal supergravity
\item RGE: renormalization group equation
\item RPV: $R$-parity violating
\item WIMP: weakly-interacting massive particle
\end{itemize}
}


\begin{thebibliography}{10%
0}

\bibitem{Newton:1756aa}
I.~Newton, {\em {Four Letters from Sir Isaac Newton to Doctor Bentley
  Containing Some Arguments in Proof of a Deity}}.
\newblock R.~and J.~Dodsley, London, 1756.
\newblock Letter 2, p.~14.

\bibitem{Golfand:1971iw}
Y.~A. Golfand and E.~P. Likhtman, ``{Extension of the Algebra of Poincare Group
  Generators and Violation of p Invariance},''
{\em JETP Lett.} {\bfseries 13} (1971) 323--326.

\bibitem{Volkov:1973ix}
D.~V. Volkov and V.~P. Akulov, ``{Is the Neutrino a Goldstone Particle?},''
\href{http://dx.doi.org/10.1016/0370-2693(73)90490-5}{{\em Phys. Lett.}
  {\bfseries B46} (1973) 109--110}.

\bibitem{Wess:1974tw}
J.~Wess and B.~Zumino, ``{Supergauge Transformations in Four-Dimensions},''
\href{http://dx.doi.org/10.1016/0550-3213(74)90355-1}{{\em Nucl. Phys.}
  {\bfseries B70} (1974) 39--50}.

\bibitem{Weinberg:1975gm}
S.~Weinberg, ``{Implications of Dynamical Symmetry Breaking},''
\href{http://dx.doi.org/10.1103/PhysRevD.13.974}{{\em Phys. Rev.} {\bfseries
  D13} (1976) 974--996}.

\bibitem{Susskind:1978ms}
L.~Susskind, ``{Dynamics of Spontaneous Symmetry Breaking in the Weinberg-Salam
  Theory},''
\href{http://dx.doi.org/10.1103/PhysRevD.20.2619}{{\em Phys. Rev.} {\bfseries
  D20} (1979) 2619--2625}.

\bibitem{'tHooft:1980xb}
G.~{'t Hooft}. {in Recent Developments in Gauge Theories. Proceedings, NATO
  Advanced Study Institute, Cargese, France, New York, Plenum (1980)}.

\bibitem{Davier:1979hr}
L.~Maiani. {in Proceedings: Summer School on Particle Physics, Paris, France
  (1979)}.

\bibitem{Veltman:1980mj}
M.~J.~G. Veltman, ``{The Infrared - Ultraviolet Connection},''
{\em Acta Phys. Polon.} {\bfseries B12} (1981) 437.

\bibitem{Witten:1981nf}
E.~Witten, ``{Dynamical Breaking of Supersymmetry},''
\href{http://dx.doi.org/10.1016/0550-3213(81)90006-7}{{\em Nucl. Phys.}
  {\bfseries B188} (1981) 513}.

\bibitem{Kaul:1981wp}
R.~K. Kaul, ``{Gauge Hierarchy in a Supersymmetric Model},''
\href{http://dx.doi.org/10.1016/0370-2693(82)90453-1}{{\em Phys.Lett.}
  {\bfseries B109} (1982) 19}.

\bibitem{Martin:1997ns}
S.~P. Martin, ``{A Supersymmetry Primer},''
\href{http://arxiv.org/abs/hep-ph/9709356}{{\ttfamily arXiv:hep-ph/9709356}}.

\bibitem{Drees:2004jm}
M.~Drees, R.~Godbole, and P.~Roy, {\em {Theory and phenomenology of sparticles:
  An account of four-dimensional N=1 supersymmetry in high energy physics}}.
\newblock World Scientific,
2004.
\newblock

\bibitem{Baer:2006rs}
H.~Baer and X.~Tata, {\em {Weak scale supersymmetry: From superfields to
  scattering events}}.
\newblock Cambridge University Press,
2006.
\newblock

\bibitem{Dine:2007zp}
M.~Dine, {\em {Supersymmetry and string theory: Beyond the standard model}}.
\newblock Cambridge University Press,
2007.
\newblock

\bibitem{Vissani:1997ys}
F.~Vissani, ``{Do experiments suggest a hierarchy problem?},''
  \href{http://dx.doi.org/10.1103/PhysRevD.57.7027}{{\em Phys.Rev.} {\bfseries
  D57} (1998) 7027--7030},
\href{http://arxiv.org/abs/hep-ph/9709409}{{\ttfamily arXiv:hep-ph/9709409
  [hep-ph]}}.

\bibitem{:2012gk}
{\bfseries ATLAS} Collaboration, G.~Aad {\em et al.}, ``{Observation of a new
  particle in the search for the Standard Model Higgs boson with the ATLAS
  detector at the LHC},''
  \href{http://dx.doi.org/10.1016/j.physletb.2012.08.020}{{\em Phys.Lett.}
  {\bfseries B716} (2012) 1--29},
\href{http://arxiv.org/abs/1207.7214}{{\ttfamily arXiv:1207.7214 [hep-ex]}}.

\bibitem{:2012gu}
{\bfseries CMS} Collaboration, S.~Chatrchyan {\em et al.}, ``{Observation of a
  new boson at a mass of 125 GeV with the CMS experiment at the LHC},''
  \href{http://dx.doi.org/10.1016/j.physletb.2012.08.021}{{\em Phys.Lett.}
  {\bfseries B716} (2012) 30--61},
\href{http://arxiv.org/abs/1207.7235}{{\ttfamily arXiv:1207.7235 [hep-ex]}}.

\bibitem{Drees:1985jx}
M.~Drees, ``{$N=1$ Supergravity GUTs with Noncanonical Kinetic Energy Terms},''
\href{http://dx.doi.org/10.1103/PhysRevD.33.1468}{{\em Phys.Rev.} {\bfseries
  D33} (1986) 1468}.

\bibitem{Pati:1973uk}
J.~C. Pati and A.~Salam, ``{Unified Lepton-Hadron Symmetry and a Gauge Theory
  of the Basic Interactions},''
\href{http://dx.doi.org/10.1103/PhysRevD.8.1240}{{\em Phys.Rev.} {\bfseries D8}
  (1973) 1240--1251}.

\bibitem{Georgi:1974sy}
H.~Georgi and S.~L. Glashow, ``{Unity of All Elementary Particle Forces},''
\href{http://dx.doi.org/10.1103/PhysRevLett.32.438}{{\em Phys. Rev. Lett.}
  {\bfseries 32} (1974) 438--441}.

\bibitem{Fritzsch:1974nn}
H.~Fritzsch and P.~Minkowski, ``{Unified Interactions of Leptons and
  Hadrons},''
\href{http://dx.doi.org/10.1016/0003-4916(75)90211-0}{{\em Annals Phys.}
  {\bfseries 93} (1975) 193--266}.

\bibitem{Gursey:1975ki}
F.~Gursey, P.~Ramond, and P.~Sikivie, ``{A Universal Gauge Theory Model Based
  on E6},''
\href{http://dx.doi.org/10.1016/0370-2693(76)90417-2}{{\em Phys.Lett.}
  {\bfseries B60} (1976) 177}.

\bibitem{Dimopoulos:1981zb}
S.~Dimopoulos and H.~Georgi, ``{Softly Broken Supersymmetry and SU(5)},''
\href{http://dx.doi.org/10.1016/0550-3213(81)90522-8}{{\em Nucl. Phys.}
  {\bfseries B193} (1981) 150}.

\bibitem{Dimopoulos:1981yj}
S.~Dimopoulos, S.~Raby, and F.~Wilczek, ``{Supersymmetry and the Scale of
  Unification},''
\href{http://dx.doi.org/10.1103/PhysRevD.24.1681}{{\em Phys. Rev.} {\bfseries
  D24} (1981) 1681--1683}.

\bibitem{Sakai:1981gr}
N.~Sakai, ``{Naturalness in Supersymmetric GUTs},''
\href{http://dx.doi.org/10.1007/BF01573998}{{\em Zeit. Phys.} {\bfseries C11}
  (1981) 153}.

\bibitem{Ibanez:1981yh}
L.~E. Ibanez and G.~G. Ross, ``{Low-Energy Predictions in Supersymmetric Grand
  Unified Theories},''
\href{http://dx.doi.org/10.1016/0370-2693(81)91200-4}{{\em Phys. Lett.}
  {\bfseries B105} (1981) 439}.

\bibitem{Einhorn:1981sx}
M.~B. Einhorn and D.~R.~T. Jones, ``{The Weak Mixing Angle and Unification Mass
  in Supersymmetric SU(5)},''
\href{http://dx.doi.org/10.1016/0550-3213(82)90502-8}{{\em Nucl. Phys.}
  {\bfseries B196} (1982) 475}.

\bibitem{Mohapatra:1999vv}
R.~Mohapatra, ``{Supersymmetric grand unification: An Update},''
\href{http://arxiv.org/abs/hep-ph/9911272}{{\ttfamily arXiv:hep-ph/9911272
  [hep-ph]}}.

\bibitem{Nath:2006ut}
P.~Nath and P.~Fileviez~Perez, ``{Proton stability in grand unified theories,
  in strings and in branes},''
  \href{http://dx.doi.org/10.1016/j.physrep.2007.02.010}{{\em Phys.Rept.}
  {\bfseries 441} (2007) 191--317},
\href{http://arxiv.org/abs/hep-ph/0601023}{{\ttfamily arXiv:hep-ph/0601023
  [hep-ph]}}.

\bibitem{Feng:2000bp}
J.~L. Feng and K.~T. Matchev, ``{Focus point supersymmetry: Proton decay,
  flavor and CP violation, and the Higgs boson mass},''
  \href{http://dx.doi.org/10.1103/PhysRevD.63.095003}{{\em Phys.Rev.}
  {\bfseries D63} (2001) 095003},
  \href{http://arxiv.org/abs/hep-ph/0011356}{{\ttfamily arXiv:hep-ph/0011356
  [hep-ph]}}.

\bibitem{Arvanitaki:2012ps}
A.~Arvanitaki, N.~Craig, S.~Dimopoulos, and G.~Villadoro, ``{Mini-Split},''
  \href{http://dx.doi.org/10.1007/JHEP02(2013)126}{{\em JHEP} {\bfseries 1302}
  (2013) 126},
\href{http://arxiv.org/abs/1210.0555}{{\ttfamily arXiv:1210.0555 [hep-ph]}}.

\bibitem{Goldberg:1983nd}
H.~Goldberg, ``{Constraint on the photino mass from cosmology},''
\href{http://dx.doi.org/10.1103/PhysRevLett.50.1419}{{\em Phys. Rev. Lett.}
  {\bfseries 50} (1983) 1419}.

\bibitem{Ellis:1983ew}
J.~R. Ellis, J.~S. Hagelin, D.~V. Nanopoulos, K.~A. Olive, and M.~Srednicki,
  ``{Supersymmetric relics from the big bang},''
\href{http://dx.doi.org/10.1016/0550-3213(84)90461-9}{{\em Nucl. Phys.}
  {\bfseries B238} (1984) 453--476}.

\bibitem{Mizuta:1992qp}
S.~Mizuta and M.~Yamaguchi, ``{Coannihilation effects and relic abundance of
  Higgsino dominant LSP(s)},''
  \href{http://dx.doi.org/10.1016/0370-2693(93)91717-2}{{\em Phys.Lett.}
  {\bfseries B298} (1993) 120--126},
\href{http://arxiv.org/abs/hep-ph/9208251}{{\ttfamily arXiv:hep-ph/9208251
  [hep-ph]}}.

\bibitem{Cirelli:2007xd}
M.~Cirelli, A.~Strumia, and M.~Tamburini, ``{Cosmology and Astrophysics of
  Minimal Dark Matter},''
  \href{http://dx.doi.org/10.1016/j.nuclphysb.2007.07.023}{{\em Nucl.Phys.}
  {\bfseries B787} (2007) 152--175},
\href{http://arxiv.org/abs/0706.4071}{{\ttfamily arXiv:0706.4071 [hep-ph]}}.

\bibitem{Feng:2000gh}
J.~L. Feng, K.~T. Matchev, and F.~Wilczek, ``{Neutralino dark matter in focus
  point supersymmetry},''
  \href{http://dx.doi.org/10.1016/S0370-2693(00)00512-8}{{\em Phys.Lett.}
  {\bfseries B482} (2000) 388--399},
\href{http://arxiv.org/abs/hep-ph/0004043}{{\ttfamily arXiv:hep-ph/0004043
  [hep-ph]}}.

\bibitem{Hisano:2006nn}
J.~Hisano, S.~Matsumoto, M.~Nagai, O.~Saito, and M.~Senami, ``{Non-perturbative
  effect on thermal relic abundance of dark matter},''
  \href{http://dx.doi.org/10.1016/j.physletb.2007.01.012}{{\em Phys.Lett.}
  {\bfseries B646} (2007) 34--38},
\href{http://arxiv.org/abs/hep-ph/0610249}{{\ttfamily arXiv:hep-ph/0610249
  [hep-ph]}}.

\bibitem{Birkedal:2004xn}
A.~Birkedal, K.~Matchev, and M.~Perelstein, ``{Dark matter at colliders: A
  Model independent approach},''
  \href{http://dx.doi.org/10.1103/PhysRevD.70.077701}{{\em Phys.Rev.}
  {\bfseries D70} (2004) 077701},
\href{http://arxiv.org/abs/hep-ph/0403004}{{\ttfamily arXiv:hep-ph/0403004
  [hep-ph]}}.

\bibitem{Feng:2005gj}
J.~L. Feng, S.~Su, and F.~Takayama, ``{Lower limit on dark matter production at
  the large hadron collider},''
  \href{http://dx.doi.org/10.1103/PhysRevLett.96.151802}{{\em Phys.Rev.Lett.}
  {\bfseries 96} (2006) 151802},
\href{http://arxiv.org/abs/hep-ph/0503117}{{\ttfamily arXiv:hep-ph/0503117
  [hep-ph]}}.

\bibitem{Beltran:2010ww}
M.~Beltran, D.~Hooper, E.~W. Kolb, Z.~A. Krusberg, and T.~M. Tait, ``{Maverick
  dark matter at colliders},''
  \href{http://dx.doi.org/10.1007/JHEP09(2010)037}{{\em JHEP} {\bfseries 1009}
  (2010) 037},
\href{http://arxiv.org/abs/1002.4137}{{\ttfamily arXiv:1002.4137 [hep-ph]}}.

\bibitem{Goodman:2010yf}
J.~Goodman, M.~Ibe, A.~Rajaraman, W.~Shepherd, T.~M. Tait, {\em et al.},
  ``{Constraints on Light Majorana dark Matter from Colliders},''
  \href{http://dx.doi.org/10.1016/j.physletb.2010.11.009}{{\em Phys.Lett.}
  {\bfseries B695} (2011) 185--188},
\href{http://arxiv.org/abs/1005.1286}{{\ttfamily arXiv:1005.1286 [hep-ph]}}.

\bibitem{Bai:2010hh}
Y.~Bai, P.~J. Fox, and R.~Harnik, ``{The Tevatron at the Frontier of Dark
  Matter Direct Detection},''
  \href{http://dx.doi.org/10.1007/JHEP12(2010)048}{{\em JHEP} {\bfseries 1012}
  (2010) 048},
\href{http://arxiv.org/abs/1005.3797}{{\ttfamily arXiv:1005.3797 [hep-ph]}}.

\bibitem{Chatrchyan:2012me}
{\bfseries CMS} Collaboration, S.~Chatrchyan {\em et al.}, ``{Search for dark
  matter and large extra dimensions in monojet events in $pp$ collisions at
  $\sqrt{s}=7$ TeV},'' \href{http://dx.doi.org/10.1007/JHEP09(2012)094}{{\em
  JHEP} {\bfseries 1209} (2012) 094},
\href{http://arxiv.org/abs/1206.5663}{{\ttfamily arXiv:1206.5663 [hep-ex]}}.

\bibitem{ATLAS:2012ky}
{\bfseries ATLAS} Collaboration, G.~Aad {\em et al.}, ``{Search for dark matter
  candidates and large extra dimensions in events with a jet and missing
  transverse momentum with the ATLAS detector},''
  \href{http://dx.doi.org/10.1007/JHEP04(2013)075}{{\em JHEP} {\bfseries 1304}
  (2013) 075},
\href{http://arxiv.org/abs/1210.4491}{{\ttfamily arXiv:1210.4491 [hep-ex]}}.

\bibitem{Hisano:2012wm}
J.~Hisano, K.~Ishiwata, and N.~Nagata, ``{Direct Search of Dark Matter in
  High-Scale Supersymmetry},''
  \href{http://dx.doi.org/10.1103/PhysRevD.87.035020}{{\em Phys.Rev.}
  {\bfseries D87} (2013) 035020},
\href{http://arxiv.org/abs/1210.5985}{{\ttfamily arXiv:1210.5985 [hep-ph]}}.

\bibitem{Feng:2010gw}
J.~L. Feng, ``{Dark Matter Candidates from Particle Physics and Methods of
  Detection},''
  \href{http://dx.doi.org/10.1146/annurev-astro-082708-101659}{{\em Ann. Rev.
  Astron. Astrophys.} {\bfseries 48} (2010) 495--545},
\href{http://arxiv.org/abs/1003.0904}{{\ttfamily arXiv:1003.0904
  [astro-ph.CO]}}.

\bibitem{Feng:2012rn}
J.~L. Feng, Z.~Surujon, and H.-B. Yu, ``{Confluence of Constraints in Gauge
  Mediation: The 125 GeV Higgs Boson and Goldilocks Cosmology},''
  \href{http://dx.doi.org/10.1103/PhysRevD.86.035003}{{\em Phys.Rev.}
  {\bfseries D86} (2012) 035003},
\href{http://arxiv.org/abs/1205.6480}{{\ttfamily arXiv:1205.6480 [hep-ph]}}.

\bibitem{Feng:2009te}
J.~L. Feng, J.-F. Grivaz, and J.~Nachtman, ``{Searches for Supersymmetry at
  High-Energy Colliders},''
  \href{http://dx.doi.org/10.1103/RevModPhys.82.699}{{\em Rev.Mod.Phys.}
  {\bfseries 82} (2010) 699--727},
\href{http://arxiv.org/abs/0903.0046}{{\ttfamily arXiv:0903.0046 [hep-ex]}}.

\bibitem{ATLASSUSY}
{\bfseries ATLAS} Collaboration, ``{ATLAS Supersymmetry Searches},''.
  https://twiki.cern.ch/twiki/bin/view/AtlasPublic/SupersymmetryPublicResults.

\bibitem{CMSSUSY}
{\bfseries CMS} Collaboration, ``{CMS Supersymmetry Physics Results},''.
  https://twiki.cern.ch/twiki/bin/view/CMSPublic/PhysicsResultsSUS.

\bibitem{ATLAS:2012ona}
{\bfseries ATLAS} Collaboration, ``{Search for squarks and gluinos with the
  ATLAS detector using final states with jets and missing transverse momentum
  and 5.8 fb$^{-1}$ of $\sqrt{s}$=8 TeV proton-proton collision data},'' Tech.
  Rep. ATLAS-CONF-2012-109, ATLAS-COM-CONF-2012-140, CERN, Geneva, 2012.

\bibitem{CMS-PAS-SUS-12-023}
{\bfseries CMS} Collaboration, ``Search for direct top squark pair production
  in events with a single isolated lepton, jets and missing transverse energy
  at $\sqrt{s} = 8$ TeV,'' Tech. Rep. CMS-PAS-SUS-12-023, CERN, Geneva, 2012.

\bibitem{CMSstop}
{\bfseries CMS} Collaboration, ``Search for supersymmetry in final states with
  missing transverse energy and 0, 1, 2, 3, or at least 4 b-quark jets in 8 TeV
  pp collisions using the variable AlphaT,'' Tech. Rep. CMS-PAS-SUS-12-028,
  CERN, Geneva, 2012.

\bibitem{ATLAS:2012dp}
{\bfseries ATLAS} Collaboration, G.~Aad {\em et al.}, ``{Search for pair
  production of massive particles decaying into three quarks with the ATLAS
  detector in $\sqrt{s}=7$ TeV $pp$ collisions at the LHC},''
  \href{http://dx.doi.org/10.1007/JHEP12(2012)086}{{\em JHEP} {\bfseries 1212}
  (2012) 086},
\href{http://arxiv.org/abs/1210.4813}{{\ttfamily arXiv:1210.4813 [hep-ex]}}.

\bibitem{CMSweak}
{\bfseries CMS} Collaboration, ``Search for direct EWK production of SUSY
  particles in multilepton modes with 8TeV data,'' Tech. Rep.
  CMS-PAS-SUS-12-022, CERN, Geneva, 2012.

\bibitem{ATLAS-CONF-2013-014}
{\bfseries ATLAS} Collaboration, ``Combined measurements of the mass and signal
  strength of the Higgs-like boson with the ATLAS detector using up to 25 fb-1
  of proton-proton collision data,'' Tech. Rep. ATLAS-CONF-2013-014, CERN,
  Geneva, 2013.

\bibitem{CMS-PAS-HIG-13-005}
{\bfseries CMS} Collaboration, ``Combination of standard model Higgs boson
  searches and measurements of the properties of the new boson with a mass near
  125 GeV,'' Tech. Rep. CMS-PAS-HIG-13-005, CERN, Geneva, 2013.

\bibitem{Okada:1990vk}
Y.~Okada, M.~Yamaguchi, and T.~Yanagida, ``{Upper bound of the lightest Higgs
  boson mass in the minimal supersymmetric standard model},''
\href{http://dx.doi.org/10.1143/PTP.85.1}{{\em Prog.Theor.Phys.} {\bfseries 85}
  (1991) 1--6}.

\bibitem{Haber:1990aw}
H.~E. Haber and R.~Hempfling, ``{Can the mass of the lightest Higgs boson of
  the minimal supersymmetric model be larger than m(Z)?},''
\href{http://dx.doi.org/10.1103/PhysRevLett.66.1815}{{\em Phys.Rev.Lett.}
  {\bfseries 66} (1991) 1815--1818}.

\bibitem{Ellis:1990nz}
J.~R. Ellis, G.~Ridolfi, and F.~Zwirner, ``{Radiative corrections to the masses
  of supersymmetric Higgs bosons},''
\href{http://dx.doi.org/10.1016/0370-2693(91)90863-L}{{\em Phys.Lett.}
  {\bfseries B257} (1991) 83--91}.

\bibitem{Carena:1995wu}
M.~S. Carena, M.~Quiros, and C.~Wagner, ``{Effective potential methods and the
  Higgs mass spectrum in the MSSM},''
  \href{http://dx.doi.org/10.1016/0550-3213(95)00665-6}{{\em Nucl.Phys.}
  {\bfseries B461} (1996) 407--436},
\href{http://arxiv.org/abs/hep-ph/9508343}{{\ttfamily arXiv:hep-ph/9508343
  [hep-ph]}}.

\bibitem{Carena:2000dp}
M.~S. Carena, H.~Haber, S.~Heinemeyer, W.~Hollik, C.~Wagner, {\em et al.},
  ``{Reconciling the two loop diagrammatic and effective field theory
  computations of the mass of the lightest CP - even Higgs boson in the
  MSSM},'' \href{http://dx.doi.org/10.1016/S0550-3213(00)00212-1}{{\em
  Nucl.Phys.} {\bfseries B580} (2000) 29--57},
\href{http://arxiv.org/abs/hep-ph/0001002}{{\ttfamily arXiv:hep-ph/0001002
  [hep-ph]}}.

\bibitem{Heinemeyer:1998yj}
S.~Heinemeyer, W.~Hollik, and G.~Weiglein, ``{FeynHiggs: A Program for the
  calculation of the masses of the neutral CP even Higgs bosons in the MSSM},''
  \href{http://dx.doi.org/10.1016/S0010-4655(99)00364-1}{{\em
  Comput.Phys.Commun.} {\bfseries 124} (2000) 76--89},
\href{http://arxiv.org/abs/hep-ph/9812320}{{\ttfamily arXiv:hep-ph/9812320
  [hep-ph]}}.

\bibitem{Allanach:2001kg}
B.~C. Allanach, ``{SOFTSUSY: A C++ program for calculating supersymmetric
  spectra},'' \href{http://dx.doi.org/10.1016/S0010-4655(01)00460-X}{{\em
  Comput. Phys. Commun.} {\bfseries 143} (2002) 305--331},
\href{http://arxiv.org/abs/hep-ph/0104145}{{\ttfamily arXiv:hep-ph/0104145}}.

\bibitem{Djouadi:2002ze}
A.~Djouadi, J.-L. Kneur, and G.~Moultaka, ``{SuSpect: A Fortran code for the
  supersymmetric and Higgs particle spectrum in the MSSM},''
  \href{http://dx.doi.org/10.1016/j.cpc.2006.11.009}{{\em Comput.Phys.Commun.}
  {\bfseries 176} (2007) 426--455},
\href{http://arxiv.org/abs/hep-ph/0211331}{{\ttfamily arXiv:hep-ph/0211331
  [hep-ph]}}.

\bibitem{Harlander:2008ju}
R.~Harlander, P.~Kant, L.~Mihaila, and M.~Steinhauser, ``{Higgs boson mass in
  supersymmetry to three loops},''
  \href{http://dx.doi.org/10.1103/PhysRevLett.101.039901,
  10.1103/PhysRevLett.100.191602}{{\em Phys.Rev.Lett.} {\bfseries 100} (2008)
  191602},
\href{http://arxiv.org/abs/0803.0672}{{\ttfamily arXiv:0803.0672 [hep-ph]}}.

\bibitem{Kant:2010tf}
P.~Kant, R.~Harlander, L.~Mihaila, and M.~Steinhauser, ``{Light MSSM Higgs
  boson mass to three-loop accuracy},''
  \href{http://dx.doi.org/10.1007/JHEP08(2010)104}{{\em JHEP} {\bfseries 1008}
  (2010) 104},
\href{http://arxiv.org/abs/1005.5709}{{\ttfamily arXiv:1005.5709 [hep-ph]}}.

\bibitem{Cao:2012fz}
J.-J. Cao, Z.-X. Heng, J.~M. Yang, Y.-M. Zhang, and J.-Y. Zhu, ``{A SM-like
  Higgs near 125 GeV in low energy SUSY: a comparative study for MSSM and
  NMSSM},'' \href{http://dx.doi.org/10.1007/JHEP03(2012)086}{{\em JHEP}
  {\bfseries 1203} (2012) 086},
\href{http://arxiv.org/abs/1202.5821}{{\ttfamily arXiv:1202.5821 [hep-ph]}}.

\bibitem{Hall:2011aa}
L.~J. Hall, D.~Pinner, and J.~T. Ruderman, ``{A Natural SUSY Higgs Near 126
  GeV},'' \href{http://dx.doi.org/10.1007/JHEP04(2012)131}{{\em JHEP}
  {\bfseries 1204} (2012) 131},
\href{http://arxiv.org/abs/1112.2703}{{\ttfamily arXiv:1112.2703 [hep-ph]}}.

\bibitem{Gabbiani:1996hi}
F.~Gabbiani, E.~Gabrielli, A.~Masiero, and L.~Silvestrini, ``{A Complete
  analysis of FCNC and CP constraints in general SUSY extensions of the
  standard model},'' \href{http://dx.doi.org/10.1016/0550-3213(96)00390-2}{{\em
  Nucl.Phys.} {\bfseries B477} (1996) 321--352},
\href{http://arxiv.org/abs/hep-ph/9604387}{{\ttfamily arXiv:hep-ph/9604387
  [hep-ph]}}.

\bibitem{Beringer:1900zz}
{\bfseries Particle Data Group} Collaboration, J.~Beringer {\em et al.},
  ``{Review of Particle Physics (RPP)},''
\href{http://dx.doi.org/10.1103/PhysRevD.86.010001}{{\em Phys.Rev.} {\bfseries
  D86} (2012) 010001}.

\bibitem{Dine:1981za}
M.~Dine, W.~Fischler, and M.~Srednicki, ``{Supersymmetric Technicolor},''
\href{http://dx.doi.org/10.1016/0550-3213(81)90582-4}{{\em Nucl. Phys.}
  {\bfseries B189} (1981) 575--593}.

\bibitem{Dimopoulos:1981au}
S.~Dimopoulos and S.~Raby, ``{Supercolor},''
\href{http://dx.doi.org/10.1016/0550-3213(81)90430-2}{{\em Nucl. Phys.}
  {\bfseries B192} (1981) 353}.

\bibitem{Nappi:1982hm}
C.~R. Nappi and B.~A. Ovrut, ``{Supersymmetric Extension of the SU(3) x SU(2) x
  U(1) Model},''
\href{http://dx.doi.org/10.1016/0370-2693(82)90418-X}{{\em Phys. Lett.}
  {\bfseries B113} (1982) 175}.

\bibitem{AlvarezGaume:1981wy}
L.~Alvarez-Gaume, M.~Claudson, and M.~B. Wise, ``{Low-Energy Supersymmetry},''
\href{http://dx.doi.org/10.1016/0550-3213(82)90138-9}{{\em Nucl. Phys.}
  {\bfseries B207} (1982) 96}.

\bibitem{Dine:1994vc}
M.~Dine, A.~E. Nelson, and Y.~Shirman, ``{Low-energy dynamical supersymmetry
  breaking simplified},''
  \href{http://dx.doi.org/10.1103/PhysRevD.51.1362}{{\em Phys. Rev.} {\bfseries
  D51} (1995) 1362--1370},
\href{http://arxiv.org/abs/hep-ph/9408384}{{\ttfamily arXiv:hep-ph/9408384}}.

\bibitem{Dine:1995ag}
M.~Dine, A.~E. Nelson, Y.~Nir, and Y.~Shirman, ``{New tools for low-energy
  dynamical supersymmetry breaking},''
  \href{http://dx.doi.org/10.1103/PhysRevD.53.2658}{{\em Phys. Rev.} {\bfseries
  D53} (1996) 2658--2669},
\href{http://arxiv.org/abs/hep-ph/9507378}{{\ttfamily arXiv:hep-ph/9507378}}.

\bibitem{Randall:1998uk}
L.~Randall and R.~Sundrum, ``{Out of this world supersymmetry breaking},''
  \href{http://dx.doi.org/10.1016/S0550-3213(99)00359-4}{{\em Nucl.Phys.}
  {\bfseries B557} (1999) 79--118},
\href{http://arxiv.org/abs/hep-th/9810155}{{\ttfamily arXiv:hep-th/9810155
  [hep-th]}}.

\bibitem{Giudice:1998xp}
G.~F. Giudice, M.~A. Luty, H.~Murayama, and R.~Rattazzi, ``{Gaugino mass
  without singlets},'' {\em JHEP} {\bfseries 9812} (1998) 027,
\href{http://arxiv.org/abs/hep-ph/9810442}{{\ttfamily arXiv:hep-ph/9810442
  [hep-ph]}}.

\bibitem{Moroi:1995yh}
T.~Moroi, ``{The Muon Anomalous Magnetic Dipole Moment in the Minimal
  Supersymmetric Standard Model},''
  \href{http://dx.doi.org/10.1103/PhysRevD.53.6565}{{\em Phys. Rev.} {\bfseries
  D53} (1996) 6565--6575},
\href{http://arxiv.org/abs/hep-ph/9512396}{{\ttfamily arXiv:hep-ph/9512396}}.

\bibitem{Dimopoulos:1995kn}
S.~Dimopoulos and S.~D. Thomas, ``{Dynamical relaxation of the supersymmetric
  CP violating phases},''
  \href{http://dx.doi.org/10.1016/0550-3213(96)00065-X}{{\em Nucl.Phys.}
  {\bfseries B465} (1996) 23--33},
\href{http://arxiv.org/abs/hep-ph/9510220}{{\ttfamily arXiv:hep-ph/9510220
  [hep-ph]}}.

\bibitem{Dine:1996xk}
M.~Dine, Y.~Nir, and Y.~Shirman, ``{Variations on minimal gauge mediated
  supersymmetry breaking},''
  \href{http://dx.doi.org/10.1103/PhysRevD.55.1501}{{\em Phys.Rev.} {\bfseries
  D55} (1997) 1501--1508},
\href{http://arxiv.org/abs/hep-ph/9607397}{{\ttfamily arXiv:hep-ph/9607397
  [hep-ph]}}.

\bibitem{Moroi:1998km}
T.~Moroi, ``{Electric dipole moments in gauge mediated models and a solution to
  the SUSY CP problem},''
  \href{http://dx.doi.org/10.1016/S0370-2693(98)01576-7}{{\em Phys. Lett.}
  {\bfseries B447} (1999) 75--82},
\href{http://arxiv.org/abs/hep-ph/9811257}{{\ttfamily arXiv:hep-ph/9811257}}.

\bibitem{Bennett:2006fi}
{\bfseries Muon g-2} Collaboration, G.~Bennett {\em et al.}, ``{Final Report of
  the Muon E821 Anomalous Magnetic Moment Measurement at BNL},''
  \href{http://dx.doi.org/10.1103/PhysRevD.73.072003}{{\em Phys.Rev.}
  {\bfseries D73} (2006) 072003},
\href{http://arxiv.org/abs/hep-ex/0602035}{{\ttfamily arXiv:hep-ex/0602035
  [hep-ex]}}.

\bibitem{Davier:2010nc}
M.~Davier, A.~Hoecker, B.~Malaescu, and Z.~Zhang, ``{Reevaluation of the
  Hadronic Contributions to the Muon g-2 and to alpha(MZ)},''
  \href{http://dx.doi.org/10.1140/epjc/s10052-012-1874-8,
  10.1140/epjc/s10052-010-1515-z}{{\em Eur.Phys.J.} {\bfseries C71} (2011)
  1515},
\href{http://arxiv.org/abs/1010.4180}{{\ttfamily arXiv:1010.4180 [hep-ph]}}.

\bibitem{Jegerlehner:2011ti}
F.~Jegerlehner and R.~Szafron, ``{$\rho^0 - \gamma$ mixing in the neutral
  channel pion form factor $F_{\pi}^{e}$ and its role in comparing $e^+ e^-$
  with $\tau$ spectral functions},''
  \href{http://dx.doi.org/10.1140/epjc/s10052-011-1632-3}{{\em Eur.Phys.J.}
  {\bfseries C71} (2011) 1632},
\href{http://arxiv.org/abs/1101.2872}{{\ttfamily arXiv:1101.2872 [hep-ph]}}.

\bibitem{Feng:2001tr}
J.~L. Feng and K.~T. Matchev, ``{Supersymmetry and the anomalous magnetic
  moment of the muon},''
  \href{http://dx.doi.org/10.1103/PhysRevLett.86.3480}{{\em Phys.Rev.Lett.}
  {\bfseries 86} (2001) 3480--3483},
\href{http://arxiv.org/abs/hep-ph/0102146}{{\ttfamily arXiv:hep-ph/0102146
  [hep-ph]}}.

\bibitem{ALEPH:2005ab}
{\bfseries ALEPH Collaboration, DELPHI Collaboration, L3 Collaboration, OPAL
  Collaboration, SLD Collaboration, LEP Electroweak Working Group, SLD
  Electroweak Group, SLD Heavy Flavour Group} Collaboration, S.~Schael {\em et
  al.}, ``{Precision electroweak measurements on the $Z$ resonance},''
  \href{http://dx.doi.org/10.1016/j.physrep.2005.12.006}{{\em Phys.Rept.}
  {\bfseries 427} (2006) 257--454},
\href{http://arxiv.org/abs/hep-ex/0509008}{{\ttfamily arXiv:hep-ex/0509008
  [hep-ex]}}.

\bibitem{Ellis:1986yg}
J.~R. Ellis, K.~Enqvist, D.~V. Nanopoulos, and F.~Zwirner, ``{Observables in
  Low-Energy Superstring Models},''
\href{http://dx.doi.org/10.1142/S0217732386000105}{{\em Mod.Phys.Lett.}
  {\bfseries A1} (1986) 57}.

\bibitem{Barbieri:1987fn}
R.~Barbieri and G.~Giudice, ``{Upper Bounds on Supersymmetric Particle
  Masses},''
\href{http://dx.doi.org/10.1016/0550-3213(88)90171-X}{{\em Nucl.Phys.}
  {\bfseries B306} (1988) 63}.

\bibitem{Feldman:2011ud}
D.~Feldman, G.~Kane, E.~Kuflik, and R.~Lu, ``{A new (string motivated) approach
  to the little hierarchy problem},''
  \href{http://dx.doi.org/10.1016/j.physletb.2011.08.063}{{\em Phys.Lett.}
  {\bfseries B704} (2011) 56--61},
\href{http://arxiv.org/abs/1105.3765}{{\ttfamily arXiv:1105.3765 [hep-ph]}}.

\bibitem{Ross:1992tz}
G.~G. Ross and R.~Roberts, ``{Minimal supersymmetric unification
  predictions},''
\href{http://dx.doi.org/10.1016/0550-3213(92)90302-R}{{\em Nucl.Phys.}
  {\bfseries B377} (1992) 571--592}.

\bibitem{deCarlos:1993yy}
B.~de~Carlos and J.~Casas, ``{One loop analysis of the electroweak breaking in
  supersymmetric models and the fine tuning problem},''
  \href{http://dx.doi.org/10.1016/0370-2693(93)90940-J}{{\em Phys.Lett.}
  {\bfseries B309} (1993) 320--328},
\href{http://arxiv.org/abs/hep-ph/9303291}{{\ttfamily arXiv:hep-ph/9303291
  [hep-ph]}}.

\bibitem{Anderson:1994tr}
G.~W. Anderson and D.~J. Castano, ``{Naturalness and superpartner masses or
  when to give up on weak scale supersymmetry},''
  \href{http://dx.doi.org/10.1103/PhysRevD.52.1693}{{\em Phys.Rev.} {\bfseries
  D52} (1995) 1693--1700},
\href{http://arxiv.org/abs/hep-ph/9412322}{{\ttfamily arXiv:hep-ph/9412322
  [hep-ph]}}.

\bibitem{Romanino:1999ut}
A.~Romanino and A.~Strumia, ``{Are heavy scalars natural in minimal
  supergravity?},'' \href{http://dx.doi.org/10.1016/S0370-2693(00)00806-6}{{\em
  Phys.Lett.} {\bfseries B487} (2000) 165--170},
\href{http://arxiv.org/abs/hep-ph/9912301}{{\ttfamily arXiv:hep-ph/9912301
  [hep-ph]}}.

\bibitem{Ciafaloni:1996zh}
P.~Ciafaloni and A.~Strumia, ``{Naturalness upper bounds on gauge mediated soft
  terms},'' \href{http://dx.doi.org/10.1016/S0550-3213(97)00138-7}{{\em
  Nucl.Phys.} {\bfseries B494} (1997) 41--53},
\href{http://arxiv.org/abs/hep-ph/9611204}{{\ttfamily arXiv:hep-ph/9611204
  [hep-ph]}}.

\bibitem{Bhattacharyya:1996dw}
G.~Bhattacharyya and A.~Romanino, ``{Naturalness constraints on gauge mediated
  supersymmetry breaking models},''
  \href{http://dx.doi.org/10.1103/PhysRevD.55.7015}{{\em Phys.Rev.} {\bfseries
  D55} (1997) 7015--7019},
\href{http://arxiv.org/abs/hep-ph/9611243}{{\ttfamily arXiv:hep-ph/9611243
  [hep-ph]}}.

\bibitem{Chan:1997bi}
K.~L. Chan, U.~Chattopadhyay, and P.~Nath, ``{Naturalness, weak scale
  supersymmetry and the prospect for the observation of supersymmetry at the
  Tevatron and at the CERN LHC},''
  \href{http://dx.doi.org/10.1103/PhysRevD.58.096004}{{\em Phys.Rev.}
  {\bfseries D58} (1998) 096004},
\href{http://arxiv.org/abs/hep-ph/9710473}{{\ttfamily arXiv:hep-ph/9710473
  [hep-ph]}}.

\bibitem{Barbieri:1998uv}
R.~Barbieri and A.~Strumia, ``{About the fine tuning price of LEP},''
  \href{http://dx.doi.org/10.1016/S0370-2693(98)00577-2}{{\em Phys.Lett.}
  {\bfseries B433} (1998) 63--66},
\href{http://arxiv.org/abs/hep-ph/9801353}{{\ttfamily arXiv:hep-ph/9801353
  [hep-ph]}}.

\bibitem{Chankowski:1998xv}
P.~H. Chankowski, J.~R. Ellis, M.~Olechowski, and S.~Pokorski, ``{Haggling over
  the fine tuning price of LEP},''
  \href{http://dx.doi.org/10.1016/S0550-3213(99)00025-5}{{\em Nucl.Phys.}
  {\bfseries B544} (1999) 39--63},
\href{http://arxiv.org/abs/hep-ph/9808275}{{\ttfamily arXiv:hep-ph/9808275
  [hep-ph]}}.

\bibitem{Baer:2012cf}
H.~Baer, V.~Barger, P.~Huang, D.~Mickelson, A.~Mustafayev, {\em et al.},
  ``{Radiative natural supersymmetry: Reconciling electroweak fine-tuning and
  the Higgs boson mass},''
\href{http://arxiv.org/abs/1212.2655}{{\ttfamily arXiv:1212.2655 [hep-ph]}}.

\bibitem{Anderson:1994dz}
G.~W. Anderson and D.~J. Castano, ``{Measures of fine tuning},''
  \href{http://dx.doi.org/10.1016/0370-2693(95)00051-L}{{\em Phys.Lett.}
  {\bfseries B347} (1995) 300--308},
\href{http://arxiv.org/abs/hep-ph/9409419}{{\ttfamily arXiv:hep-ph/9409419
  [hep-ph]}}.

\bibitem{Ibanez:1983di}
L.~E. Ibanez and C.~Lopez, ``{N=1 Supergravity, the Weak Scale and the
  Low-Energy Particle Spectrum},''
\href{http://dx.doi.org/10.1016/0550-3213(84)90581-9}{{\em Nucl.Phys.}
  {\bfseries B233} (1984) 511}.

\bibitem{Ibanez:1984vq}
L.~E. Ibanez, C.~Lopez, and C.~Munoz, ``{The Low-Energy Supersymmetric Spectrum
  According to N=1 Supergravity Guts},''
\href{http://dx.doi.org/10.1016/0550-3213(85)90393-1}{{\em Nucl.Phys.}
  {\bfseries B256} (1985) 218--252}.

\bibitem{Abe:2007kf}
H.~Abe, T.~Kobayashi, and Y.~Omura, ``{Relaxed fine-tuning in models with
  non-universal gaugino masses},''
  \href{http://dx.doi.org/10.1103/PhysRevD.76.015002}{{\em Phys.Rev.}
  {\bfseries D76} (2007) 015002},
\href{http://arxiv.org/abs/hep-ph/0703044}{{\ttfamily arXiv:hep-ph/0703044
  [HEP-PH]}}.

\bibitem{Martin:2007gf}
S.~P. Martin, ``{Compressed supersymmetry and natural neutralino dark matter
  from top squark-mediated annihilation to top quarks},''
  \href{http://dx.doi.org/10.1103/PhysRevD.75.115005}{{\em Phys.Rev.}
  {\bfseries D75} (2007) 115005},
\href{http://arxiv.org/abs/hep-ph/0703097}{{\ttfamily arXiv:hep-ph/0703097
  [HEP-PH]}}.

\bibitem{Antusch:2012gv}
S.~Antusch, L.~Calibbi, V.~Maurer, M.~Monaco, and M.~Spinrath, ``{Naturalness
  of the Non-Universal MSSM in the Light of the Recent Higgs Results},''
  \href{http://dx.doi.org/10.1007/JHEP01(2013)187}{{\em JHEP} {\bfseries 01}
  (2013) 187},
\href{http://arxiv.org/abs/1207.7236}{{\ttfamily arXiv:1207.7236 [hep-ph]}}.

\bibitem{Dimopoulos:1995mi}
S.~Dimopoulos and G.~F. Giudice, ``{Naturalness constraints in supersymmetric
  theories with nonuniversal soft terms},''
  \href{http://dx.doi.org/10.1016/0370-2693(95)00961-J}{{\em Phys. Lett.}
  {\bfseries B357} (1995) 573--578},
\href{http://arxiv.org/abs/hep-ph/9507282}{{\ttfamily arXiv:hep-ph/9507282}}.

\bibitem{Pomarol:1995xc}
A.~Pomarol and D.~Tommasini, ``{Horizontal symmetries for the supersymmetric
  flavor problem},'' \href{http://dx.doi.org/10.1016/0550-3213(96)00074-0}{{\em
  Nucl.Phys.} {\bfseries B466} (1996) 3--24},
\href{http://arxiv.org/abs/hep-ph/9507462}{{\ttfamily arXiv:hep-ph/9507462
  [hep-ph]}}.

\bibitem{Dine:1990jd}
M.~Dine, A.~Kagan, and S.~Samuel, ``{Naturalness in supersymmetry, or raising
  the supersymmetry breaking scale},''
\href{http://dx.doi.org/10.1016/0370-2693(90)90847-Y}{{\em Phys.Lett.}
  {\bfseries B243} (1990) 250--256}.

\bibitem{Cohen:1996vb}
A.~G. Cohen, D.~B. Kaplan, and A.~E. Nelson, ``{The more minimal supersymmetric
  standard model},''
  \href{http://dx.doi.org/10.1016/S0370-2693(96)01183-5}{{\em Phys. Lett.}
  {\bfseries B388} (1996) 588--598},
\href{http://arxiv.org/abs/hep-ph/9607394}{{\ttfamily arXiv:hep-ph/9607394}}.

\bibitem{Dvali:1996rj}
G.~R. Dvali and A.~Pomarol, ``{Anomalous U(1) as a mediator of supersymmetry
  breaking},'' \href{http://dx.doi.org/10.1103/PhysRevLett.77.3728}{{\em Phys.
  Rev. Lett.} {\bfseries 77} (1996) 3728--3731},
\href{http://arxiv.org/abs/hep-ph/9607383}{{\ttfamily arXiv:hep-ph/9607383}}.

\bibitem{Binetruy:1996uv}
P.~Binetruy and E.~Dudas, ``{Gaugino condensation and the anomalous U(1)},''
  \href{http://dx.doi.org/10.1016/S0370-2693(96)01305-6}{{\em Phys.Lett.}
  {\bfseries B389} (1996) 503--509},
\href{http://arxiv.org/abs/hep-th/9607172}{{\ttfamily arXiv:hep-th/9607172
  [hep-th]}}.

\bibitem{Mohapatra:1996in}
R.~Mohapatra and A.~Riotto, ``{Supersymmetric models with anomalous U(1)
  mediated supersymmetry breaking},''
  \href{http://dx.doi.org/10.1103/PhysRevD.55.4262}{{\em Phys.Rev.} {\bfseries
  D55} (1997) 4262--4267},
\href{http://arxiv.org/abs/hep-ph/9611273}{{\ttfamily arXiv:hep-ph/9611273
  [hep-ph]}}.

\bibitem{Nelson:1997bt}
A.~E. Nelson and D.~Wright, ``{Horizontal, anomalous U(1) symmetry for the more
  minimal supersymmetric standard model},''
  \href{http://dx.doi.org/10.1103/PhysRevD.56.1598}{{\em Phys.Rev.} {\bfseries
  D56} (1997) 1598--1604},
\href{http://arxiv.org/abs/hep-ph/9702359}{{\ttfamily arXiv:hep-ph/9702359
  [hep-ph]}}.

\bibitem{Agashe:1998zz}
K.~Agashe and M.~Graesser, ``{Supersymmetry breaking and the supersymmetric
  flavor problem: An Analysis of decoupling the first two generation
  scalars},'' \href{http://dx.doi.org/10.1103/PhysRevD.59.015007}{{\em
  Phys.Rev.} {\bfseries D59} (1999) 015007},
\href{http://arxiv.org/abs/hep-ph/9801446}{{\ttfamily arXiv:hep-ph/9801446
  [hep-ph]}}.

\bibitem{Kaplan:1998jk}
D.~E. Kaplan, F.~Lepeintre, A.~Masiero, A.~E. Nelson, and A.~Riotto, ``{Fermion
  masses and gauge mediated supersymmetry breaking from a single U(1)},''
  \href{http://dx.doi.org/10.1103/PhysRevD.60.055003}{{\em Phys.Rev.}
  {\bfseries D60} (1999) 055003},
\href{http://arxiv.org/abs/hep-ph/9806430}{{\ttfamily arXiv:hep-ph/9806430
  [hep-ph]}}.

\bibitem{Hisano:1998tm}
J.~Hisano, K.~Kurosawa, and Y.~Nomura, ``{Large squark and slepton masses for
  the first two generations in the anomalous U(1) SUSY breaking models},''
  \href{http://dx.doi.org/10.1016/S0370-2693(98)01512-3}{{\em Phys.Lett.}
  {\bfseries B445} (1999) 316--322},
\href{http://arxiv.org/abs/hep-ph/9810411}{{\ttfamily arXiv:hep-ph/9810411
  [hep-ph]}}.

\bibitem{Kaplan:1999iq}
D.~E. Kaplan and G.~D. Kribs, ``{Phenomenology of flavor mediated supersymmetry
  breaking},'' \href{http://dx.doi.org/10.1103/PhysRevD.61.075011}{{\em
  Phys.Rev.} {\bfseries D61} (2000) 075011},
\href{http://arxiv.org/abs/hep-ph/9906341}{{\ttfamily arXiv:hep-ph/9906341
  [hep-ph]}}.

\bibitem{Everett:2000hb}
L.~L. Everett, P.~Langacker, M.~Plumacher, and J.~Wang, ``{Alternative
  supersymmetric spectra},''
  \href{http://dx.doi.org/10.1016/S0370-2693(00)00187-8}{{\em Phys.Lett.}
  {\bfseries B477} (2000) 233--241},
\href{http://arxiv.org/abs/hep-ph/0001073}{{\ttfamily arXiv:hep-ph/0001073
  [hep-ph]}}.

\bibitem{Feng:1998iq}
J.~L. Feng, C.~F. Kolda, and N.~Polonsky, ``{Solving the supersymmetric flavor
  problem with radiatively generated mass hierarchies},''
  \href{http://dx.doi.org/10.1016/S0550-3213(99)00026-7}{{\em Nucl. Phys.}
  {\bfseries B546} (1999) 3--18},
\href{http://arxiv.org/abs/hep-ph/9810500}{{\ttfamily arXiv:hep-ph/9810500}}.

\bibitem{Bagger:1999ty}
J.~Bagger, J.~L. Feng, and N.~Polonsky, ``{Naturally heavy scalars in
  supersymmetric grand unified theories},''
  \href{http://dx.doi.org/10.1016/S0550-3213(99)00577-5}{{\em Nucl. Phys.}
  {\bfseries B563} (1999) 3--20},
\href{http://arxiv.org/abs/hep-ph/9905292}{{\ttfamily arXiv:hep-ph/9905292}}.

\bibitem{Bagger:1999sy}
J.~A. Bagger, J.~L. Feng, N.~Polonsky, and R.-J. Zhang, ``{Superheavy
  supersymmetry from scalar mass A-parameter fixed points},''
  \href{http://dx.doi.org/10.1016/S0370-2693(99)01501-4}{{\em Phys. Lett.}
  {\bfseries B473} (2000) 264--271},
\href{http://arxiv.org/abs/hep-ph/9911255}{{\ttfamily arXiv:hep-ph/9911255}}.

\bibitem{Krippendorf:2012ir}
S.~Krippendorf, H.~P. Nilles, M.~Ratz, and M.~W. Winkler, ``{The heterotic
  string yields natural supersymmetry},''
  \href{http://dx.doi.org/10.1016/j.physletb.2012.04.043}{{\em Phys.Lett.}
  {\bfseries B712} (2012) 87--92},
\href{http://arxiv.org/abs/1201.4857}{{\ttfamily arXiv:1201.4857 [hep-ph]}}.

\bibitem{Badziak:2012rf}
M.~Badziak, E.~Dudas, M.~Olechowski, and S.~Pokorski, ``{Inverted Sfermion Mass
  Hierarchy and the Higgs Boson Mass in the MSSM},''
  \href{http://dx.doi.org/10.1007/JHEP07(2012)155}{{\em JHEP} {\bfseries 1207}
  (2012) 155},
\href{http://arxiv.org/abs/1205.1675}{{\ttfamily arXiv:1205.1675 [hep-ph]}}.

\bibitem{Barger:2006dh}
V.~Barger, P.~Langacker, H.-S. Lee, and G.~Shaughnessy, ``{Higgs Sector in
  Extensions of the MSSM},''
  \href{http://dx.doi.org/10.1103/PhysRevD.73.115010}{{\em Phys.Rev.}
  {\bfseries D73} (2006) 115010},
\href{http://arxiv.org/abs/hep-ph/0603247}{{\ttfamily arXiv:hep-ph/0603247
  [hep-ph]}}.

\bibitem{Ellwanger:2009dp}
U.~Ellwanger, C.~Hugonie, and A.~M. Teixeira, ``{The Next-to-Minimal
  Supersymmetric Standard Model},''
  \href{http://dx.doi.org/10.1016/j.physrep.2010.07.001}{{\em Phys.Rept.}
  {\bfseries 496} (2010) 1--77},
\href{http://arxiv.org/abs/0910.1785}{{\ttfamily arXiv:0910.1785 [hep-ph]}}.

\bibitem{Kitano:2006gv}
R.~Kitano and Y.~Nomura, ``{Supersymmetry, naturalness, and signatures at the
  LHC},'' \href{http://dx.doi.org/10.1103/PhysRevD.73.095004}{{\em Phys.Rev.}
  {\bfseries D73} (2006) 095004},
\href{http://arxiv.org/abs/hep-ph/0602096}{{\ttfamily arXiv:hep-ph/0602096
  [hep-ph]}}.

\bibitem{Baer:2010ny}
H.~Baer, S.~Kraml, A.~Lessa, S.~Sekmen, and X.~Tata, ``{Effective Supersymmetry
  at the LHC},'' \href{http://dx.doi.org/10.1007/JHEP10(2010)018}{{\em JHEP}
  {\bfseries 1010} (2010) 018},
\href{http://arxiv.org/abs/1007.3897}{{\ttfamily arXiv:1007.3897 [hep-ph]}}.

\bibitem{Cohen:1996sq}
A.~G. Cohen, D.~B. Kaplan, F.~Lepeintre, and A.~E. Nelson, ``{B factory physics
  from effective supersymmetry},''
  \href{http://dx.doi.org/10.1103/PhysRevLett.78.2300}{{\em Phys.Rev.Lett.}
  {\bfseries 78} (1997) 2300--2303},
\href{http://arxiv.org/abs/hep-ph/9610252}{{\ttfamily arXiv:hep-ph/9610252
  [hep-ph]}}.

\bibitem{Feng:1999mn}
J.~L. Feng, K.~T. Matchev, and T.~Moroi, ``{Multi - TeV scalars are natural in
  minimal supergravity},''
  \href{http://dx.doi.org/10.1103/PhysRevLett.84.2322}{{\em Phys.Rev.Lett.}
  {\bfseries 84} (2000) 2322--2325},
  \href{http://arxiv.org/abs/hep-ph/9908309}{{\ttfamily arXiv:hep-ph/9908309
  [hep-ph]}}.

\bibitem{Feng:1999zg}
J.~L. Feng, K.~T. Matchev, and T.~Moroi, ``{Focus points and naturalness in
  supersymmetry},'' \href{http://dx.doi.org/10.1103/PhysRevD.61.075005}{{\em
  Phys.Rev.} {\bfseries D61} (2000) 075005},
  \href{http://arxiv.org/abs/hep-ph/9909334}{{\ttfamily arXiv:hep-ph/9909334
  [hep-ph]}}.

\bibitem{Horton:2009ed}
D.~Horton and G.~Ross, ``{Naturalness and Focus Points with Non-Universal
  Gaugino Masses},''
  \href{http://dx.doi.org/10.1016/j.nuclphysb.2009.12.031}{{\em Nucl.Phys.}
  {\bfseries B830} (2010) 221--247},
\href{http://arxiv.org/abs/0908.0857}{{\ttfamily arXiv:0908.0857 [hep-ph]}}.

\bibitem{Younkin:2012ui}
J.~E. Younkin and S.~P. Martin, ``{Non-universal gaugino masses, the
  supersymmetric little hierarchy problem, and dark matter},''
  \href{http://dx.doi.org/10.1103/PhysRevD.85.055028}{{\em Phys.Rev.}
  {\bfseries D85} (2012) 055028},
\href{http://arxiv.org/abs/1201.2989}{{\ttfamily arXiv:1201.2989 [hep-ph]}}.

\bibitem{Yanagida:2013ah}
T.~T. Yanagida and N.~Yokozaki, ``{Focus Point in Gaugino Mediation:
  Reconsideration of the Fine-tuning Problem},''
\href{http://arxiv.org/abs/1301.1137}{{\ttfamily arXiv:1301.1137 [hep-ph]}}.

\bibitem{Asano:2011kj}
M.~Asano, T.~Moroi, R.~Sato, and T.~T. Yanagida, ``{Focus Point Assisted by
  Right-Handed Neutrinos},''
  \href{http://dx.doi.org/10.1016/j.physletb.2012.01.030}{{\em Phys.Lett.}
  {\bfseries B708} (2012) 107--111},
\href{http://arxiv.org/abs/1111.3506}{{\ttfamily arXiv:1111.3506 [hep-ph]}}.

\bibitem{Feng:2012jfa}
J.~L. Feng and D.~Sanford, ``{A Natural 125 GeV Higgs Boson in the MSSM from
  Focus Point Supersymmetry with A-Terms},''
  \href{http://dx.doi.org/10.1103/PhysRevD.86.055015}{{\em Phys.Rev.}
  {\bfseries D86} (2012) 055015},
\href{http://arxiv.org/abs/1205.2372}{{\ttfamily arXiv:1205.2372 [hep-ph]}}.

\bibitem{Choi:2005hd}
K.~Choi, K.~S. Jeong, T.~Kobayashi, and K.-i. Okumura, ``{Little SUSY hierarchy
  in mixed modulus-anomaly mediation},''
  \href{http://dx.doi.org/10.1016/j.physletb.2005.11.078}{{\em Phys.Lett.}
  {\bfseries B633} (2006) 355--361},
\href{http://arxiv.org/abs/hep-ph/0508029}{{\ttfamily arXiv:hep-ph/0508029
  [hep-ph]}}.

\bibitem{Kitano:2005wc}
R.~Kitano and Y.~Nomura, ``{A Solution to the supersymmetric fine-tuning
  problem within the MSSM},''
  \href{http://dx.doi.org/10.1016/j.physletb.2005.10.003}{{\em Phys.Lett.}
  {\bfseries B631} (2005) 58--67},
\href{http://arxiv.org/abs/hep-ph/0509039}{{\ttfamily arXiv:hep-ph/0509039
  [hep-ph]}}.

\bibitem{Lebedev:2005ge}
O.~Lebedev, H.~P. Nilles, and M.~Ratz, ``{A Note on fine-tuning in mirage
  mediation},''
\href{http://arxiv.org/abs/hep-ph/0511320}{{\ttfamily arXiv:hep-ph/0511320
  [hep-ph]}}.

\bibitem{Feng:2000zu}
J.~L. Feng, K.~T. Matchev, and F.~Wilczek, ``{Prospects for indirect detection
  of neutralino dark matter},''
  \href{http://dx.doi.org/10.1103/PhysRevD.63.045024}{{\em Phys.Rev.}
  {\bfseries D63} (2001) 045024},
\href{http://arxiv.org/abs/astro-ph/0008115}{{\ttfamily arXiv:astro-ph/0008115
  [astro-ph]}}.

\bibitem{Fox:2002bu}
P.~J. Fox, A.~E. Nelson, and N.~Weiner, ``{Dirac gaugino masses and supersoft
  supersymmetry breaking},'' {\em JHEP} {\bfseries 0208} (2002) 035,
\href{http://arxiv.org/abs/hep-ph/0206096}{{\ttfamily arXiv:hep-ph/0206096
  [hep-ph]}}.

\bibitem{Kribs:2007ac}
G.~D. Kribs, E.~Poppitz, and N.~Weiner, ``{Flavor in supersymmetry with an
  extended R-symmetry},''
  \href{http://dx.doi.org/10.1103/PhysRevD.78.055010}{{\em Phys.Rev.}
  {\bfseries D78} (2008) 055010},
\href{http://arxiv.org/abs/0712.2039}{{\ttfamily arXiv:0712.2039 [hep-ph]}}.

\bibitem{Kribs:2012gx}
G.~D. Kribs and A.~Martin, ``{Supersoft Supersymmetry is Super-Safe},''
  \href{http://dx.doi.org/10.1103/PhysRevD.85.115014}{{\em Phys.Rev.}
  {\bfseries D85} (2012) 115014},
\href{http://arxiv.org/abs/1203.4821}{{\ttfamily arXiv:1203.4821 [hep-ph]}}.

\bibitem{Kane:1998im}
G.~L. Kane and S.~King, ``{Naturalness implications of LEP results},''
  \href{http://dx.doi.org/10.1016/S0370-2693(99)00190-2}{{\em Phys.Lett.}
  {\bfseries B451} (1999) 113--122},
\href{http://arxiv.org/abs/hep-ph/9810374}{{\ttfamily arXiv:hep-ph/9810374
  [hep-ph]}}.

\bibitem{BasteroGil:1999gu}
M.~Bastero-Gil, G.~L. Kane, and S.~King, ``{Fine tuning constraints on
  supergravity models},''
  \href{http://dx.doi.org/10.1016/S0370-2693(00)00002-2}{{\em Phys.Lett.}
  {\bfseries B474} (2000) 103--112},
\href{http://arxiv.org/abs/hep-ph/9910506}{{\ttfamily arXiv:hep-ph/9910506
  [hep-ph]}}.

\bibitem{Ellis:1984bm}
J.~R. Ellis, C.~Kounnas, and D.~V. Nanopoulos, ``{No Scale Supersymmetric
  Guts},''
\href{http://dx.doi.org/10.1016/0550-3213(84)90555-8}{{\em Nucl.Phys.}
  {\bfseries B247} (1984) 373--395}.

\bibitem{Anderson:1996bg}
G.~Anderson, C.~Chen, J.~Gunion, J.~D. Lykken, T.~Moroi, {\em et al.},
  ``{Motivations for and implications of nonuniversal GUT scale boundary
  conditions for soft SUSY breaking parameters},'' {\em eConf} {\bfseries
  C960625} (1996) SUP107,
\href{http://arxiv.org/abs/hep-ph/9609457}{{\ttfamily arXiv:hep-ph/9609457
  [hep-ph]}}.

\bibitem{Baer:2007uz}
H.~Baer, A.~Box, E.-K. Park, and X.~Tata, ``{Implications of compressed
  supersymmetry for collider and dark matter searches},''
  \href{http://dx.doi.org/10.1088/1126-6708/2007/08/060}{{\em JHEP} {\bfseries
  0708} (2007) 060},
\href{http://arxiv.org/abs/0707.0618}{{\ttfamily arXiv:0707.0618 [hep-ph]}}.

\bibitem{Martin:2008aw}
S.~P. Martin, ``{Exploring compressed supersymmetry with same-sign top quarks
  at the Large Hadron Collider},''
  \href{http://dx.doi.org/10.1103/PhysRevD.78.055019}{{\em Phys.Rev.}
  {\bfseries D78} (2008) 055019},
\href{http://arxiv.org/abs/0807.2820}{{\ttfamily arXiv:0807.2820 [hep-ph]}}.

\bibitem{LeCompte:2011cn}
T.~J. LeCompte and S.~P. Martin, ``{Large Hadron Collider reach for
  supersymmetric models with compressed mass spectra},''
  \href{http://dx.doi.org/10.1103/PhysRevD.84.015004}{{\em Phys.Rev.}
  {\bfseries D84} (2011) 015004},
\href{http://arxiv.org/abs/1105.4304}{{\ttfamily arXiv:1105.4304 [hep-ph]}}.

\bibitem{LeCompte:2011fh}
T.~J. LeCompte and S.~P. Martin, ``{Compressed supersymmetry after 1/fb at the
  Large Hadron Collider},''
  \href{http://dx.doi.org/10.1103/PhysRevD.85.035023}{{\em Phys.Rev.}
  {\bfseries D85} (2012) 035023},
\href{http://arxiv.org/abs/1111.6897}{{\ttfamily arXiv:1111.6897 [hep-ph]}}.

\bibitem{Rolbiecki:2012gn}
K.~Rolbiecki and K.~Sakurai, ``{Constraining compressed supersymmetry using
  leptonic signatures},'' \href{http://dx.doi.org/10.1007/JHEP10(2012)071}{{\em
  JHEP} {\bfseries 1210} (2012) 071},
\href{http://arxiv.org/abs/1206.6767}{{\ttfamily arXiv:1206.6767 [hep-ph]}}.

\bibitem{Belanger:2012mk}
G.~Belanger, M.~Heikinheimo, and V.~Sanz, ``{Model-Independent Bounds on
  Squarks from Monophoton Searches},''
  \href{http://dx.doi.org/10.1007/JHEP08(2012)151}{{\em JHEP} {\bfseries 1208}
  (2012) 151},
\href{http://arxiv.org/abs/1205.1463}{{\ttfamily arXiv:1205.1463 [hep-ph]}}.

\bibitem{Dreiner:2012gx}
H.~K. Dreiner, M.~Kramer, and J.~Tattersall, ``{How low can SUSY go? Matching,
  monojets and compressed spectra},''
  \href{http://dx.doi.org/10.1209/0295-5075/99/61001}{{\em Europhys.Lett.}
  {\bfseries 99} (2012) 61001},
\href{http://arxiv.org/abs/1207.1613}{{\ttfamily arXiv:1207.1613 [hep-ph]}}.

\bibitem{Martin:2007hn}
S.~P. Martin, ``{The Top squark-mediated annihilation scenario and direct
  detection of dark matter in compressed supersymmetry},''
  \href{http://dx.doi.org/10.1103/PhysRevD.76.095005}{{\em Phys.Rev.}
  {\bfseries D76} (2007) 095005},
\href{http://arxiv.org/abs/0707.2812}{{\ttfamily arXiv:0707.2812 [hep-ph]}}.

\bibitem{Strassler:2006im}
M.~J. Strassler and K.~M. Zurek, ``{Echoes of a hidden valley at hadron
  colliders},'' \href{http://dx.doi.org/10.1016/j.physletb.2007.06.055}{{\em
  Phys.Lett.} {\bfseries B651} (2007) 374--379},
\href{http://arxiv.org/abs/hep-ph/0604261}{{\ttfamily arXiv:hep-ph/0604261
  [hep-ph]}}.

\bibitem{Strassler:2006qa}
M.~J. Strassler, ``{Possible effects of a hidden valley on supersymmetric
  phenomenology},''
\href{http://arxiv.org/abs/hep-ph/0607160}{{\ttfamily arXiv:hep-ph/0607160
  [hep-ph]}}.

\bibitem{Fan:2011yu}
J.~Fan, M.~Reece, and J.~T. Ruderman, ``{Stealth Supersymmetry},''
  \href{http://dx.doi.org/10.1007/JHEP11(2011)012}{{\em JHEP} {\bfseries 1111}
  (2011) 012},
\href{http://arxiv.org/abs/1105.5135}{{\ttfamily arXiv:1105.5135 [hep-ph]}}.

\bibitem{Hall:1983id}
L.~J. Hall and M.~Suzuki, ``{Explicit R-Parity Breaking in Supersymmetric
  Models},''
\href{http://dx.doi.org/10.1016/0550-3213(84)90513-3}{{\em Nucl.Phys.}
  {\bfseries B231} (1984) 419}.

\bibitem{Ellis:1984gi}
J.~R. Ellis, G.~Gelmini, C.~Jarlskog, G.~G. Ross, and J.~Valle,
  ``{Phenomenology of Supersymmetry with Broken R-Parity},''
\href{http://dx.doi.org/10.1016/0370-2693(85)90157-1}{{\em Phys.Lett.}
  {\bfseries B150} (1985) 142}.

\bibitem{Barger:1989rk}
V.~D. Barger, G.~Giudice, and T.~Han, ``{Some New Aspects of Supersymmetry
  R-Parity Violating Interactions},''
\href{http://dx.doi.org/10.1103/PhysRevD.40.2987}{{\em Phys.Rev.} {\bfseries
  D40} (1989) 2987}.

\bibitem{Allanach:1999ic}
B.~Allanach, A.~Dedes, and H.~K. Dreiner, ``{Bounds on R-parity violating
  couplings at the weak scale and at the GUT scale},''
  \href{http://dx.doi.org/10.1103/PhysRevD.60.075014}{{\em Phys.Rev.}
  {\bfseries D60} (1999) 075014},
\href{http://arxiv.org/abs/hep-ph/9906209}{{\ttfamily arXiv:hep-ph/9906209
  [hep-ph]}}.

\bibitem{Barbier:2004ez}
R.~Barbier, C.~Berat, M.~Besancon, M.~Chemtob, A.~Deandrea, {\em et al.},
  ``{R-parity violating supersymmetry},''
  \href{http://dx.doi.org/10.1016/j.physrep.2005.08.006}{{\em Phys.Rept.}
  {\bfseries 420} (2005) 1--202},
\href{http://arxiv.org/abs/hep-ph/0406039}{{\ttfamily arXiv:hep-ph/0406039
  [hep-ph]}}.

\bibitem{Csaki:2011ge}
C.~Csaki, Y.~Grossman, and B.~Heidenreich, ``{MFV SUSY: A Natural Theory for
  R-Parity Violation},''
  \href{http://dx.doi.org/10.1103/PhysRevD.85.095009}{{\em Phys.Rev.}
  {\bfseries D85} (2012) 095009},
\href{http://arxiv.org/abs/1111.1239}{{\ttfamily arXiv:1111.1239 [hep-ph]}}.

\bibitem{Krnjaic:2012aj}
G.~Krnjaic and D.~Stolarski, ``{Gauging the Way to MFV},''
  \href{http://dx.doi.org/10.1007/JHEP04(2013)064}{{\em JHEP} {\bfseries
  JHEP04} (2013) 064},
\href{http://arxiv.org/abs/1212.4860}{{\ttfamily arXiv:1212.4860 [hep-ph]}}.

\bibitem{Bhattacherjee:2013gr}
B.~Bhattacherjee, J.~L. Evans, M.~Ibe, S.~Matsumoto, and T.~T. Yanagida,
  ``{Natural SUSY's Last Hope: R-parity Violation via UDD Operators},''
\href{http://arxiv.org/abs/1301.2336}{{\ttfamily arXiv:1301.2336 [hep-ph]}}.

\bibitem{Franceschini:2013ne}
R.~Franceschini and R.~Mohapatra, ``{New Patterns of Natural R-Parity Violation
  with Supersymmetric Gauged Flavor},''
  \href{http://dx.doi.org/10.1007/JHEP04(2013)098}{{\em JHEP} {\bfseries 1304}
  (2013) 098},
\href{http://arxiv.org/abs/1301.3637}{{\ttfamily arXiv:1301.3637 [hep-ph]}}.

\bibitem{Csaki:2013we}
C.~Csaki and B.~Heidenreich, ``{A Complete Model for R-parity Violation},''
\href{http://arxiv.org/abs/1302.0004}{{\ttfamily arXiv:1302.0004 [hep-ph]}}.

\end{thebibliography}
\end{document}

